\documentclass[preprint,3p,number]{elsarticle}

\usepackage{graphicx}

\newcommand{\subfloat}[2][(a)]{\begin{minipage}{8.2cm}#2\begin{center}#1\end{center}\end{minipage}}

\usepackage{amsmath,amssymb,wasysym}
\biboptions{sort&compress}

\usepackage{hyperref}
\usepackage{array} 

\usepackage{lineno}

\usepackage{color}

\newcommand{\picu}{}
\newcommand{\degree}{\ensuremath{^\circ}}

\newcommand{\geant}{\textsc{Geant4}}

\def\etal{{\it et al.}}    
\def\ie{{\it i.e.}}
\def\vs{{\it vs.}}
\def\etc{{\it etc.}}
\def\eg{{\it e.g.}}

\def\~{{$\tilde{\phantom{a}}$}}
\def\babar{{\sc BaBar}}
\def\lsim{\mathrel {\vcenter {\baselineskip 0pt \kern 0pt
    \hbox{$<$} \kern 0pt \hbox{$\sim$} }}}
\def\gsim{\mathrel {\vcenter {\baselineskip 0pt \kern 0pt
    \hbox{$>$} \kern 0pt \hbox{$\sim$} }}}
\def\ave#1{\left\langle#1\right\rangle} 

\begin{document}
\begin{frontmatter}
  \title{Undulator-Based Production of Polarized
    Positrons%
\begin{picture}(0,0)
\put(-6.7,1.5){{\normalsize SLAC-PUB-13605, DESY
        09-061, IPPP/09/38, DCPT/09/76}}
\end{picture}%
\tnoteref{thtitel}}
  \tnotetext[thtitel]{This work was supported in part by DOE contract No. DE-AC03-76SF00515, DOE grants and Nos. DE-FG05-91ER40627, DE-FG02-91ER40671, DE-FG02-03ER41283 and DE-FG02-04ER41353 by NSF grant No. PHY-0202078 (USA), by European Commission contract No. RIDS-011899 (Germany), by the STFC (United Kingdom), and by ISF contract No. 342/05 (Israel).}

\author[10]{G.~Alexander} 
\author[6]{J.~Barley} 
\author[8]{Y.~Batygin} 
\author[7]{S.~Berridge} 
\author[8]{V.~Bharadwaj} 
\author[8]{G.~Bower} 
\author[7]{W.~Bugg} 
\author[8]{F.-J.~Decker} 
\author[2]{R.~Dollan} 
\author[7]{Y.~Efremenko} 
\author[5]{K.~Fl\"ottmann} 
\author[5,11]{V.~Gharibyan} 
\author[8]{C.~Hast} 
\author[8]{R.~Iverson} 
\author[2]{H.~Kolanoski} 
\author[1]{J.W.~Kovermann} 
\author[12]{K.~Laihem} 
\author[2]{T.~Lohse} 
\author[9]{K.T.~McDonald} 
\author[6]{A.A.~Mikhailichenko} 
\author[4]{G.A.~Moortgat-Pick} 
\author[5]{P.~Pahl} 
\author[8]{R.~Pitthan} 
\author[5]{R.~P\"oschl} 
\author[10]{E.~Reinherz-Aronis} 
\author[12]{S.~Riemann} 
\author[12]{A.~Sch\"alicke\corref{cor1}} 
\author[5]{K.P.~Sch\"uler} 
\author[2]{T.~Schweizer} 
\author[3]{D.~Scott} 
\author[8]{J.C.~Sheppard} 
\author[1]{A.~Stahl} 
\author[8]{Z.~Szalata} 
\author[8]{D.R.~Walz} 
\author[8]{A.~Weidemann} 

\address[1]{RWTH Aachen, D-52056 Aachen, Germany} 
\address[2]{Humboldt-Universit{\"a}t zu Berlin, 12489 Berlin, Germany} 
\address[3]{STFC Daresbury Laboratory and The Cockcroft Institute, Daresbury, Warrington, Cheshire, WA5 0HB, UK} 
\address[4]{University of Durham, Durham, DH1 3LE, UK} 
\address[5]{DESY, D-22607 Hamburg, Germany} 
\address[6]{Cornell University, Ithaca, NY 14853, USA} 
\address[7]{University of Tennessee, Knoxville, TN 37996, USA} 
\address[8]{SLAC, Menlo Park, CA 94025, USA} 
\address[9]{Princeton University, Princeton, NJ 08544, USA} 
\address[{10}]{Tel-Aviv University, Tel Aviv 69978, Israel} 
\address[{11}]{YerPhI, Yerevan 375036, Armenia}  
\address[{12}]{DESY, D-15738 Zeuthen, Germany} 
\cortext[cor1]{Corresponding author, Tel.: +49\,33762\,77422; fax:
  +49\,33762\,77330; E-mail address: andreas.schaelicke@desy.de}

  \setlength{\unitlength}{1cm}




  \begin{abstract}
    Full exploitation of the physics potential of a future International Linear Collider will require the use of polarized electron and positron beams. 
Experiment E166 at the Stanford Linear Accelerator Center (SLAC) has demonstrated a scheme in which an electron beam passes through a helical undulator to generate photons (whose first-harmonic spectrum extended to 7.9\,MeV) with circular polarization, which are then converted in a thin target to generate longitudinally polarized positrons and electrons. 
The experiment was carried out with a one-meter-long, 400-period, pulsed helical undulator in the Final Focus Test Beam (FFTB) operated at 46.6\,GeV. 
Measurements of the positron polarization have been performed at five positron energies from 4.5 to 7.5\,MeV. 
In addition, the electron polarization has been determined at 6.7\,MeV, and the effect of operating the undulator with a ferrofluid was also investigated. 
To compare the measurements with expectations, detailed simulations were made with an upgraded version of \geant\ that includes the dominant polarization-dependent interactions of electrons, positrons, and photons with matter. 
 The measurements agree with calculations, corresponding to 80\,\%
polarization for positrons near 6\,MeV and 90\,\% for electrons near 7\,MeV. 
  \end{abstract}
  

\end{frontmatter}
\newpage
{\footnotesize
 \setcounter{tocdepth}{2}
 \tableofcontents
}

\section{Introduction}

Full exploitation of the physics potential of a future linear collider (such as the International Linear Collider, ILC\cite{ILCRDR} and the Compact Linear Collider, CLIC\cite{CLIC}) will require the development of polarized positron beams\cite{ilcphysics}. 
High polarization of both electron and positron beams is optimal for addressing both expected and unforeseen challenges in searches for new physics: fixing the chirality of the couplings of the interacting particles, maximizing the precision of measurement of polarization-dependent observables, and providing a powerful tool for analyzing signals of new physics in a model-independent way. 

\begin{figure*}[h]
   \begin{center}
  \includegraphics[width=0.95\textwidth]{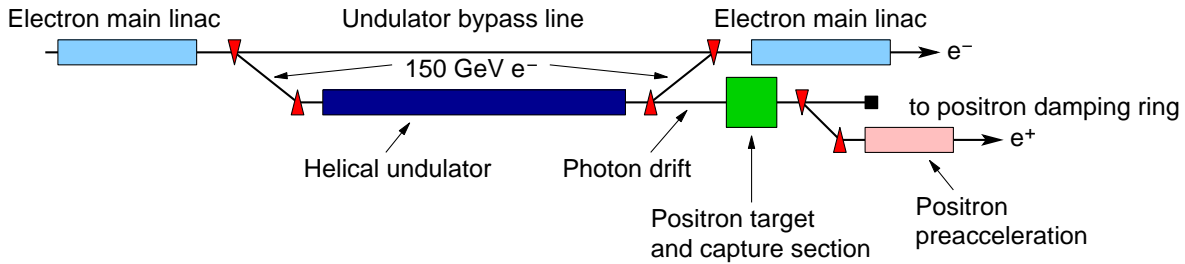}   
  \caption{A concept for undulator-based production of polarized positrons at a linear collider.}
  \label{concept}
   \end{center}
\end{figure*}

Polarized positrons can be produced via the pair-production process initiated by circularly polarized photons\cite{Olsen59},
which will permit much higher intensity beams of polarized positrons than could be obtained from decays of
radioactive nuclei\cite{Zitzewitz}. 
In the proposed scheme of Balakin and Mikhailichenko\cite{balakin} a helical undulator\cite{Wingerson} is employed to generate photons of several MeV with circular polarization\cite{Alferov76}. 
A possible implementation of this scheme at a linear collider is sketched in Fig.~\ref{concept}, 
in which an electron
beam of $\approx 150$\,GeV energy passes through a helical undulator to
produce a beam of circularly polarized photons of energies up to 10\,MeV.  These
MeV photons are incident on a thin target, in which there is good polarization
transfer to the positrons (and electrons) that are pair-produced.  The low-energy
positrons are collected for injection into one arm of the linear collider, while
the high-energy electron beam (which is largely undisturbed by its passage through
the undulator) is directed into the other arm.

In an alternative scheme the circularly polarized photons are produced by laser backscattering off an electron beam\cite{Bessonov,omori:114801}, as has been proposed for the Japanese Linear Collider Project (JLC)\cite{Hirose,Okugi}.

\begin{figure*}
  \centering
   \includegraphics[width=0.95\textwidth]{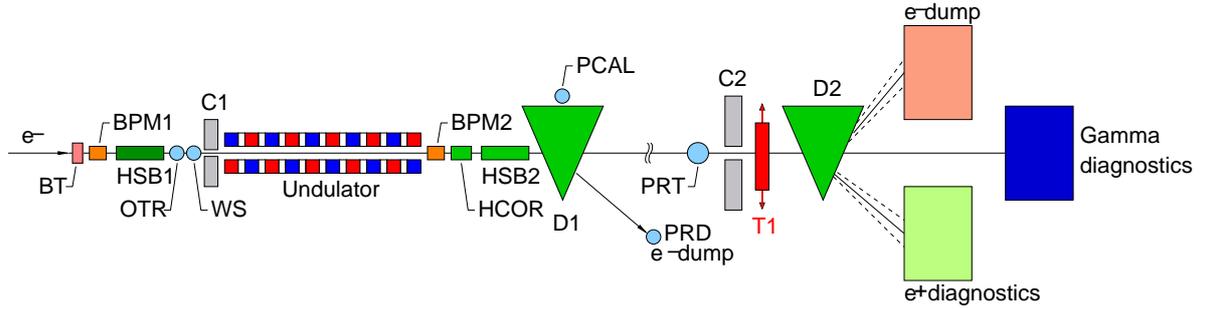}    
  \caption[Conceptual layout of the experiment.]{Conceptual layout (not to scale) of the experiment to demonstrate the production of polarized positrons in the SLAC FFTB. 
The electrons enter from the left and are dumped using magnet {\sf D1} after traversing the undulator. 
The positron-production target {\sf T1} and the positron and photon diagnostics are located 35\,m downstream of the undulator. 
{\sf BPM} = beam-position monitor; 
{\sf HSB} = ``hard'' soft bend; 
{\sf OTR} = optical-transition-radiation beam-profile monitor; 
{\sf BT} = beam-current toroid; 
{\sf WS} = wire scanner; 
{\sf C} = aperture-limiting collimators; 
{\sf HCOR} = horizontal-steering magnet; 
{\sf D1} = FFTB primary-beam-dump bend-magnet string; 
{\sf PCAL} = positron calorimeter;
{\sf PRD} = dumpline beam-profile monitor; 
{\sf PRT} = production-target beam-profile monitor; 
{\sf D2} = positron spectrometer.}
  \label{fig:explayc2k}
\end{figure*}

\begin{figure}[ht]
\begin{center}
\includegraphics*[width=8cm]{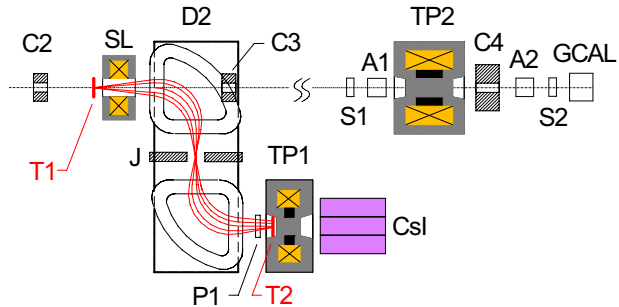}    
\caption[Schematic of the photon and positron diagnostics.]{\label{prlfig2} 
Schematic of the photon and positron diagnostics. 
{\sf A1}, {\sf A2} = aerogel Cherenkov detectors, 
{\sf C2}--{\sf C4} = collimators,
{\sf D2} = dipole spectrometer magnet,
{\sf CsI} = $3 \times 3$ array of CsI crystals,
{\sf GCAL} = Si-W calorimeter,
{\sf J} = movable W jaws,
{\sf P1}, {\sf S1}, {\sf S2}= Si-diode detectors,
{\sf SL} = solenoid lens,
{\sf T1} = positron-production target, 
{\sf T2} = reconversion target, 
{\sf TP1} = positron-transmission-polarimeter solenoid,
{\sf TP2} = photon-transmission-polarimeter solenoid.
The detectors were encased in lead and tungsten shielding (not shown).
}
\end{center}
\end{figure}

Experiment E166\cite{Alexander:2003fh,prl} was performed to demonstrate that undulator-based production of
polarized positrons can produce beams of sufficient quality for use in
future linear colliders. 
Data were collected during two run periods in June and September 2005.
A conceptual layout of the experiment is shown in Fig.~\ref{fig:explayc2k} and details of the
photon and positron diagnostics are given in Fig.~\ref{prlfig2}.

The low-emittance, 46.6\,GeV electron beam of the Final Focus Test Beam (FFTB)\cite{fftb}
 at the Stanford Linear Accelerator Center (SLAC)\cite{slac_linac}
 was passed through the undulator at 10\,Hz to produce circularly polarized photons
whose energy spectrum had its first-harmonic cutoff near 7.9\,MeV.
After the undulator, the beam electrons were deflected downwards to a dump by magnet {\sf D1}.
Positrons (and electrons) with energies in the range of a few MeV were produced by conversion of
the undulator photons in a 0.8-mm-thick tungsten-alloy target {\sf T1}.
 
The helical undulator \cite{CBN05-2} was one meter long with a period of 2.54\,mm and a strength parameter $K \approx 0.17$, as defined in Eq.~(\ref{eq_u2}).
The helical coil was made by winding copper wires on a stainless-steel vacuum chamber with an aperture of 0.9\,mm.
The undulator was tested at currents up to 2300\,A at 30\,Hz repetition rate. 
During the data runs, it was operated at 10\,Hz and 2300\,A and delivered more than $4 \cdot 10^7$ pulses without a single failure.
The transformer oil in the undulator housing was replaced with a ferrofluid during part of the experiment, which yielded a 11\% enhancement of the photon flux\cite{Ferrofluid}.

The undulator performance was characterized by measuring the total photon flux as a function of excitation current and the transmission of the photon beam through a 15-cm-long iron-core solenoid (transmission-polarimeter magnet {\sf TP2}). 
A set of Si-W detectors {\sf S1--2} and {\sf GCAL} and aerogel Cherenkov detectors {\sf A1-2} was used to measure the incident and transmitted photon flux and energy. 
The transmission asymmetry with respect to reversal of the polarity of magnet {\sf TP2} was affected by the spectral-transmission properties of magnetized iron and the polarization-dependent term of Compton scattering in the iron. 
These processes were modeled using a version of \textsc{Geant3} that was modified to include the spin-dependent scattering effects. 
The measured fluxes and asymmetries agreed well with expectations based on MERMAID\cite{ref:mermaid}
calculations for the undulator strength and on the theoretical formulae for helical-undulator radiation.

The positrons created in target {\sf T1} were transported by a focusing solenoid {\sf SL} to the entrance of a spectrometer {\sf D2} that separated the positrons from the incident photon beam and selected a narrow band of positron energies. 
The spectrometer, whose bend was in the horizontal plane, consisted of a pair of dipole magnets and adjustable
tungsten-alloy jaws.
The polarization of the positrons was measured by photon transmission polarimetry\cite{Schopper}
after the positrons had produced photons in a thin tungsten ``reconversion target'' {\sf T2}.
These photons were passed through a 7.5-cm-long magnetized iron cylinder {\sf TP1}. 
The photons leaving the analyzer magnet were much less collimated than in the case of the undulator photons. 
Therefore, a wider angular acceptance was obtained by use of a 3$\times$3 array of {\sf CsI} crystals with a total cross section of about 180\,mm$\times$180\,mm. 
To simulate the positron polarimeter, \geant\cite{Agostinelli:2002hh,Allison:2006ve} was extended to include the relevant spin-dependent effects. 
These polarization extensions\cite{g4pol,Geant4:PhysRefMan} are part of \geant\ from v8.2 onwards.

Positron polarizations were measured at five energy settings of spectrometer {\sf D2}. 
In addition, an electron-polarization measurement was made at a single energy setting by reversing the polarity of the spectrometer. 
Over the measured energy range of 4--8\,MeV, the positron (and electron) polarization was about 80\,\% with a relative measurement error of about 10\,\% to 15\,\%, dominated by the systematic uncertainties\cite{prl}. 
The measured results agree well with expectations from detailed simulation of all aspects of the experiment. 

The remainder of this paper describes in detail the experimental technique, data analysis and simulation, and the results of measurements of flux and transmission asymmetry of the undulator photons and of polarization of the positrons created from these photons.

\section{Experimental Method}\label{sec:methods}

This section summarizes the principles of the methods used in this experiment
for production of polarized photons and positrons, and for measurement of the longitudinal polarization of these particles.

\subsection{Production of Polarized Photons in a Helical Undulator}\label{sec_undulator_physics}

Polarized positrons were produced in the present experiment
by conversion in a thin target of circularly polarized photons with energy of a few MeV.  
The photons were produced by scattering of virtual photons of a helical undulator 
\cite{Wingerson} with period $\lambda_u$
off an electron beam of energy $E_e = \gamma m c^2$,
where $m$ is the mass of an electron, $c$ is the speed of light, and the electron beam
is coaxial with the undulator. 
The highest energy photons take on the polarization of the undulator field, so that a helical undulator leads to circularly polarized photons.

The intensity of undulator photons depends on the intensity of the virtual photons of the undulator, and hence on the square of its magnetic field strength.  
It is conventional to measure the field strength of an undulator in terms of a dimensionless parameter $K$  defined as,
\begin{linenomath}
  \begin{equation}
    K = \frac{e B_0 \lambda_{\rm u} }{2 \pi \,m c^2}
    \cong 0.0934\, (B_0 / 1\,{\rm T}) (\lambda _u / 1\,{\rm mm}),
    \label{eq_u2}
  \end {equation}
\end{linenomath}
in which $e$ is the magnitude of the charge of an electron,
and B$_0$ is the magnetic field on the axis of the undulator, which field is constant in
magnitude while rotating through $360^\circ$ during each period $\lambda_{\rm u}$. 
The value of $K$ in the present experiment was small, about 0.17, because of practical limitations
to the current in the (pulsed) undulator.

For small $K$-values, the number of photons $d N_\gamma / dL$ emitted per meter of an undulator and per beam electron is
\begin{linenomath}
  \begin{equation}
    \frac{{dN_\gamma  }}{{dL}}\; \approx \;\frac{4}{3}\frac{{\pi \alpha
      }} {{\lambda _u}} K^2 \; 
    \cong \;\frac{{30.6}}{{\lambda _u / 1\,{\rm mm} }} K^2
    \;\; {\rm photons/m/}e^- ,
    \label{eq_Ngamma}
  \end {equation} 
\end{linenomath}
where $\alpha$ is the fine-structure constant.
The photon-number spectrum is relatively flat up to the maximum energy $E_{1}$ for scattering of
a single virtual photon (dipole radiation of a beam electron), 
\begin{linenomath}
\begin{align}
  E_{1} &=  \frac{2 E_e^2 \lambda_{\rm C} / \lambda_{\rm u} m c^2}{1 + K^2 + 2 \gamma \lambda_{\rm C} / \lambda_{\rm u}} 
\approx \frac{2 E_e^2 \lambda_{\rm C} / \lambda_{\rm u} m c^2}{ 1 + K^2} \nonumber\\
  &\cong  23.7\, \mbox{MeV} \frac{ \left(E_e/ 50\,\mbox{GeV} \right)^2 }
                                   { (\lambda_{\rm u} / 1\,\mbox{mm})  (1 + K^2)}\, ,
  \label{p2}
\end{align}
\end{linenomath}
where $\lambda_{\rm C} = h / m c = 2.4 \cdot 10^{-12}$\,m is the Compton wavelength of the electron. 
The kinematic relation between energy and angle of emission of a real photon due to the
scattering of $n$ virtual photons ($n^{th}$-order-multipole radiation) is, for small angles
$\theta$ with respect to the electron beam,
\begin{linenomath}
  \begin{equation}
    E_\gamma(n,\theta) = \frac{n E_{1} }{
      1 + (\gamma \theta)^2 / (1 + K^2)}\;\; .
    \label{p3}
  \end{equation}
\end{linenomath}
As seen from Eq.~(\ref{p3}), the upper half of the energy spectrum at any order $n$
is emitted into a cone of angle $\theta = \sqrt{1 + K^2} / \gamma$.
The emission of photons due to higher-multipole radiation (with correspondingly higher energies) is suppressed for low values of $K$. 

\begin{figure*}
  \subfloat[(a)]{\includegraphics[width=8cm]{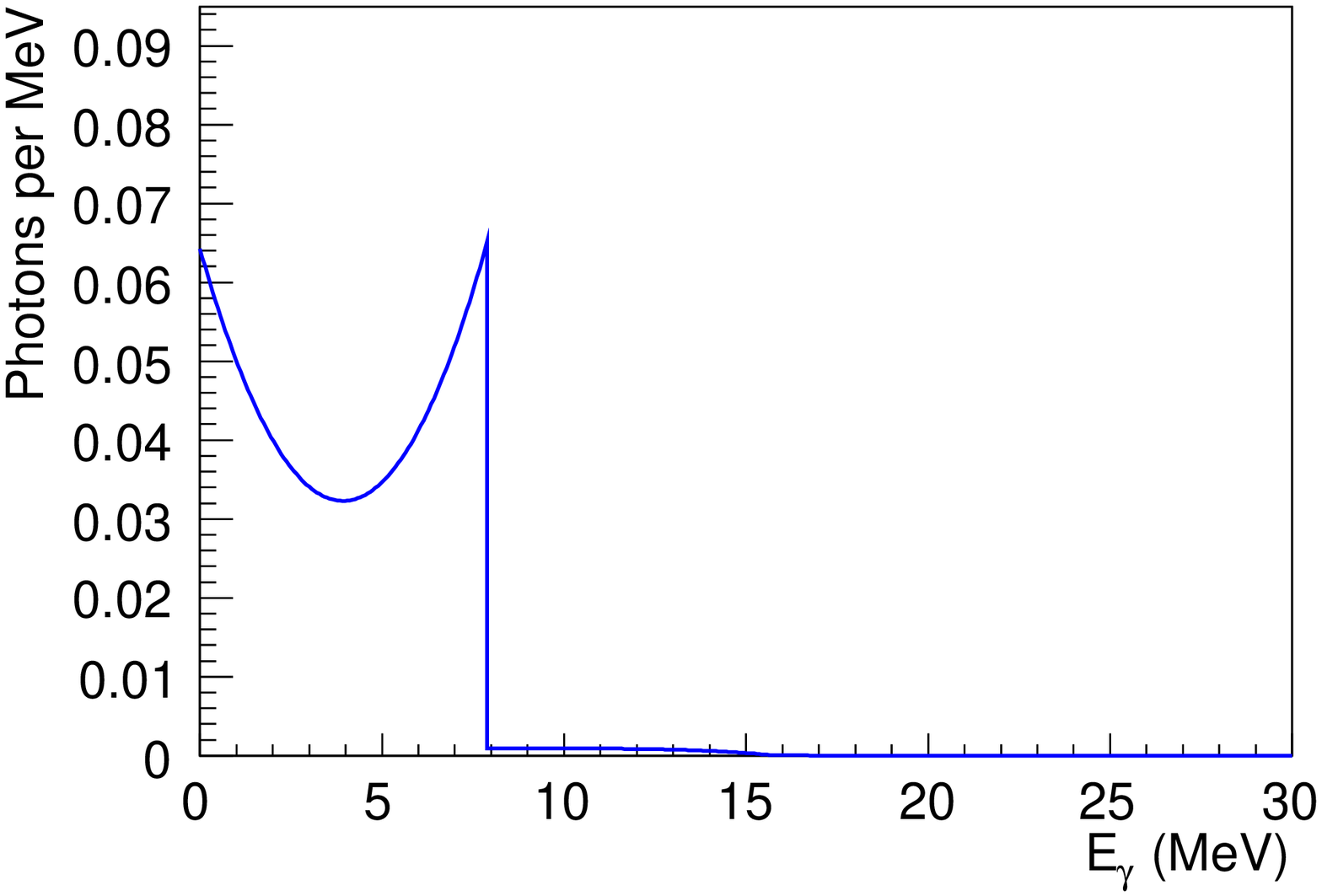}}   
  \hfill
  \subfloat[(b)]{\includegraphics[width=8cm]{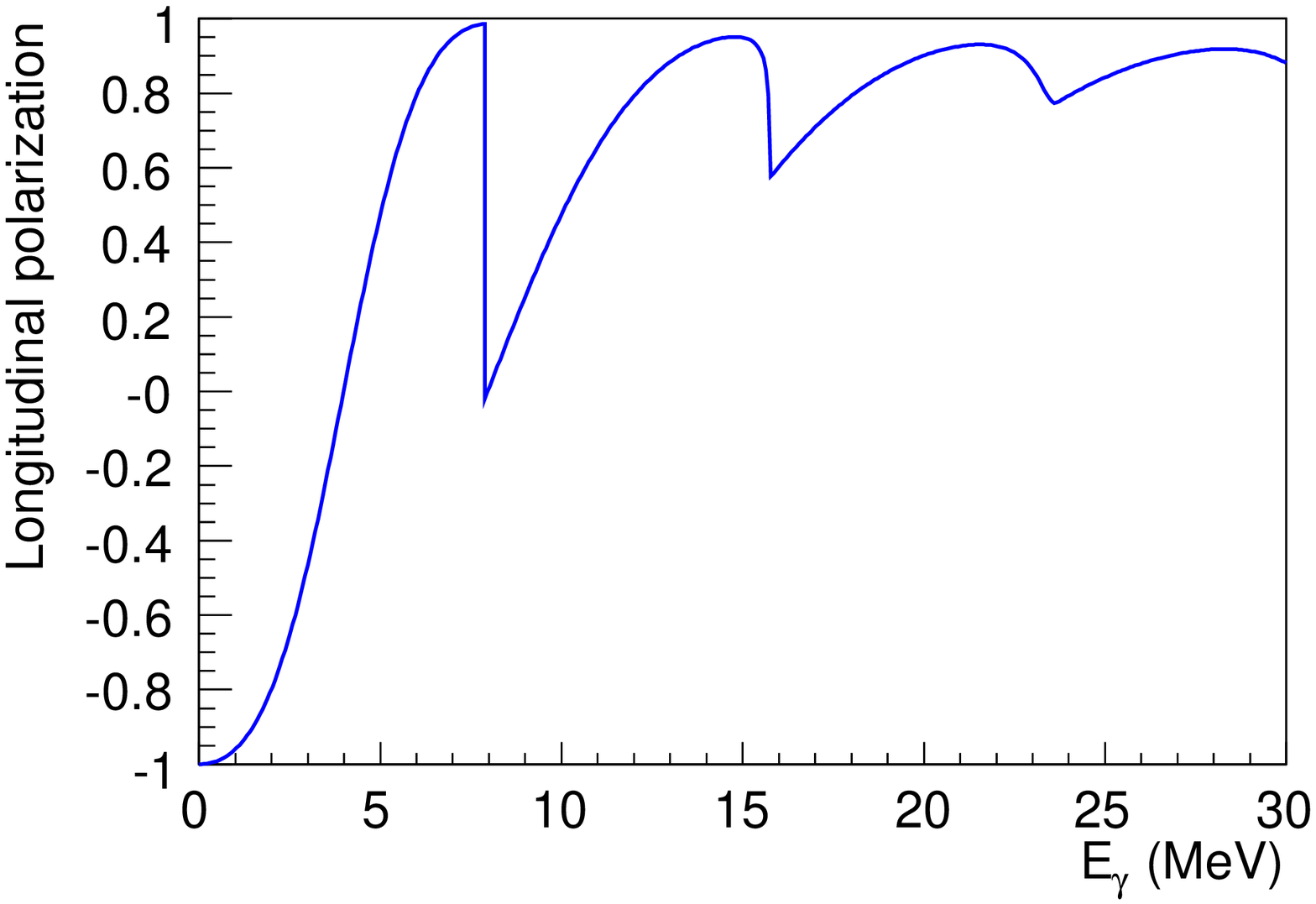}}   
  \caption[The number spectrum  $N_\gamma(E)$ and the longitudinal polarization $P_\gamma(E)$ of the undulator radiation.]{(a) The number spectrum $N_\gamma(E)$ of undulator radiation as a function of photon energy $E$, integrated over angle, for electron energy $E_e = 46.6$\,GeV, undulator period $\lambda_{\rm u} = 2.54$\,mm, and undulator-strength parameter $K = 0.17$. 
The peak energy $E_{1}$ of the first-harmonic (dipole) radiation is 7.89\,MeV. 
(b) The longitudinal polarization $P_\gamma(E)$ of the undulator
radiation as a function of photon energy for an undulator with a right-handed helical winding.}
  \label{fig:Photon}
\end{figure*}

The photon-number spectrum $N_\gamma(E)$ is illustrated in Fig.\:\ref{fig:Photon}(a)
for the experimental parameters: $E_e = 46.6$\,GeV, $\lambda_{\rm u} = 2.54$\,mm, $K = 0.17$, and
$E_1 = 7.89$\,MeV.  The number of photons generated is 0.35 per beam electron.

For the undulator photons produced at $\theta = 0$ the longitudinal
polarization is maximal, $P_\gamma=+1$ 
(for an undulator with a right-handed helical winding),
but falls off for larger angles (which correspond to lower energies).  
This behavior is illustrated in Fig.\:\ref{fig:Photon}(b) for the experimental parameters. 
The polarization of higher-harmonic radiation approaches unity at the corresponding higher cutoff energies, but the rates there are very low.

The average polarization of all undulator photons is nearly zero, but since higher-energy photons have higher polarization, the energy-weighted average polarization is 50\,\%.

Detailed descriptions of helical-undulator radiation can be found in\cite{Alferov76,Kincaid77,Blewett77,Mikh02,Sheplcxx}.

\subsection{Generation of Polarized Positrons}

\begin{figure}
  \begin{center}
    \includegraphics[width=7.5cm]{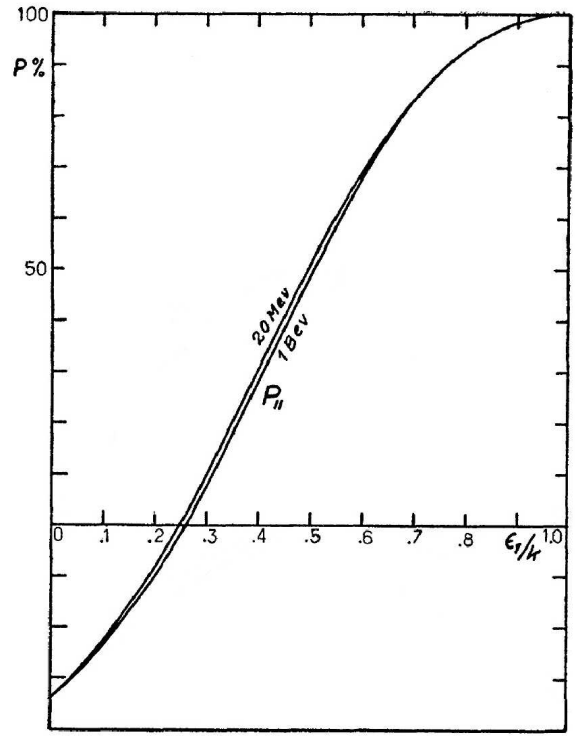}  
    \caption[Longitudinal polarization of positrons produced in pair
    production from circularly polarized photons.]{ Longitudinal polarization of positrons (or electrons) produced by conversion of monochromatic circularly polarized photons in an infinitely thin target, as a function of the ratio $E_{e^+}/E_{\gamma}$ of positron to photon energies; from \cite{Olsen59}.}
    \label{fig:Olsen}
  \end{center}
\end{figure}

When a circularly polarized photon creates an electron-positron pair in a thin target, the polarization state of the photon is transferred to the outgoing leptons, as discussed by Olsen and Maximon in 1959\cite{Olsen59}. 
Positrons with an energy close to the energy of the incoming photons are 100\,\% longitudinally polarized, while positrons with a lower energy have a lower longitudinal polarization (see Fig.\:\ref{fig:Olsen}).
The sign of the positron polarization is opposite to that of the photon for positron energies less than
25\,\% of the photon energy.

\begin{figure} 
  \begin{center}
    \includegraphics[width=8cm]{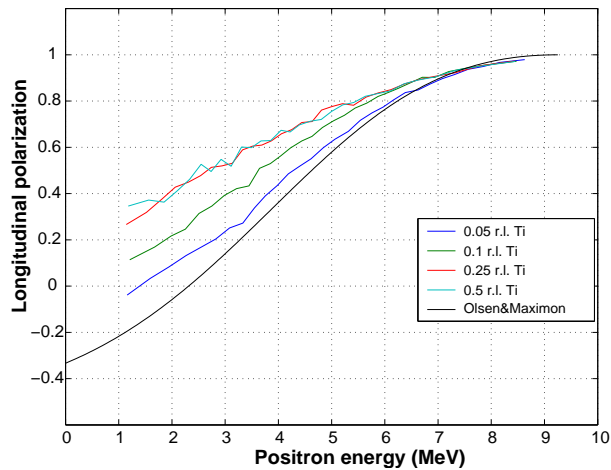}    
    \caption[Longitudinal polarization of positrons produced by conversion photons in  targets of various thickness.]{Longitudinal polarization of positrons  produced by conversion of 10\,MeV circularly polarized photons in  targets of various thickness in radiation lengths, as a function of the positron energy.}
    \label{fig:Olsen2}
  \end{center}
\end{figure}

The probability for the production of positrons is roughly independent of the fractional energy $E_{e^+}/E_{\gamma}$ in the pair-production process, so that positrons with all energies up to the photon energy are produced (with initial polarization as shown in Fig.\:\ref{fig:Olsen}). 
However, even in a thin target, low-energy positrons are stopped due to the ionization loss (which rises sharply for energies below 1\,MeV), while high-energy positrons lose a fraction of their energy due to Bremsstrahlung.  
The energy loss by Bremsstrahlung is accompanied by a slight loss of polarization; however, the energy loss is more significant than the polarization loss. 
As a result, the low-energy portion of the positron spectrum is repopulated with positrons from the higher-energy portion, and the polarization of positrons of a given energy is higher in targets of up to $\approx 0.5$ radiation length than in an infinitely thin target\cite{floettmann}, as shown in Fig.\:\ref{fig:Olsen2}.

For targets thicker than about 0.5 radiation length the polarization decreases again. Hence, positrons are nearly unpolarized in a conventional thick-target positron source even if the incoming photons are polarized.

The basic processes of polarization transfer in electromagnetic
cascades (showers) are well known, but detailed understanding of the
interplay of all processes in a shower is best obtained via numerical
simulation with a Monte Carlo code. 
At the beginning of the experiment there was no code available that included all relevant processes.
Therefore, a major effort was expended to implement polarization effects into the \geant\ code\cite{g4pol}. 
Details of the resulting simulations are reported in Sec.~\ref{sec_simulation}.

\subsection{Photon Polarimetry}\label{sec_phot_pol_meth}

\begin{figure}
  \begin{center}
    \includegraphics*[width=8cm]{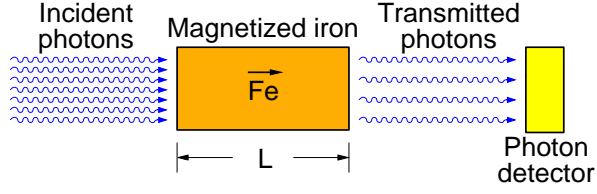}   
    \caption[The concept of transmission polarimetry.]{The concept of transmission polarimetry, in which the survival rate is measured for photons that pass through a magnetized iron absorber.}
    \label{fig:trans_pol}
  \end{center}
\end{figure}

Measurements of the circular polarization of energetic photons are most commonly based on the spin dependence of their interaction with polarized atomic electrons\cite{Fagg,YuanWu63}.   For photons of energy 1--10\,MeV this interaction
is dominantly Compton scattering.
In this experiment, transmission polarimetry was used, \ie,  measuring the transmission of unscattered photons through a thick, magnetized iron absorber\cite{Gunst,Schopper,Fukuda03}, as sketched in Fig.\:\ref{fig:trans_pol}.

\begin{figure}
  \begin{center}
    \includegraphics*[width=8cm]{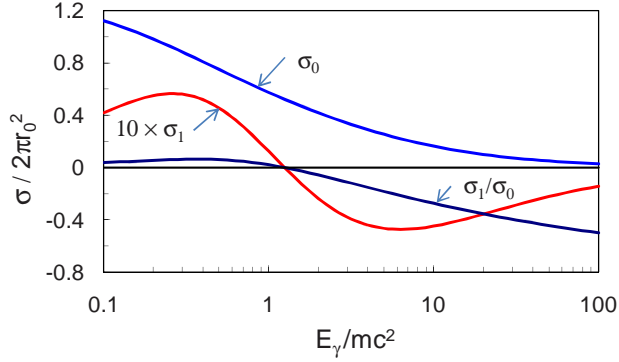} 
    \caption[The total cross sections $\sigma_0$ and $\sigma_1$ for polarized Compton scattering.]{The total cross sections $\sigma_0$ and $\sigma_1$ of Eq.~(\ref{eq:sigma}) for Compton scattering of longitudinally polarized photons of energy $E_\gamma$ off unpolarized and longitudinally polarized electrons, respectively;  
$r_0$ is the classical electron radius.}
    \label{fig:sigma_pol}
  \end{center}
\end{figure}

The Compton-scattering cross section for photons of energy $E$ in the MeV range off atomic electrons is taken to be that off free electrons,
\begin{linenomath}
  \begin{equation}
    \sigma(E) = \sigma_0(E) + P_\gamma(E) P_{e^-}^{\rm Fe} \sigma_1(E),
    \label{eq:sigma}
  \end{equation}
\end{linenomath}
where $\sigma_0$ is the unpolarized (Klein-Nishina) cross section,
$P_\gamma(E)$ is the net longitudinal polarization of the photons, $
P_{e^-}^{\rm Fe}$ is the net longitudinal polarization of the atomic
electrons, and $\sigma_1$ is the polarized cross section \cite{Gunst}. 
Figure~\ref{fig:sigma_pol} illustrates the energy dependence of the cross sections $\sigma_0$ and $\sigma_1$.

\begin{figure*}
  \subfloat[(a)]{\includegraphics[width=8cm]{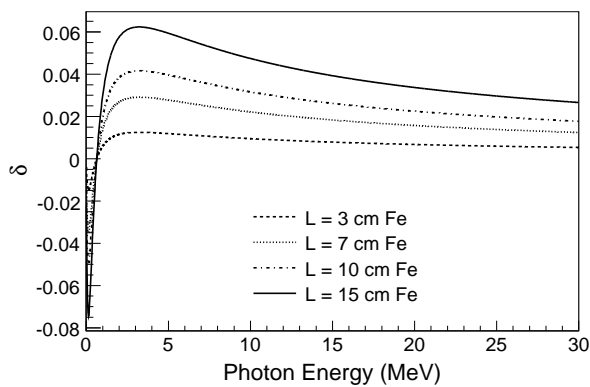}}   
  \hfill
  \subfloat[(b)]{\includegraphics[width=8cm]{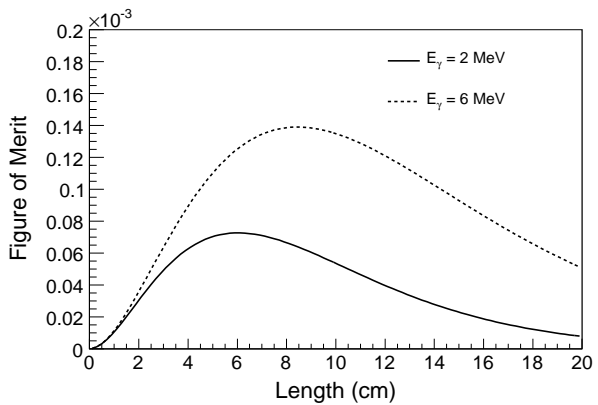}}    
  \caption[The asymmetry $\delta(E,L)$ and the figure of merit $\delta^2 T$. ]{(a) The asymmetry $\delta(E,L)$ defined in Eq.\:(\ref{eq:analpower2}) for transmission polarimetry of fully polarized photons on MeV energy $E$ in various lengths $L$ of saturated iron.
(b) The figure of merit $\delta^2 T$ for 2\,MeV photons (solid) and for
6\,MeV photons (dashed) as a function of the length of a transmission iron polarimeter.}
    \label{fig:fom_trans} 
  \end{figure*}

The average longitudinal polarization $P_{e^-}^{\rm Fe}$ of atomic
electrons in an iron-core solenoid can be related to the average (longitudinal) magnetic field $\langle B - B_0 \rangle$ of the iron, where $B_0$ is the part of the total magnetic field $B$ directly due to the current in the energizing coils,
by 
\begin{linenomath}
  \begin{equation}
     P_{e^-}^{\rm Fe} = \frac{\langle M_s \rangle}{n_ e^{\rm Fe} \mu_B}
                  = 0.03727 \langle B\,[{\rm T}] -B_0\,[{\rm T}] \rangle,
    \label{eq:p22a}
  \end{equation}
\end{linenomath}
where $M_s = 2 (g' - 1) M / g' = (0.958 \pm 0.002) M$ \cite{Ullman} is the dominant spin part of the total magnetization $M = (B - B_0) / \mu_0$, 
$g' = 1.919 \pm 0.002$ is the magnetomechanical ratio as measured in Einstein-de-Haas-type experiments\cite{Scott}, and $\mu_B$ is the Bohr magneton. 
These expressions are also summarized in Table~\ref{tab_analmag_three}.
Naively,  $P_{e^-}^{\rm Fe} = \pm 2 / 26$ for saturated iron, but the number of aligned Bohr magnetons per atom is more accurately determined to be $2.218 \pm 0.001$ for high-purity iron \cite{Cleaves,Bozorth} so that the maximum electron spin polarization in this material is $\pm 0.958 \cdot 2.218 / 26 = \pm 8.19$\,\%.
\begin{table}
  \small
  \centering      
  \caption{Parameters and expressions relevant to electron polarization of magnetized iron.}
  \label{tab_analmag_three}
   \vspace{1mm}
\footnotesize
  \begin{tabular*}{8.5cm}{l@{\extracolsep\fill}l}
    \hline
    \hline
    Parameter                                 &  Expression \\
    \hline
    Electron polarization $P_{e^-}^{\rm Fe}$  & $\langle M_s \rangle /(n_e^{\rm Fe} \, \mu_B)$  \\ [-1mm]
                                              & $= 0.03727 \langle B\,({\rm T}) -B_0\,({\rm T}) \rangle$ \\
    Magnetization $M$ (A/m)                         & $(B-B_0)/\mu_0$  \\    
    Spin fraction  $M_s/M$                    & $2 (g'-1)/g' = 0.958 \pm 0.002$\\
    Magnetomechanical ratio $g'$              & $1.919 \pm  0.002$ \\
    Electron number density  $n_e^{\rm Fe}$   & $N_{\rm A}\, \rho\,  Z / A = 2.206\cdot10^{30}$\,m$^{-3}$\\
    Bohr magneton $\mu_B$                     & $9.272 \cdot 10^{-24}$ J/T \\
    Vacuum permeability  $\mu_0$              & $4 \pi \cdot 10^{-7}$ T-m/A \\
    \hline
    \hline
  \end{tabular*}
\end{table}
The 
transmission probability $T^\pm(E,L)$ for photons of energy $E$ and helicity $P_\gamma$ through a piece of magnetized iron whose length is $L$ can be written as
\begin{linenomath}
  \begin{equation}
    T^\pm(E,L) = e^{-n_{e^-}^{\rm Fe} L \sigma^\pm} 
           = e^{-n_{e^-}^{\rm Fe} L (\sigma_0+\sigma_{\rm phot}+\sigma_{\rm pair})} e^{\pm n_{e^-}^{\rm Fe} L P_{e^-}^{\rm Fe} P_\gamma \sigma_1},
    \label{p10}
  \end{equation}
\end{linenomath}
which takes also the cross sections for photoelectric effect, $\sigma_{\rm phot}$, and pair production, $\sigma_{\rm pair}$, into account.
The $+(-)$ in $T^\pm$ applies if the electron spin in the iron is
parallel (antiparallel) to the spin direction of the incident photons, and $n_{e^-}^{\rm Fe}$ denotes the number density of atoms in iron.
The asymmetry
\begin{linenomath}
\begin{equation}
  \delta(E,L) = \frac{T^- - T^+ }{ T^- + T^+}
  \label{eq:analpower}
\end{equation}
\end{linenomath}
in transmission of photons through the iron absorber when the sign of $ P_{e^-}^{\rm Fe}$ is reversed, corresponding to a reversal of the magnetization of the iron, is
\begin{linenomath}
\begin{equation}
  \delta(E,L) = \tanh \left( - n_{e^-}^{\rm Fe} L\, P_{e^-}^{\rm Fe} P_\gamma\, \sigma_1 \right) 
  \approx - n_{e^-}^{\rm Fe} L \sigma_1\, P_{e^-}^{\rm Fe} P_\gamma.
  \label{eq:analpower2}
\end{equation}
\end{linenomath}
The sign convention in Eq.~\ref{eq:analpower}  
  results in a positive asymmetry $\delta$, given that the
  polarization-dependent Compton scattering cross section $\sigma_1$
  is negative at the energies of this experiment. 
This asymmetry is shown in Fig.\:\ref{fig:fom_trans}(a)
for fully polarized photons incident on various lengths of saturated iron, using
photon-iron cross sections from \cite{NIST}. 
The peak asymmetry is in the range of 1--6\,\% for photon energies in the range of several MeV and for lengths of iron of 3--15\,cm.
For the photon polarimeter, where
the rates were high, the length~$L$ of magnet {\sf TP2} was chosen to be 15\,cm to increase the size of the asymmetry.

An ``analyzing power'' $A_\gamma$ for transmission polarimetry can be defined as
\begin{linenomath}
  \begin{equation}
    A_\gamma(E,L) \equiv \frac{\delta(E,L) }{ P_{e^-}^{\rm Fe} P_\gamma(E)},
    \label{eq:p21}
  \end{equation}
\end{linenomath}
where $A_\gamma \approx n_{e^-}^{\rm Fe} L \sigma_1$ for small asymmetries such that the final form of Eq.\:(\ref{eq:analpower2}) holds.
Then, a measurement of the asymmetry $\delta$, plus knowledge of the
electron polarization $ P_{e^-}^{\rm Fe}$ in the magnetized iron and of the analyzing power $A_\gamma$, would determine the
photon polarization to be
\begin{linenomath}
  \begin{equation}
    P_\gamma = \frac{\delta }{ P_{e^-}^{\rm Fe} A_\gamma}\, .
    \label{eq:p22}
  \end{equation}
\end{linenomath}

However, in the case of a broad distribution of photon energies,
Eq.~(\ref{eq:p21}) becomes a convolution over energy-dependent detector
efficiency, analyzing power, and photon polarization \cite{KTM}.
Correspondingly, the detectors in the photon line measure an
effective polarization dominated by the high energy part of the
undulator spectrum. To gauge the understanding of the polarization 
of the photon beam, the observed asymmetries, Eq.~(\ref{eq:analpower}),
will be compared with simulations.

\subsection{Positron Polarimetry}\label{sec:posipol}

The polarization of positrons could be determined in principle by observation of any polarization-dependent interaction of the positrons. 
For example, good precision can be obtained measuring Bhabha scattering in a thin, magnetized iron foil when the final-state electron and positron are detected in coincidence\cite{Corriveau}. 
However, such a method is not applicable to the present experiment in which the positrons occur in pulses only a few-picosecond wide, such that coincidences are difficult to identify.  
Under these conditions, the simplest technique is the method of
transmission polarimetry, in which the positrons are ``reconverted''
into photons which inherit the positron polarization (either by
annihilation\cite{Page, Mcmaster} or by
Bremsstrahlung\cite{Olsen59,Gluckstern,Fronsdahl}), and the photons
subsequently pass through a thick iron
absorber\cite{Culligan,Goldhaber57,Macq,Bloom,Zwart}.
A measurement of the photon transmission asymmetry for magnetic fields
(in the iron) parallel and antiparallel to the positron momentum
vector then allows
the polarization of the positrons
to be inferred.

\begin{figure}
  \begin{center}
    \includegraphics[width=8cm]{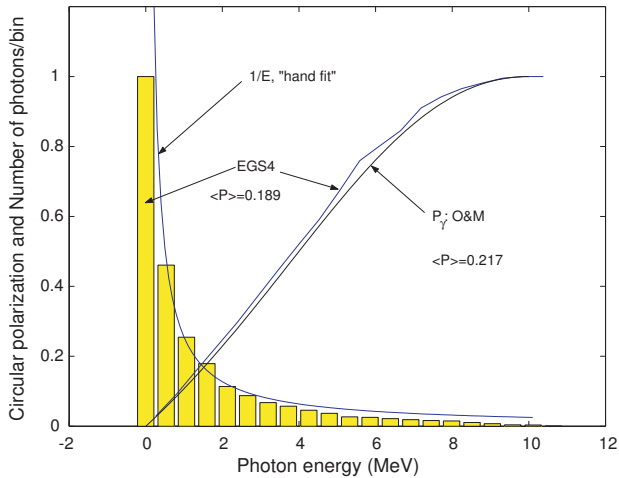}    
    \caption[Photon-number spectrum and polarization for
    Bremsstrahlung and annihilation of a 10\,MeV positron.]{Solid curves: the polarization, as a function of energy, of photons  generated by a 10\,MeV positron incident on 0.5\,mm of tungsten, according to an EGS4 simulation \cite{EGS4} and also to a calculation \cite{Olsen59}.
 Histogram: the energy spectrum of the photons according to an EGS4 simulation.  The photons are mainly due to Bremsstrahlung and their energies follow a $1/E$ ``hand fit"; photons with energy above 10\,MeV are due to annihilation.}
    \label{fig:transfer}
  \end{center}
\end{figure}

The transfer of polarization from positrons to photons (``reconversion'') in a thin foil is illustrated in Fig.\:\ref{fig:transfer}.
The average polarization of the photons from a 10\,MeV positron is only 21\,\% of that of the positron.
 
An asymmetry $\delta$ in the number of transmitted photons is measured by reversing the polarization $P_{e^-}^{\rm Fe}$ of the electrons in the iron absorber.  The polarization $P_{e^+}$ of the parent positrons can then be inferred according to
\begin{linenomath}  
  \begin{equation}
    P_{e^+} = \frac{\delta }{ P_{e^-}^{\rm Fe} A_{e^+}}\, ,
    \label{p23}
  \end{equation}
\end{linenomath}
in terms of an analyzing power $A_{e^+}$ that can be calculated in a simulation which combines the processes of polarization transfer from positron to photon and transmission of the photons through the iron absorber. Because the reconverted photons have a nearly isotropic angular distribution (due to the large multiple scattering of the parent positrons in the reconversion target), the computation of the analyzing power $A_{e^+}$ is more complicated than for $A_\gamma$ in the case of transmission polarimetry of a collimated photon beam.

The relative error on the measurement (\ref{p23}) of the polarization
varies inversely as the product of the asymmetry $\delta$ and the
square root of the transmission factor $T=(T^+ + T^-)/2$.  
A figure of merit for transmission polarimetry can therefore be defined as $\delta^2 T$, where larger values are better.  
This figure of merit is shown for 2\,MeV photons in Fig.\:\ref{fig:fom_trans}(b) as a function of length $L$ of the magnetized iron, which indicates that $L \approx 6$\,cm maximizes the statistical significance of the asymmetry at this energy.
The mean energy of the reconversion photons in the present experiment was 1--2\,MeV, and the length of the positron polarimeter magnet {\sf TP1} was chosen to be 7.5\,cm.

\section{Experimental Setup}\label{sec:setup}

A schematic layout of the experiment is shown in Fig.\:\ref{fig:explayc2k}.
A 46.6\,GeV electron beam generated polarized photons in the undulator. 
A set of bending magnets {\sf D1} deflected the high-energy electrons to a beam dump. 
The undulator photons were converted to electron-positron pairs in a thin target {\sf T1}. 
Positrons and photons were analyzed in the apparatus sketched in Fig.~\ref{prlfig2} (and also in
Fig.\:\ref{fig:E166Scheme}), while the low-energy electrons were dumped. 
Reversal of the spectrometer magnet {\sf D2} allowed for analysis of electron, while the positrons were in turn sent to the dump.

The remainder of this section presents details of the beamline, undulator, production target, photon and positron (or electron) diagnostics, data-acquisition system, runs types and data-file structure.

\subsection{Beamline Layout and Alignment}\label{sec:beamline}

The experiment required a high-energy, low-emittance electron beam to pass through a small-aperture, helical undulator to generate the polarized photon beam.  
Therefore, the experiment was performed in the
Final Focus Test Beam (FFTB)\cite{fftb} at SLAC \cite{slac_linac}, which could operate at energies up to 54\,GeV and deliver an electron spot size of less than $50 \times 50\ \mu$m$^2$ to an area appropriate for small experiments. 

Figure\:\ref{fig:explayc2k} shows a schematic of the layout of the FFTB beamline elements specific to the experiment. 
Table\:\ref{tab:ebeam_parameter}
\begin{table*}
  \small
  \begin{center}
    \caption{Nominal and actual electron-beam parameters of the FFTB.}
    \label{tab:ebeam_parameter}
  \centering
 \vspace{1mm}
\footnotesize
    \begin{tabular*}{\textwidth}{l@{\extracolsep\fill}ccccccccc}
      \hline
      \hline
      & $E_{e^{-}}$ & $\gamma$ &  $f_{rep}$ & $N_{e^{-}}$  & $\gamma \epsilon_{x}, \ \gamma \epsilon_{y}$ &
      $\beta_{x}$, $\beta_y $ &  $\sigma_{x}$, $\sigma_{y}$ & $\sigma_{x'}$, $\sigma_{y'}$ & ${\sigma_E}/{E}$ \\
  & (GeV) & & (Hz) & ($10^9$) & ($10^{-5}$ rad) & (m)& ($\mu$m) & ($\mu$rad) & (\%) \\
      \hline
      Nominal & 50   & $9.8 \cdot 10^4$ & 30 & 20  &  3, 3     & 5.2, 5.2  & 40 & 13      & 0.3 \\
      Actual  & 46.6 & $9.1 \cdot 10^4$ & 10 & 1--4 & 2.2, 0.5  & 5.7, 19.8 & 35--40, 30--36 & 10, 1 & $\sim$ 0.2     \\
      \hline 
      \hline
    \end{tabular*}
  \end{center}
\end{table*}
lists the nominal parameters of the FFTB beamline at the point where the undulator was installed, and the actual parameters achieved during running. 
The beam energy actually used, 46.6\,GeV, was lower than nominal to insure more stable electron beam energy. 
The intensity was reduced to 1--$4 \cdot 10^9\ e$/bunch to suppress the electron beam halo relative to the core beam, and thereby to decrease the relative size of backgrounds due to interaction of the tails of the beam with the 0.71-mm-diameter, 6.35-cm-long protection collimator {\sf C1}.   
The reduced bunch intensity also permitted spot sizes slightly smaller than nominal.

\begin{figure*}
  \centering
 
  \begin{picture}(346,260)
   \put(0,0){\includegraphics[width=346pt]{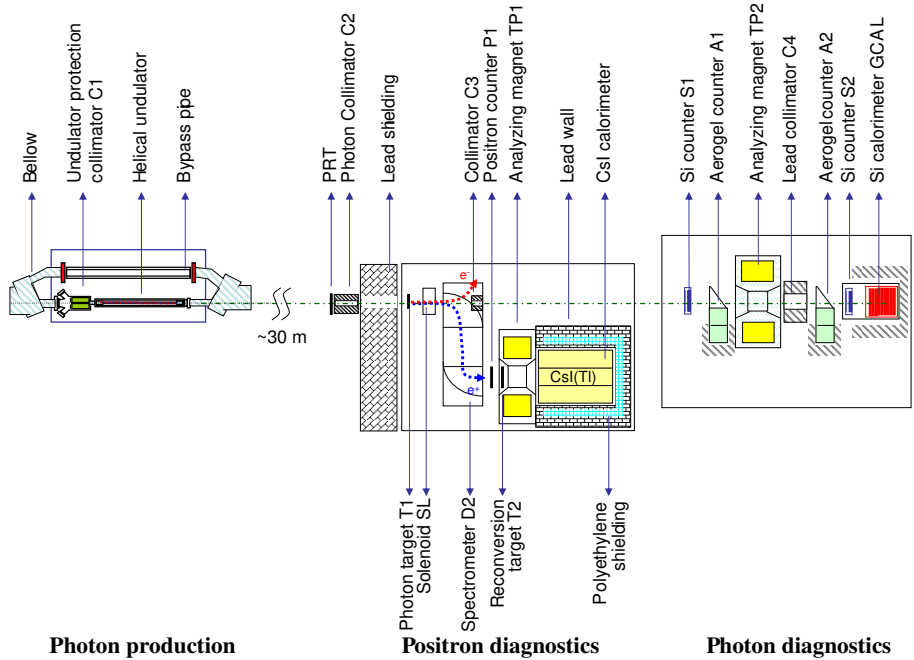}}
   \end{picture}
  \caption[Schematic of the experimental apparatus along the $\gamma$-line.]{Schematic of the experimental apparatus along the $\gamma$-line, including photon production in the undulator, conversion to $e^\pm$ in target {\sf T1}, and subsequent diagnostics of photons and positrons.  
During the initial setup of the electron beam the bypass pipe was placed in the beamline rather than the undulator.
}
  \label{fig:E166Scheme}
\end{figure*}

The undulator was preceded and followed by weak vertical-deflection magnets {\sf HSB1} and {\sf HSB2} that deflected the primary electron beam downward by $0.02\degree$ each.  
This generated only a soft spectrum of synchrotron radiation along the undulator-photon line while decoupling this line from photons traveling with the primary electron beam upstream of the undulator.  After {\sf HSB2}, a string of seven permanent magnets {\sf D1} deflected the primary electron beam downwards by an additional $1\degree$ and into the FFTB beam dump 45\,m downstream of the undulator, as shown schematically in Fig.~\ref{fig:explayc2k}.

The electron-beam parameters were tuned using the beam-position
monitors {\sf BPM1} and {\sf BPM2}, the wire scanner {\sf WS} and 
optical-transition-radiation monitor {\sf OTR} to measure beam size,
and toroid {\sf BT} to measure the bunch charge. 

The major challenge in operating the beam was to pass it through the
extremely small aperture of collimator {\sf C1} (0.71~mm in diameter,
6.35~cm long), and then to maintain the beam tune so that the beam
passed cleanly through the undulator during data-taking runs of 30--60
minutes each.  
The beam was initially set up using a 2.5-cm-diameter bypass beam pipe, shown in Fig.~\ref{fig:E166Scheme}, instead of the undulator, and the beam orbit was recorded using the {\sf BPM}s. 
Then the undulator was moved onto the beam orbit and aligned to a precision of a few $\mu$m using an array of five motion stages to minimize beam losses.

The undulator photons drifted in the $\gamma$-line, defined by the direction of the electron beam between magnets {\sf HSB1} and {\sf HSB2}, for about 35\,m to the diagnostic apparatus
(Figs.~\ref{prlfig2} and \ref{fig:E166Scheme}),
where they were either converted to electron-positron pairs in a thin target {\sf T1} (see Sec.~3.3)
or analyzed (see Sec.~3.4). 
The undulator photons passed through tungsten collimator {\sf C2}
(3\,mm in aperture, 10.16\,cm long) located 32.49\,m downstream of the
undulator center, which defined the transverse extent of the photon
beam thereafter.  
The full width at half maximum of the undulator-photon beam was approximately 0.8\,mm at collimator {\sf C2}, based on Eq.~(\ref{p3}) for the photon angular distribution and on the electron-beam parameters given in Table~\ref{tab:ebeam_parameter}.
Initial alignment of the electron beam (via horizontal- and vertical-correction magnets upstream), such that the $\gamma$-line passed through collimator {\sf C2} with maximal flux in detector {\sf S1}, was accomplished using the bypass beam pipe and a beam of Bremsstrahlung photons generated by the 25-$\mu$m-thick  (1-$\mu$m-thick in September 2005 run) Ti foil of the optical-transition-radiation monitor {\sf OTR}.

Due to operational difficulties with the undulator motion stages, the
electron beam was resteered slightly during much of the data
collection to minimize backgrounds from collisions with collimator
{\sf C1} and with the undulator itself.  
This steering resulted in partial loss of intensity in the
$\gamma$-line due to reduced transmission through collimator {\sf C2}.  

\subsection{Undulator}\label{sec_undulator}

A single undulator, {\sf U1}, was used throughout the experiment, although two complete undulator systems
(Table\:\ref{tab:UndPar}) were fabricated and sent to SLAC.
\begin{table}
  \small
  \centering
  \caption{Parameters of the two fabricated undulator systems
    \cite{CBN05-2}. Only the system given in the first column (U1) was
    actually used in the experiment.} 
\vspace{1mm}
\footnotesize
  \begin{tabular*}{7cm}{l@{\extracolsep\fill}cc}
    \hline
    \hline
    Parameter (Units) & {\sf U1} & {\sf U2} \\
    \hline
    Energy (GeV)            & 46.6        &   46.6 \\
    Length (mm)             & 1000    & 1000  \\
    Period (mm)             & 2.54        &    2.43 \\ 
    Number of periods       & 394      & 406 \\
    Aperture (mm)  & 0.87       &   1.07 \\ 
    Winding direction  &  left-handed & left-handed \\
    Axial field (T)         & $\sim 0.71$ & $\sim 0.54$ \\
    $K$                     & $\sim 0.17$ & $\sim 0.12$ \\
    $E_1$ (MeV)             & $\sim 7.9$  & $\sim 8.4$  \\
    Photons/$e^-$           &  0.35       & 0.18 \\
    $\Delta E/e^-$ (MeV)    & 1.65        & 0.88 \\
    Voltage (V)		    & $\sim 656$  & $\sim 592$ \\
    Current (kA)            & 2.3         & 2.3  \\  
    Pulse width ($\mu$s)    & 12          & 13   \\
    $\Delta T$/pulse ($\degree$C) & $\sim1.7$ & $\sim 1.3$ \\
    Inductance ($\mu$H)     & $\sim 1.4$  & $\sim 1.5$ \\
    Resistance ($\Omega$)   & $\sim 0.22$ & $\sim 0.26$ \\
    Oil flow (l/min)        &  13.25       &  13.25 \\
    Press.\ drop (bar)      &  $\sim 0.76$ & $\sim 0.76$ \\ 
    \hline
    \hline
  \end{tabular*}
  \label{tab:UndPar}
\end{table}

\subsubsection{Undulator Fabrication}

Six undulator coils were wound, three for each tube diameter, and each of these windings was tested at full current.

\begin{figure}
  \centering
  \includegraphics[width=8cm]{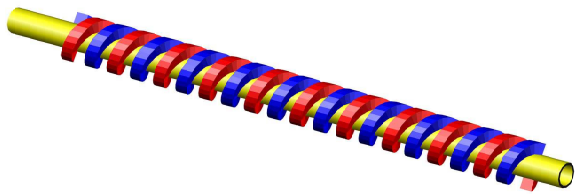} 
  \caption{Sketch of the left-handed, bifilar windings of the undulator.}
  \label{fig:HELIX1-}
\end{figure}

The undulator conductors were bifilar-helical windings with currents running in opposite directions (Fig.\:\ref{fig:HELIX1-}).
This method of helical-field generation, proposed in\cite{Wingerson,Chodorow}, was used successfully some years ago\cite{Mikhailichenko} for an undulator with a period of 6\,mm and $K\sim 0.35$.

The undulator conductor was oxygen-free CDA 10200 copper wire with
square cross section $0.6\times0.6$\,mm$^2$ and  corner radius  $\leq 0.06$\,mm. 
The wires were wound on hypodermic 304-L stainless-steel tubes with nominal OD's of 1.07\,mm (19-XTW) and  1.27\,mm (18-XTW). 
All tubes had nominal wall thicknesses of 0.10\,mm. 
Each tube was wrapped in Kapton insulation with a thickness of
12.7\,$\mu$m. 
Four copper wires were wound onto a tube at a time: two of which were
square cross section (bare) conductors, and the other two were of
round cross section with 0.483\,mm diameter and used as spacers. 
After completion of winding, the spacer wires were removed; it was found
that the remaining conductors adhered to the tube without slippage.
This procedure resulted in a period of 2.54\,mm for the windings on
0.87-mm-diameter tubes, and 2.43\,mm for those on 1.07-mm-diameter
tubes.

All undulators were wound left-handed, \ie, counterclockwise as seen
by a beam electron. 
This resulted in negative polarization $P_\gamma$ for the high-energy
part of the undulator spectrum (see Sec.\ \ref{sec_undulator_physics}
and Fig.\ \ref{fig:Photon}). 
Visual inspection of the undulators under a microscope allowed removal of tiny pieces of
copper chips created in the winding process. 

\begin{figure}
  \centering
  \includegraphics[width=8cm]{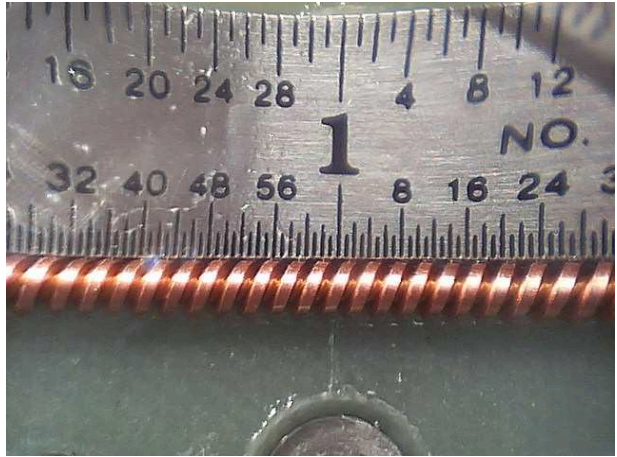} 
  \caption[Photograph of the windings of undulator {\sf U1}.]{Photograph of the windings of undulator {\sf U1}. 
The rectangular cross-section wire has dimensions of $0.6\times0.6$\,mm$^2$, and a period of 2.54\,mm. 
The smallest scale division is 1/64 of an inch (0.397\,mm). }
  \label{fig:IMGA0637}
\end{figure}

Even though the bending radius was of the order of the wire size, the keystone effect was not significant.
A magnified view of windings is shown in Fig.\:\ref{fig:IMGA0637}. 
\begin{figure}
  \centering
  \includegraphics[width=8cm]{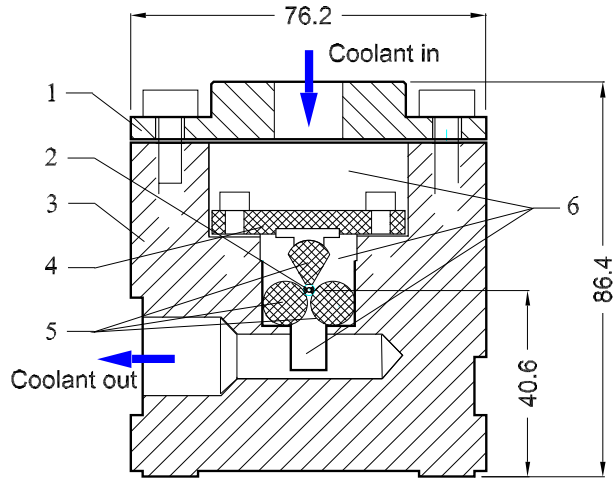} 
  \caption[Cross-section of an undulator system (dimensions in mm).]{Cross-section of an undulator system (dimensions in mm). 
Two G-10 rods (5) were placed in the corners of a long channel in an aluminum bar (3). 
A third rod (5) compressed the windings (2) to the other two rods via a spring-loaded bar (4). 
The Al undulator cover (1) was sealed with an indium gasket.   
The interior volume (6) was filled with oil or ferrofluid.
}
  \label{fig:Figure4}
\end{figure}
These windings were rather flexible and so were constrained by three cylindrical G-10 rods. 
Two of the rods sat in the lower corners of a U-shaped groove milled into an aluminum block and provided insulated support for the entire helix, as shown in Fig.\:\ref{fig:Figure4}. 
The third rod compressed the helical windings from above against the other two. 
The aluminum block had overall dimensions of  $7.62\times7.62\times114.3$\,cm$^3$.  
The U-shaped groove was made to a tolerance of 12.7\,$\mu$m throughout its length. 
This accuracy was achieved in a few milling steps after all flanges (with Al/stainless steel transitions) had been welded to the housing.
The dimensions were checked commercially with a semi-automatic coordinate-measuring machine.  

Two undulator bodies were milled simultaneously while attached to a baseplate by special holders which allowed expansion in the longitudinal direction. 
After fabrication, all Al parts were black anodized to minimize contact of the Al surface with oil. 
This procedure did not change the dimensions appreciably. 

\begin{figure}
  \centering
  \includegraphics*[width=8cm]{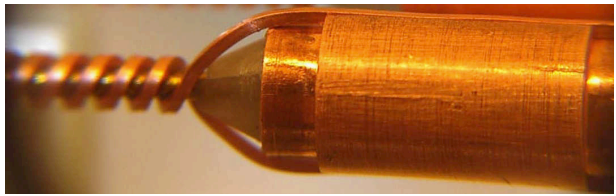} 
  \caption{The downstream end of the undulator winding, showing the ``jumper'' that closed the circuit
though a copper cylinder.}
  \label{fig:endjumper}
\end{figure}

The routing of the leads to the helical winding, and of the ``jumper'' that closed the circuit at the end of the winding (Fig.~\ref{fig:endjumper}), resulted in regions of net transverse magnetic field on the electron beam at the two ends of the undulator.  In future fabrications, the extent of these regions should be minimized, and the directions
of the transverse magnetic fields should be opposing.

 The aperture and straightness of the windings of undulators {\sf U1} and {\sf U2} was tested by stretching a 0.4-mm-diameter stainless-steel wire through the undulator and noting absence of electrical contact between wire and tube. 
For an electron-beam size of 40\,$\mu$m, 0.4\,mm is about $10\,\sigma$, sufficient for successful beam passage through the undulator. 
Further tests at SLAC measured the aperture of the undulator used in the experiment to be at least 0.71\,mm.

\subsubsection{Undulator Operation}

Pure transformer oil was used as a coolant. 
The oil flowed in a circuit that included a stainless-steel oil pump, heat exchanger, reservoir, pressure gauges and valves. 
The oil flowed into the undulator case from the top center and exited at the lowest point in the groove below the G-10 rods (Fig.\:\ref{fig:Figure4}), also in the center. 
The oil pressure inside the case was about 2.4\,bar, which expanded the chamber by a small amount. 

The pumping/cooling system was constructed as a single mobile unit with oil pump, flow meters, heat exchanger, and 3-phase control electronics. 
This system could be operated locally or remotely and contained a set of thermal interlocks which ensured operational readiness of the system. 
A pressure transducer, PS~302-200GV, was attached to the line through a pressure snub, PS-4E, together with a DP25-SR strain gauge meter, also attached to the readiness interlock. 
The pressure transducer could be attached either to the outgoing or incoming line. 
The oil line was also equipped with dial-type pressure gauges installed near the undulator.

The undulator was connected to the cooling loop by oil-resistant, flexible tubes. 
The power supply was located in a rack together with the pulser in the FFTB tunnel.

\begin{figure}
  \centering
  \includegraphics[width=8cm]{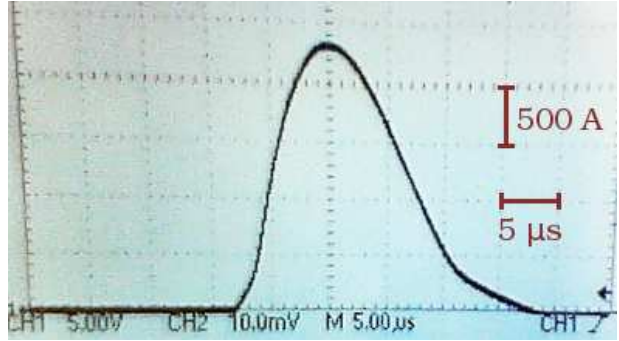} 
  \caption{Current observed in the undulator shunt during a 2.3\,kA pulse of
     $12\,\mu$s duration.}
  \label{fig:fig3_shunt}
\end{figure}

The pulser is similar to the one used for the positron-production upgrade of CESR\cite{Barley}. 
The undulator was tested at 2.3\,kA and 30\,Hz for one hour. 
This current represents the upper limit of the power supply (EMS800-2.5-5-D). 
Thus, in case of failure of all the electronics, the power supply would not deliver sufficient current to damage the undulator. 
The undulator was operated at 2.3\,kA and 10\,Hz 
with an average pulsed-current density of about 6.39\,kA/mm$^2$, and a pulse duration of about 12\,$\mu$s. 
The pulse waveform is shown in Fig.\:\ref{fig:fig3_shunt}.

The current density in the undulator winding was calculated with the 3D code MERMAID \cite{ref:mermaid}. 
This code calculated the current by solving for the electrostatic potential inside the conductor taking into account the actual geometry of the thick conductor. 
As the dimensions of the conductor were comparable to the bending radius, the current had a tendency to flow in layers closer to the center yielding a field enhancement on the axis. 
The current density varied by as much as a factor of four over the conductor cross-section. 
Wires with rectangular cross-section yielded a 15\,\% higher axial field compared to wires with circular cross-section.

The calculated field at the axis was $B_{\bot} \approx 0.71$\,T at $I=2.3$\,kA. 
As the measured period was $\lambda_{\rm u} \approx 2.54$\,mm, the undulator factor $K$ defined in Eq.\:(\ref{eq_u2}) was approximately 0.17. 
However, since the undulator pulse width was only about 12\,$\mu$s at half height (Fig.\:\ref{fig:fig3_shunt}), the skin depth was formally about 0.3\,mm. 
Thus, the current had a tendency to be expelled from the interior of the conductor, which affected the magnetic field; but it is difficult to calculate the magnitude of this effect. 
An estimate of the undulator $K$ from direct measurements of the photon flux is presented in Sec.~\ref{sec_undulator_performance}.

Towards the end of the data-taking run, the coolant oil was replaced with a ferrofluid (EMG 900) whose magnetic properties allowed it to serve as a return yoke for the magnetic field, thus increasing the effective undulator magnetic field. 
Calculations predicted a photon enhancement of about 20\,\%\cite{Ferrofluid}. 
Data taken using the ferrofluid (Sec.\:\ref{sec:ferro}) showed an enhancement of about one half this
amount.
The thermo-hydraulic properties of the ferrofluid are similar to those of oil, so no circulation or cooling problems were observed. 
The presence of the ferrofluid also increased the pulse rise time, which improved operation of the power-supply thyristors.
At the end of the experiment the ferrofluid was found to be somewhat more radioactive than the normal oil. 

The successful running of the present experiment verified the predicted undulator parameters and confirmed the engineering design principles.

\subsubsection{Beam Deflection by the Undulator}\label{sec:steering}

A small deflection of the electron beam was observed when the undulator was energized in time with the beam.
This had minimal impact on the present experiment, but such a deflection would be undesirable at a linear collider
in which the electron beam passes through the undulator before colliding with the positron beam.

\begin{figure}
  \centering
  \includegraphics[width=8cm]{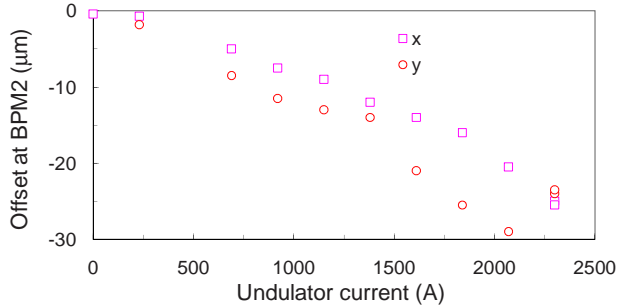} 
  \caption{Offsets in $\mu$m of the electron beam at {\sf BPM2} as a function of undulator current. }
  \label{fig:Undulatorperformance5}
\end{figure}

The deflection was first observed on an Al$_2$O$_3$(Cr) screen $\approx 42.9$\,m downstream of the undulator, just upstream of the beam dump, where the beam spot was offset by $\approx 1$\,mm for beam pulses in and out of time with the undulator.
This angular deflection of about 25\,$\mu$rad was corroborated by observation of small offsets of the electron beam in {\sf BPM2}, just downstream of the undulator, that varied with the current in the undulator, as shown in Fig.\:\ref{fig:Undulatorperformance5}.
The corresponding angular deflection at full current was about 25\,$\mu$rad, assuming that the deflection occurred at the upstream end of the undulator, about 1.55\,m from {\sf BPM2}.
Also, the {\sf PRT} counters preceding collimator {\sf C2} showed a different displacement of background photons (from interactions of the electron beam with collimator {\sf C1} and/or the undulator) when the undulator was in and out of time with the electron beam.
 
It was concluded that the deflection was due to the routing of the leads at the upstream and downstream ends of the undulator, the latter of which is shown in Fig.\:\ref{fig:endjumper}, whose current may have introduced the $0.0039$\,T$\cdot$m kick corresponding to the 25\,$\mu$rad angular deflection.  
More careful design of the undulator leads should mitigate this issue. 
Imperfect alignment of the electron beam with respect to the undulator axis may also have contributed to the beam deflection.

\subsection{Production Target {\sf T1}}

The left-handed undulator generated negative-helicity photons that 
passed through a tungsten collimator {\sf C2} with a 3-mm-diameter aperture  before they struck the production target {\sf T1} where positrons were generated by pair production (see Figs.\:\ref{prlfig2} and \ref{fig:E166Scheme}).
The full width at half maximum of the photon beam was 0.9\,mm at target {\sf T1}.
A five-position target holder provided for four targets, 0.2- and 0.5-radiation-length (r.l.) tungsten and titanium, and a no-target position. 
The tungsten was a sintered composite, W-4Ni-3Cu-3Fe, while the composition of the titanium alloy used was not specified (probably Ti-6Al-4V). 
Data were taken primarily with the 0.2\,r.l.\,(0.81\,mm) tungsten target, which provided both a good yield of positrons and good polarization transfer from the undulator photons, as shown in Fig.~\ref{fig:Olsen2}.

\subsection{$\gamma$-Line}\label{sec_photon_line}

\subsubsection{Overview}

The $\gamma$-line and its associated photon diagnostics are shown in Figs.\:\ref{prlfig2} and \ref{fig:E166Scheme}. 
Prior to impinging on the production target {\sf T1}, undulator photons passed through a
3-mm-diameter, 10.16-cm-long, tungsten collimator {\sf C2}, whose aperture defined the downstream $\gamma$-line.
A set of four Si-W detectors ({\sf PRT}) arrayed outside the aperture of collimator {\sf C2} was used
to monitor the alignment of the $\gamma$-line.
The photon flux from the undulator, after passing through collimator {\sf C2} and target {\sf T1}, was measured independently by two counters, an aerogel Cherenkov counter {\sf A1} with an energy threshold of $\approx$\,4\,MeV, and a Si-W counter {\sf S1} whose simulated response was utilized to provide normalization of the undulator-photon flux. 
A transmission polarimeter consisting of a 15-cm-long iron magnet {\sf TP2} (see Sec.\:\ref{sec_analyzer_magnet}), aerogel counters {\sf A2}, and Si-W counter {\sf S2} detected effects of the polarization of the undulator photons. 
In addition, a Si-W calorimeter {\sf GCAL} measured the total energy of the photon beam after the
transmission polarimeter.

\begin{figure}
  \centering
  \includegraphics[width=8cm]{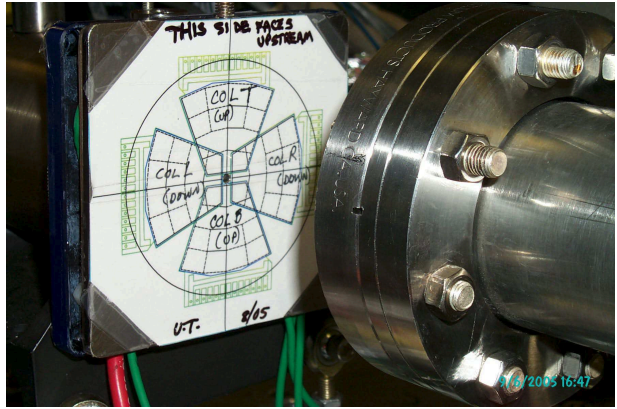} 
  \caption{Placement of the quadrant detectors {\sf PRT1--4} upstream of collimator {\sf C2}.}
  \label{fig:P0008160}
\end{figure}

Photons from the undulator were measured by the Si-W detectors {\sf S1} and {\sf PRT} whose absolute calibration was determined by \geant\ simulation as described in Section\:\ref{sec:detectors}. 
The four {\sf PRT} detectors were located immediately upstream of collimator {\sf C2} with edges tangent to the 3-mm-diameter collimator aperture, as shown in Fig.\:\ref{fig:P0008160}. 

The shape and orientation of the individual  detectors {\sf PRT}top, {\sf PRT}right, {\sf PRT}bottom, and {\sf PRT}left are shown in the photograph with the labels {\sf ColT}, {\sf ColR}, {\sf ColB}, and {\sf ColL}, respectively. 
The photon beam emerged from the $\gamma$-line vacuum window and entered the collimator from the right to the left. 
These detectors sampled that portion of the flux that does not enter the collimator aperture. 
They were installed between the June and September running periods to provide a measure of the centering of the undulator beam. 
The photons that exited collimator {\sf C2} struck the positron-production target {\sf T1} producing electron-positron pairs, and the unconverted photons were subsequently measured by detectors {\sf A1} and {\sf S1} prior to incidence on the polarimeter magnet {\sf TP2}, after which the photon-transmission asymmetry was determined. 
With suitable correction, the sum of the fluxes through detectors {\sf A1} and {\sf S1} yielded the number of photons  produced by the undulator per beam electron.

\subsubsection{Particle Detectors}\label{sec:detectors}

The particle detectors were required to deal with a rather wide range of energies and intensities. 
In the $\gamma$-line they measured undulator photons with intensities of 10$^7$--10$^9$ photons/pulse.
In contrast, in the positron line typically a few thousand positrons per pulse were measured, leading to deposited energies of a few hundred MeV in the CsI calorimeter.
Since asymmetry measurements of the form $A=(L-R)/(L+R)$ were utilized in the analysis of both photon and positron polarization, absolute calibration of detectors was not essential for the polarization analysis. 
However, to evaluate many quantities of interest, \eg, the number of undulator photons generated per beam electron, or to compare measured transmission with simulations, it was necessary to have a good understanding of the detector response functions.

\paragraph*{Aerogel Counters:}
The aerogel counters designated {\sf A1} and {\sf A2} in Figs.\:\ref{prlfig2} and \ref{fig:E166Scheme} were identical counters designed to detect undulator photons incident on, and transmitted through, 15\,cm of magnetized iron ({\sf TP2}). 
The sensitive elements were 2-cm-thick blocks of aerogel. 
The index of refraction of the aerogel was measured by interferometric techniques to be 1.0095\,$\pm$\,0.0001, resulting in a Cherenkov threshold of 3.78\,MeV. 
Cherenkov photons were reflected vertically by an aluminum mirror and passed through an air-filled light pipe to a photomultiplier tube. 
To prevent possible false asymmetries the photomultipliers were carefully shielded from background radiation and external magnetic fields.  

\paragraph*{Si-W Detectors:}
The other detectors in the $\gamma$-line were based on silicon detector technology. 
The charge-sensitive elements were 300-$\mu$m-thick, reverse-biased, high-resistivity silicon layers (manufactured by Hamamatsu) mounted on 900-$\mu$m-thick G-10 supports.  To enhance the sensitivity
of these detectors to photons, each Si layer was preceded by a layer of tungsten (Densalloy\,SD170
\cite{Densalloy}.

It takes 3.66\,eV for an electron or positron to create an electron-hole pair in silicon \cite{Scholze}, and therefore 1\,keV of energy deposited in a silicon detector liberated 4.38$\cdot$10$^{-5}$\,pC of electrons.
These electrons were collected in LeCroy 2249W ADCs after appropriate attenuation, required since the flux of
$\approx 10^9$ photons per pulse in the $\gamma$-line led to large signals in the Si-W detectors.

The Silicon detectors in the $\gamma$-line were:
\begin{itemize}
\item{\sf S1, S2:}
These detectors were identical devices which counted incident and transmitted undulator photons (as did the Cherenkov counters {\sf A1} and {\sf A2}). 
They consisted of a 555-$\mu$m-thick (0.13\,r.l.) tungsten converter, and a single Si/G-10 detector. 
Under normal operating conditions the {\sf S1} signal was attenuated by 46--60\,db and that of {\sf S2} by 20\,db.

\item{\sf GCAL:}
This device was a 9-element calorimeter, each element consisting of a 3.7-mm-thick ($\approx$\,0.9\,r.l.)
tungsten plate and a Si/G-10 detector, and provided a measure of the total energy of the transmitted photons. 
The {\sf GCAL} signal was normally attenuated by 40\,db.

\item{\sf PRT:}
This was a set of four counters placed at the upstream edge of collimator {\sf C2}, as shown in Figs.~\ref{fig:explayc2k}, \ref{fig:E166Scheme} and \ref{fig:P0008160}.
Each consisted of a 3.7-mm-thick tungsten converter and a Si/G-10 detector.  Together they
comprised a quadrant detector to aid in steering of the undulator-photon beam.
They also provided an estimate of the fraction of undulator photons that did not enter the collimator aperture. 
The {\sf PRT} signal was normally attenuated by 40\,db.

\end{itemize}

\begin{figure}
    \centering
    \includegraphics[width=8cm]{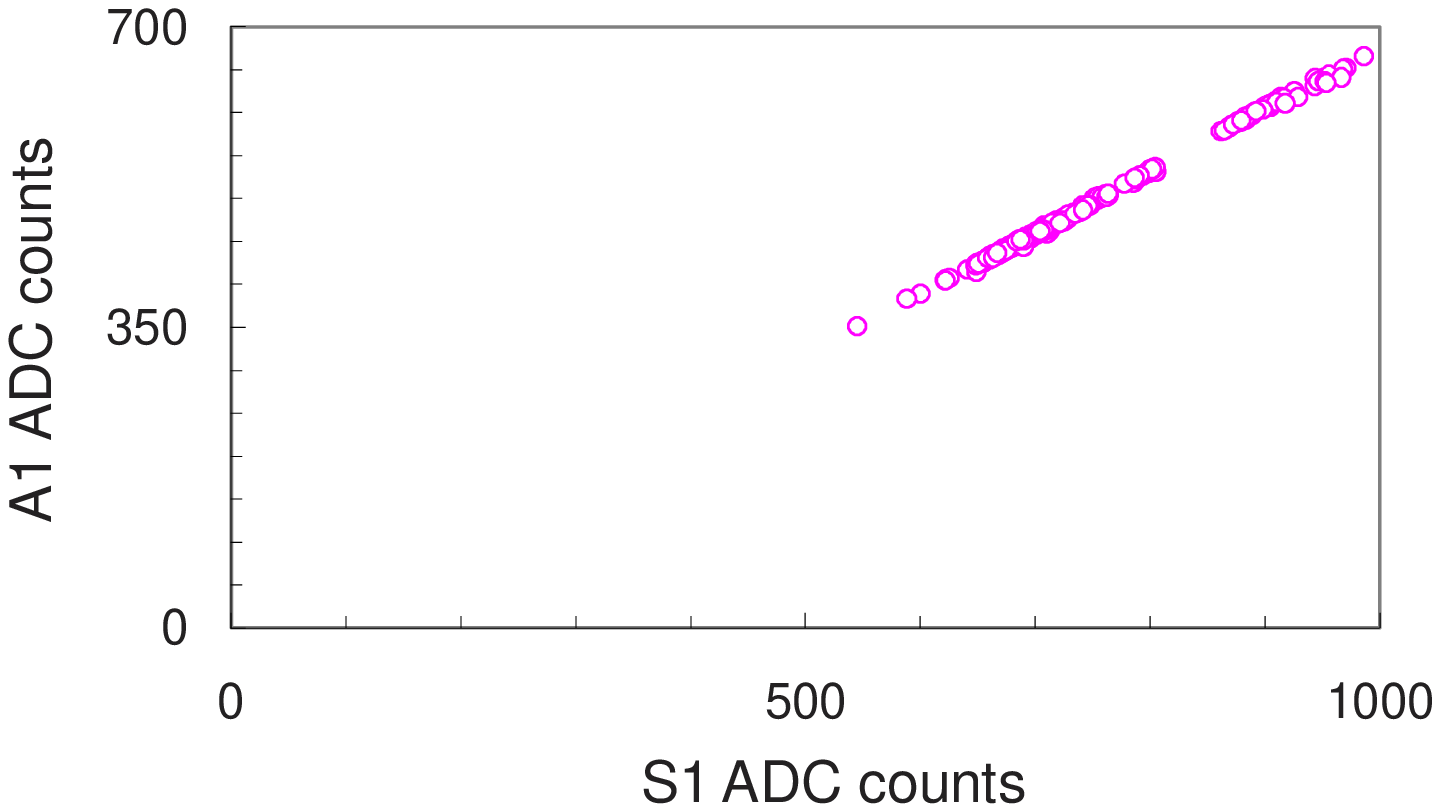}  
    \includegraphics[width=8cm]{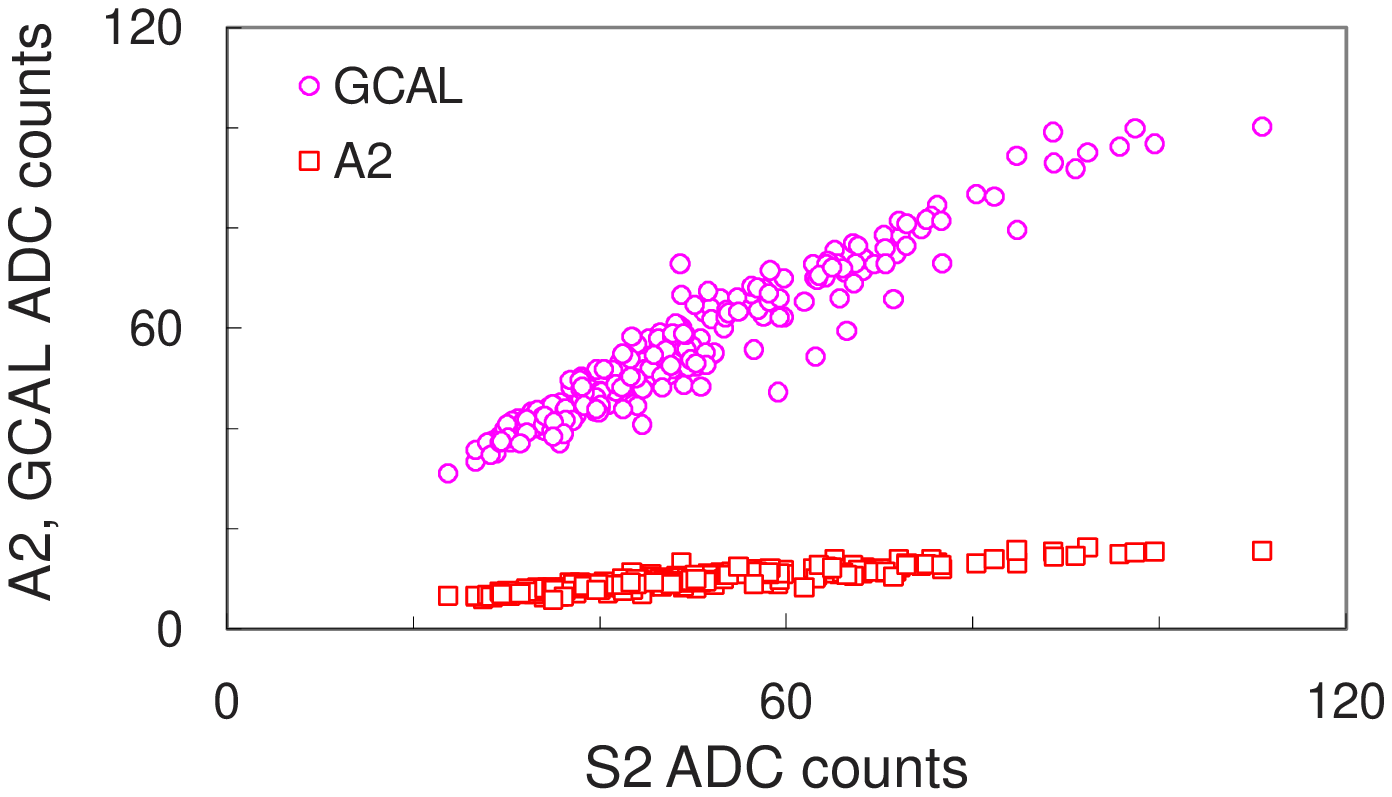} 
  \caption[Comparison of signals in photon counters {\sf A1} and {\sf S1}, and in counters {\sf A2},
{\sf S2} and {\sf GCAL}.]{Comparison of signals in photon counters {\sf A1} and {\sf S1} (a), and in counters {\sf A2},
{\sf S2} and {\sf GCAL} (b).}
  \label{fig:und_phot1_and_trans1}
\end{figure}

\paragraph*{Counter Comparisons:}
The undulator flux that passed through collimator {\sf C2} was independently measured by counters {\sf A1} and {\sf S1} upstream of the iron-core solenoid {\sf TP2},
and the transmitted flux was monitored by counters {\sf A2}, {\sf S2} and calorimeter {\sf GCAL}. 
While these detectors measure somewhat different features of the photon spectrum and had different sensitivities, good correlation was observed between signals in {\sf A1} and {\sf S1}, as shown in 
Fig.\:\ref{fig:und_phot1_and_trans1}\,(a) for a sample of 290 runs during September 2005, and between {\sf A2}, {\sf S2} and {\sf GCAL}, as shown in Fig.\:\ref{fig:und_phot1_and_trans1}\,(b).

\subsubsection{{\sf PCAL} Background Positron Monitors}\label{sec_background_monitors}

A set of three 23-r.l.-long Si-W calorimeter units,  
{\sf PCAL}c, {\sf PCAL}d and {\sf PCAL}e, was placed above the dump magnet {\sf D1}, about 10\,m downstream of the undulator, as shown schematically in Fig.~\ref{fig:explayc2k},
to intercept energetic positrons from interactions of the tails of the primary electron beam with collimator {\sf C1} or with the undulator {\sf U1}.
The sensitivity of these calorimeter was a few hundred MeV per ADC count; thus, they were sensitive to interactions of individual beam particles.
The {\sf PCAL}s were used to monitor the steering of the electron beam through the undulator aperture to minimize backgrounds in the downstream photon and positron detectors.

\subsubsection{Other Background Detectors}
Assorted other silicon detectors, CsI crystals and scintillator paddles were placed in the experimental area for background monitoring purposes; they were not essential to the operation of the experiment and they will not be discussed further.

\subsection{Positron Transport and Diagnostics}\label{sec:pos_diag}

Positrons were generated from undulator photons by pair production in a 0.81-mm-long tungsten production target ({\sf T1}), and subsequently deflected with a double-dipole spectrometer ({\sf D2}) into the positron analyzer, as shown in Figs.\:\ref{prlfig2} and \ref{fig:E166Scheme}.
A solenoid lens ({\sf SL}) behind the production target increased the useful positron flux. 
A second tungsten target ({\sf T2}) in front of the positron-polarimeter magnet ({\sf TP1}) reconverted positrons to photons, so that the positron polarization could be inferred from that of the photons via transmission polarimetry.  The transmitted, reconverted photons were detected in an array
of CsI crystals.

\begin{figure}
  \centering
  \includegraphics[width=8cm]{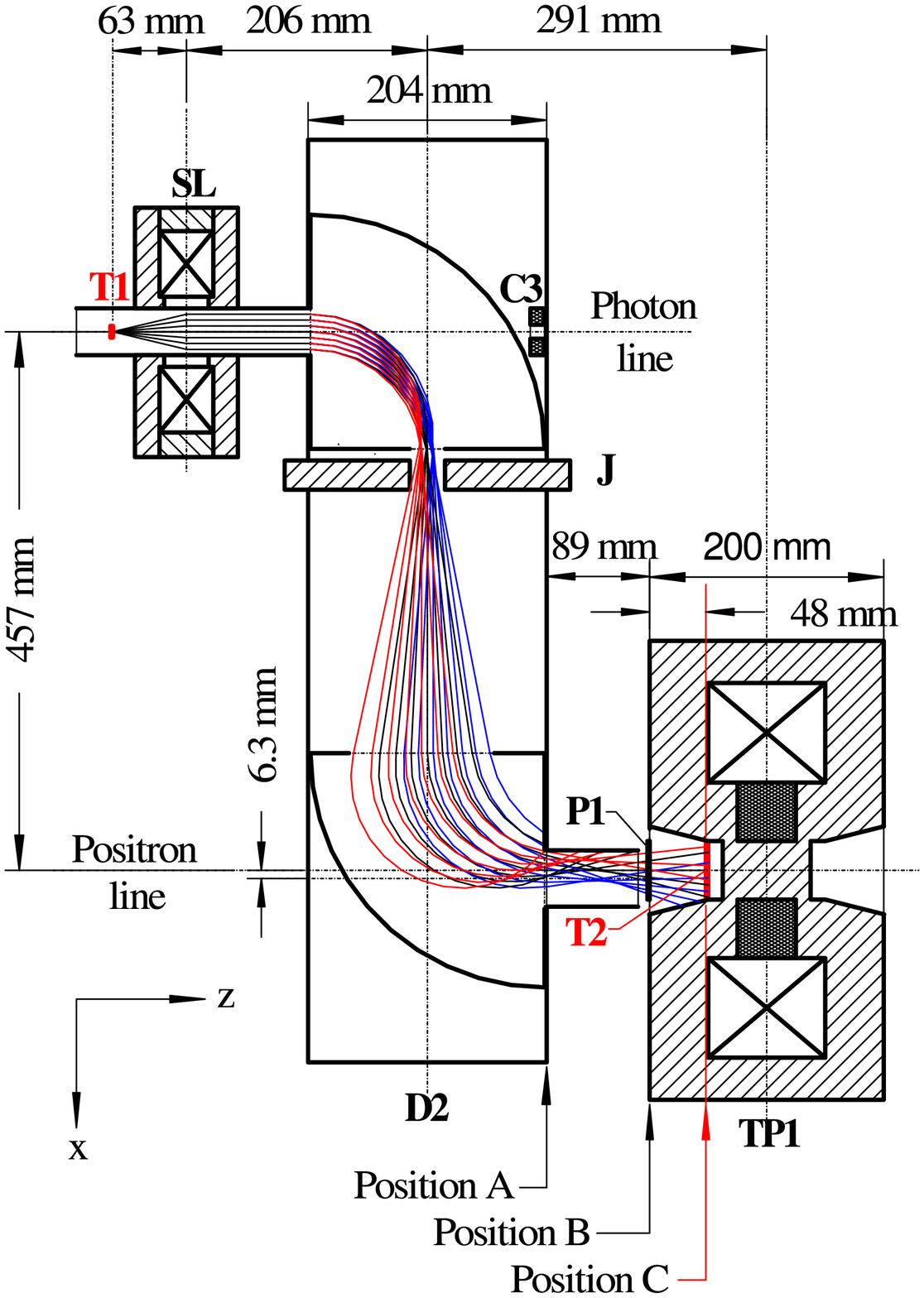} 
  \caption[Top view of the positron/electron transport system.]{Top
    view of the positron/electron transport system showing the
    production target {\sf T1}, the solenoid lens {\sf SL}, the
    collimator {\sf C3} in the $\gamma$-line, the energy-selection
    jaws {\sf J} and the vacuum chamber of spectrometer {\sf D2}, the positron
    detector {\sf P1}, and the positron reconversion target {\sf
      T2}, together with nominal positron trajectories
      for the design energy selection of $\pm5$\,\%. 
}
  \label{fig:spectrometerBIS}
\end{figure}

Additional details of the positron transport are shown in Fig.~\ref{fig:spectrometerBIS}.

During the experiment data were taken as a function of both the solenoid lens current $I_{\rm L}$ and the dipole spectrometer current $I_{\rm S}$.  
The five sets of currents used for the positron analysis are summarized in Table~\ref{tab:transport_sim}.
The lens currents $I_{\rm L}$ were chosen to obtain maximal positron yield in detector {\sf P1} for each spectrometer current $I_{\rm S}$. 
Electron analysis was performed only for $I_{\rm S} = 160$\,A by reversing the spectrometer current (but not reversing the corresponding solenoid current).
Data were collected with ferrofluid in the undulator only for positrons with $I_{\rm S} = 180$\,A.
\begin{table}
\small
  \begin{center}
    \caption{The currents $I_{\rm S}$ in the dipole spectrometer  and $I_{\rm L}$ in the solenoid lens used for positron analysis in the experiment and in the \geant\ simulations (Sec.~\ref{sec_simulation}).
The lens currents $I_{\rm L}$ were chosen to obtain maximal positron yield for each spectrometer current $I_{\rm S}$.
Electron analysis was performed only for $I_{\rm S} = 160$\,A by reversing this current.
The central positron energy $E_{e^{+}}$ selected by the spectrometer as simulated with \geant\ is compared with the values calculated from a calibration with a $^{90}$Sr $\beta$-source.
The range of energies in column five reflects the uncertainty in the determination of the endpoint of the $\beta$ spectrum.
}
    \label{tab:transport_sim}
 \vspace{1mm}
\footnotesize
    \begin{tabular*}{8cm}{c@{\extracolsep\fill}cccc}
      \hline
      \hline
$I_{\rm S}$ (A) & $I_{\rm L}$ (A) & $I_{\rm L}$ (A) & $E_{e^{+}}$ (MeV) & $E_{e^{+}}$ (MeV)   \\
  expt. & expt. & \geant &  \geant\ & $^{90}$Sr $\beta$   \\
      \hline
      100    &  220 &  225   &  4.59   &  4.42 - 4.84 \\
      120    &  260 &  250   &  5.36   &  5.14 - 5.62 \\
      140    &  340 &  300   &  6.07   &  5.81 - 6.36 \\
      160    &  360 &  325   &  6.72   &  6.45 - 7.06 \\
      180    &  374 &  350   &  7.35   &  7.05 - 7.72 \\
      \hline
      \hline
    \end{tabular*}
  \end{center}
\end{table}

The central positron energies corresponding to the spectrometer currents $I_{\rm S}$ were confirmed  by placing
a $^{90}$Sr source at the position of the production target {\sf T1},
which showed that the highest current for 
which a signal was detected at counter {\sf P1} (see Figs.~\ref{prlfig2} and \ref{fig:E166Scheme}) was 52\,A.
The nominal endpoint of the $\beta$ spectrum was 2.752\,MeV, but after correcting for energy loss in source windows and spectrometer air it was determined that $E_{e^-}=2.515\,$MeV ($p_{e^-} = 2.462$\,MeV/$c$) corresponded to a spectrometer current of $I_{\rm S} = 50.5 \pm 1.5\,$A. 
The range of energies reported in column 5 of Table~\ref{tab:transport_sim} were scaled from this result
using Eq.~(\ref{eq_b_is_quadratic}) to characterize the slightly nonlinear dependence of the spectrometer
magnetic field on current.

\subsubsection{Solenoid Lens {\sf SL}}\label{sec:solenoid}

The solenoid lens {\sf SL} provided a point-to-parallel transport of positrons from the production target {\sf T1} to the entrance of the dipole spectrometer {\sf D2}.
The solenoid coils were four double pancakes with 14 radial turns
each, wound using a square copper conductor with a cross section of
$4.76\times4.76\,{\rm mm}^2$, an inner, circular water channel of
3.175\,mm diameter, and a conductor area of 13.9~mm$^2$.  
The coil thickness, including insulation, was 44.6\,mm.
The coils were housed in a 1010-iron flux return, about 20\,mm thick, with overall length of 88.6\,mm and outer radius of 106.5\,mm.  
The lens carried current $I_{\rm L}$ up to 374\,A, corresponding to a maximum current density of 27\,A/mm$^2$.
Cooling water was circulated through the coils with a pressure differential of 5\,bar, such that the operating temperature of the solenoid was only slightly above ambient.

\paragraph*{Magnetic Field:}
The magnetic field of the solenoid lens (and of the spectrometer dipoles) was initially modeled with the 3D-code MERMAID\cite{ref:mermaid}, which is based on a finite-element algorithm that takes into account the geometry and the magnetic properties of different components of a given setup.
At the conclusion of the experiment the field of the lens was mapped by the SLAC Magnetic Measurements group, who measured along the $z$-axis for different $x$-positions (using a one-dimensional Hall probe, providing only the component $B_0(z)= B_{z}(r=0,z)$.  The $(x,y,z)$ coordinate systems used here have the $z$-axis
parallel to the $\gamma$-line (which tilted downwards at an angle of about $0.3\degree$), the $x$-axis horizontal, and the $y$-axis (nearly) vertical.
A complete field map of the components $B_{r}(r,z)$, and $B_{z}(r,z)$ in the current-free space inside the solenoid was extrapolated to second order (assuming azimuthal symmetry) via Maxwell's equations from the field measured on-axis:
\begin{linenomath}
\begin{align}
  \label{eq:solenoid_Br}
  B_{r}(r,z)   &= -\frac{r}{2}\left( \frac{dB_{0}(z)}{dz}-\frac{r^{2}}{8}\frac{d^{3}B_{0}(z)}{dz^{3}}\right) , \\
  \label{eq:solenoid_Bphi}
  B_{\phi}&=  0 ,\\
  \label{eq:solenoid_Bz}
  B_{z}(r,z)   &=  B_{0}(z)-\frac{r^{2}}{4}\frac{d^{2}B_{0}(z)}{dz^{2}} .
\end{align} 
\end{linenomath}
The measured and extrapolated field maps agree well with a computation using MERMAID. 
The parametrization of Eqs.\:(\ref{eq:solenoid_Br})--(\ref{eq:solenoid_Bz}) was used in the simulations described in Sec.\:\ref{sec_simulation}.

\subsubsection{Dipole Spectrometer {\sf D2}}\label{sec_spectrometer}

The dipole spectrometer {\sf D2} was designed to select and transport a 5\,\% energy bite of positrons (or electrons) to a
second beamline offset by 46\,cm from the $\gamma$-line, such that positron polarimetry could be carried out in a low-background environment.

The spectrometer was a system of two opposite-polarity dipole magnets
separated by a 25.4\,cm drift space (see also Fig.\:\ref{fig:spectrometerBIS}) such that the total angular deflection was approximately zero while the transverse displacement was 46.4\,cm.   The resulting ``dog leg'' beam transport included an intermediate focus close to the exit of the first dipole, at which point 25.4-mm-thick tungsten jaws were placed to select the energy of the positrons via remote control of the separation (normally 30\,mm) of the jaws.

The spectrometer contained a vacuum chamber that included the production target {\sf T1}, a cylindrical
entrance pipe of 36\,mm inner diameter that passed through the solenoid lens, and a cylindrical pipe of 48\,mm inner diameter at the exit of the spectrometer.  
The entrance window, upstream of target {\sf T1}, was 25\,mm in diameter, and the exit window was 50\,mm in diameter; both windows were 25-$\mu$m-thick stainless-steel foils.
The vacuum chamber was fabricated such that the offset between the entrance and exit pipes was 464\,mm,
rather than 457\,mm as designed, which resulted in some occulting of the positron beam by the
exit pipe, as discussed further in Sec.~\ref{sec:transport_system}.

The magnet poles were machined as quadrants of a cone of radius 20\,cm and height 3.8\,cm to provide vertical focusing in the entrance and exit fringe fields.  
The smallest gap between the poles was 5.33\,cm, into which gap a vacuum chamber was inserted.
Within each dipole the central orbit was a circle of radius $R_0\approx 13.1\,{\rm cm}$. The tapered magnet gaps resulted in a field gradient with a slope factor near the central orbit  
\begin{linenomath}
\begin{equation}
  n = - \frac{R_0}{B_0} \frac{\partial B}{\partial r} \approx 0.5.
\end{equation}
\end{linenomath}
The coil of each magnet pole was wound with $6\times6$ turns of the same square copper conductor used in the solenoid lens.  
The two dipoles were energized in series, and the flux of each magnet passed through the other via iron return-yoke plates above and below the magnet poles.

The magnet poles were faced with DEN23 tungsten (Tungsten Products, Madison, AL, USA) machined to the inverse of the shape of the poles.  The nominal magnetic permeability of the tungsten was less than 1.05, but consistency between measurements and calculations of the magnetic field in the spectrometer (see below) indicated that the permeability was close to 4.
This deviation from its design value led to modifications of the beam
trajectories in the spectrometer, as discussed further in Secs.~\ref{sec_p1}
and \ref{sec:sim_veri}. 

\begin{figure}
  \centering
  \includegraphics[width=8cm]{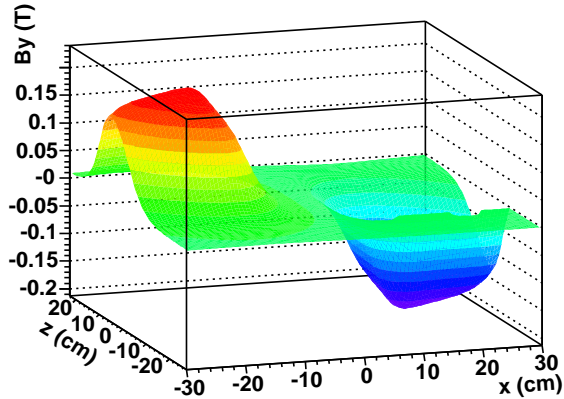}  
  \caption{Profile of the vertical magnetic-field component $B_y$ of the dipole spectrometer \vs\ $x$ and $z$ in horizontal midplane $y=0$.}
  \label{fig:Compare_fieldMap}
\end{figure}

\paragraph*{Magnetic field:}
The spectrometer magnetic field was calculated using the MERMAID program\cite{ref:mermaid}, and was also mapped by the SLAC Magnetic Measurement group at the conclusion of the experiment.  
The measured field map consisted of ten parallel $(x,y)$ planes that were grouped into a 3D lattice covering the region from the production target through the spectrometer to the polarimeter section. 
The calculated magnetic field is in good agreement with the results of field measurements. 
Figure\:\ref{fig:Compare_fieldMap}
shows the profile of $B_y$ in the $(x,z)$ plane at $y=0$.

A field map was calculated using MERMAID for a spectrometer current of $I_{\rm S}=150\,$A. 
In the absence of saturation effects the magnetic field $B_y$ would change linearly with the current. Hall-probe measurements of field \vs\ current at several locations revealed a nonlinear dependence of the form
\begin{linenomath}
\begin{equation}
  \label{eq_b_is_quadratic}
  B_y(I_{\rm S})= a \cdot I^2_S +  b \cdot I_{\rm S} + c ,
\end{equation}
\end{linenomath}
with $a =-1.684 \cdot 10^{-6}$\,T/A$^2$, $b=  1.536 \cdot 10^{-3}$\,T/A and $c=  1.841 \cdot 10^{-3}$\,T. 
At the highest current, 180\,A, the field was about 20\,\% less than that expected for linear behavior.
The field maps used for current $I_{\rm S}$ in the simulations (Sec.~\ref{sec_simulation}) were obtained by scaling the MERMAID map at 150\,A by the factor $B_y(I_{\rm S})/B_y(150\,{\rm A})$ according to Eq.~(\ref{eq_b_is_quadratic}).

\subsubsection{Positron Monitor {\sf P1}}\label{sec_p1}

 Silicon detector {\sf P1} was placed between the exit window of the spectrometer and the positron reconversion target {\sf T2} (see Fig.\:\ref{fig:spectrometerBIS}) to measure the positron flux and spatial distribution.
It consisted of a single 300-$\mu$m-thick layer of silicon that was read out in four quadrants. 
The counter's sensitivity was about 49 positrons per ADC count, and typical signals were a few hundred counts/pulse.

\begin{figure}
  \centering
  \includegraphics[width=8cm]{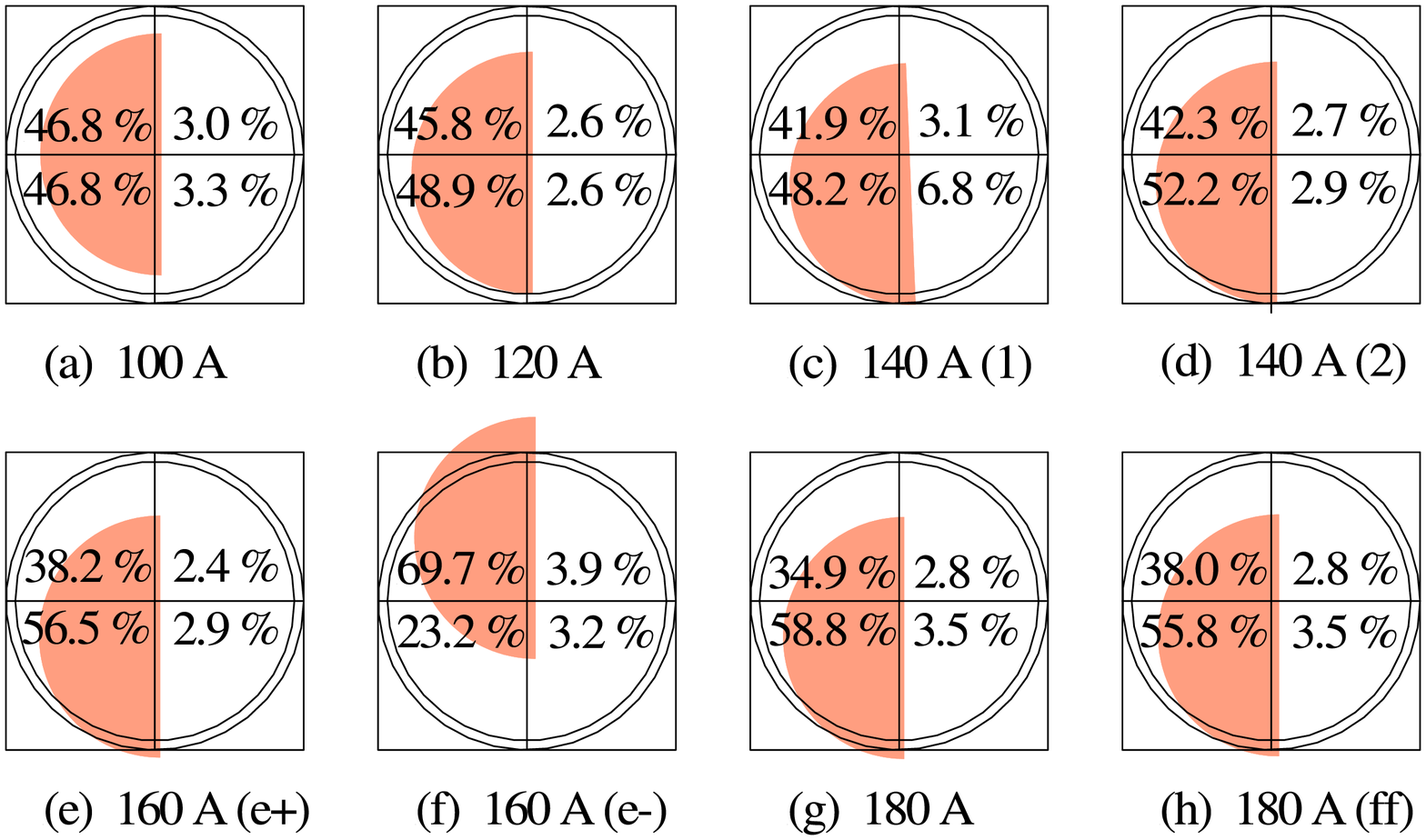}
  \caption[Representation of the distributions of positrons at the {\sf P1} quadrant detector.]{Representation of the distributions of positrons at the {\sf P1} quadrant detector listed in
Table\:\ref{tab:frac_P1_quads}, assuming that the positrons uniformly populated a half-circular region.}
  \label{fig:frac_P1_quads}
\end{figure}

Experimental data from the Si-W detector {\sf P1}, which was read out in quadrants, showed that 90--95\,\% of the positrons passed through its left half, as summarized in Table\:\ref{tab:frac_P1_quads} and illustrated graphically in Fig.\:\ref{fig:frac_P1_quads}.
\begin{table}
\centering
  \caption{Fraction in percent of particles observed in the quadrants of detector {\sf P1} for the various spectrometer currents.  ff = ferrofluid.}
  \label{tab:frac_P1_quads}
 \vspace{1mm}
\footnotesize
  \begin{tabular*}{8cm}{l@{\extracolsep\fill}cccc}
    \hline
    \hline
    $I_{\rm S}$ (A) & top/left & top/right & bottom/left & bottom/right \\
    \hline
    100	&46.8&		3.0&		46.8&		3.3\\	
    120	&45.8&		2.6&		48.9&		2.6\\	
    140(1)	&41.9&		3.1&		48.2&		6.8\\	
    140(2)	&42.3&		2.7&		52.2&		2.9\\	
    160(e$^{+}$)	&38.2&		2.4&		56.5&		2.9\\	
    160(e$^{-}$)	&69.7&		3.9&		23.2&		3.2\\	
    180	&34.9&		2.8&		58.8&		3.5\\	
    180(ff)	&38.0&		2.8&		55.8&		3.5\\	
    \hline
    \hline
  \end{tabular*}
\end{table}
This was caused by deviations of the spectrometer properties from their design
specifications, as described above and discussed further in Sec.~\ref{sec:sim_veri}.
The distribution of electrons was shifted upwards with respect to that of positrons.

\subsubsection{Positron Reconversion Target {\sf T2}}\label{sec_reconversion}

To determine the polarization of the positrons, they were converted back into photons and the
polarization of the latter was measured in a transmission polarimeter based on the iron-core magnet {\sf TP1}.
A~2-mm-thick tungsten composite 
({\sc Densimet D17k}, with a nominal composition of 90.5\,\% W, 7\,\% Ni,
  2.5\,\% Cu) target, {\sf T2}, was placed 12.5\,mm upstream of the magnet to convert most of the positrons into photons, although some positrons were only converted after they entered the iron magnet.

\subsubsection{Polarimeter Magnets}\label{sec_analyzer_magnet}

Iron-core solenoid magnets were employed for the measurement of the photon and (reconverted) positron beam polarizations according to Eqs.~(\ref{eq:p22}) and (\ref{p23}), where the average longitudinal polarization $P_{e^-}^{\rm Fe}$ of atomic
electrons in the iron is related to longitudinal magnetic field  $B$ according to Eq.~(\ref{eq:p22a}).

\begin{figure}
  \centering
  \includegraphics[width=8cm]{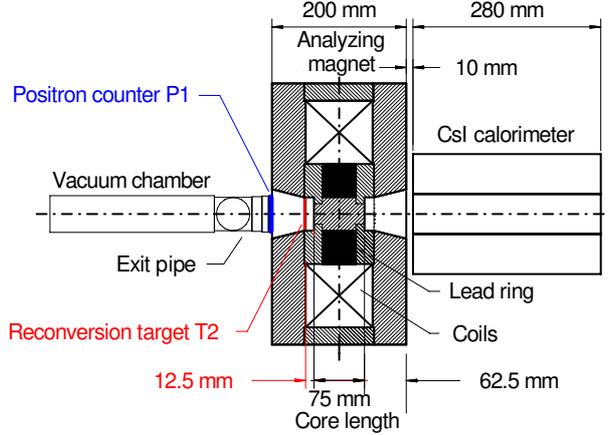} 
  \caption[Sketch of the positron polarimeter.]{Sketch of the positron polarimeter.  Positrons entered from the left, passing in succession through the exit window of the spectrometer, the Si-W counter {\sf P1}, the reconversion target {\sf T2}, after which the reconverted photons passed through the iron-core magnet {\sf TP1} and were absorbed in the CsI calorimeter on the right.}
  \label{fig:polarimeter_scheme}
\end{figure}

\begin{figure}
\centering
\includegraphics[width=8cm]{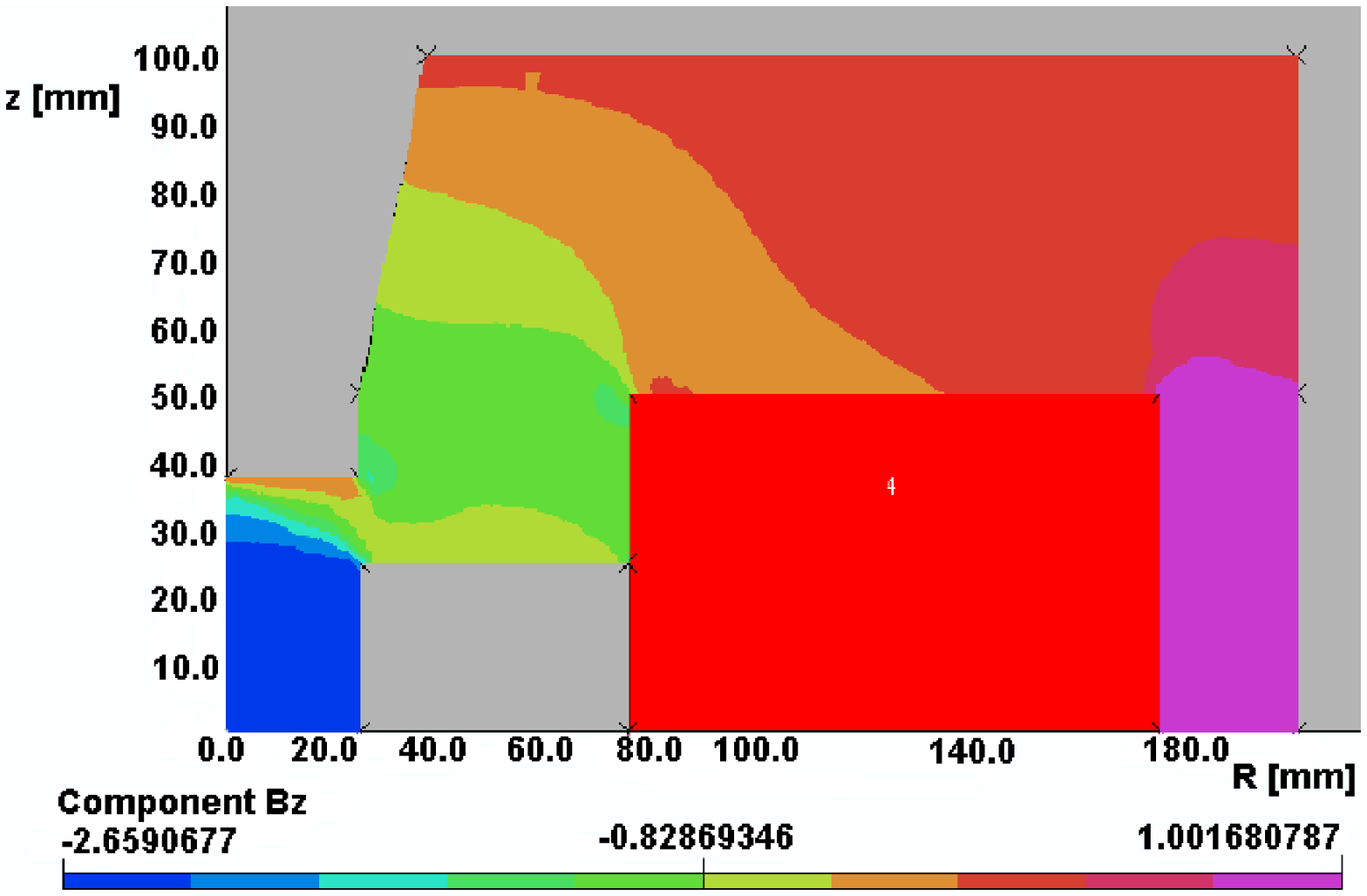} 

  \includegraphics[width=8cm]{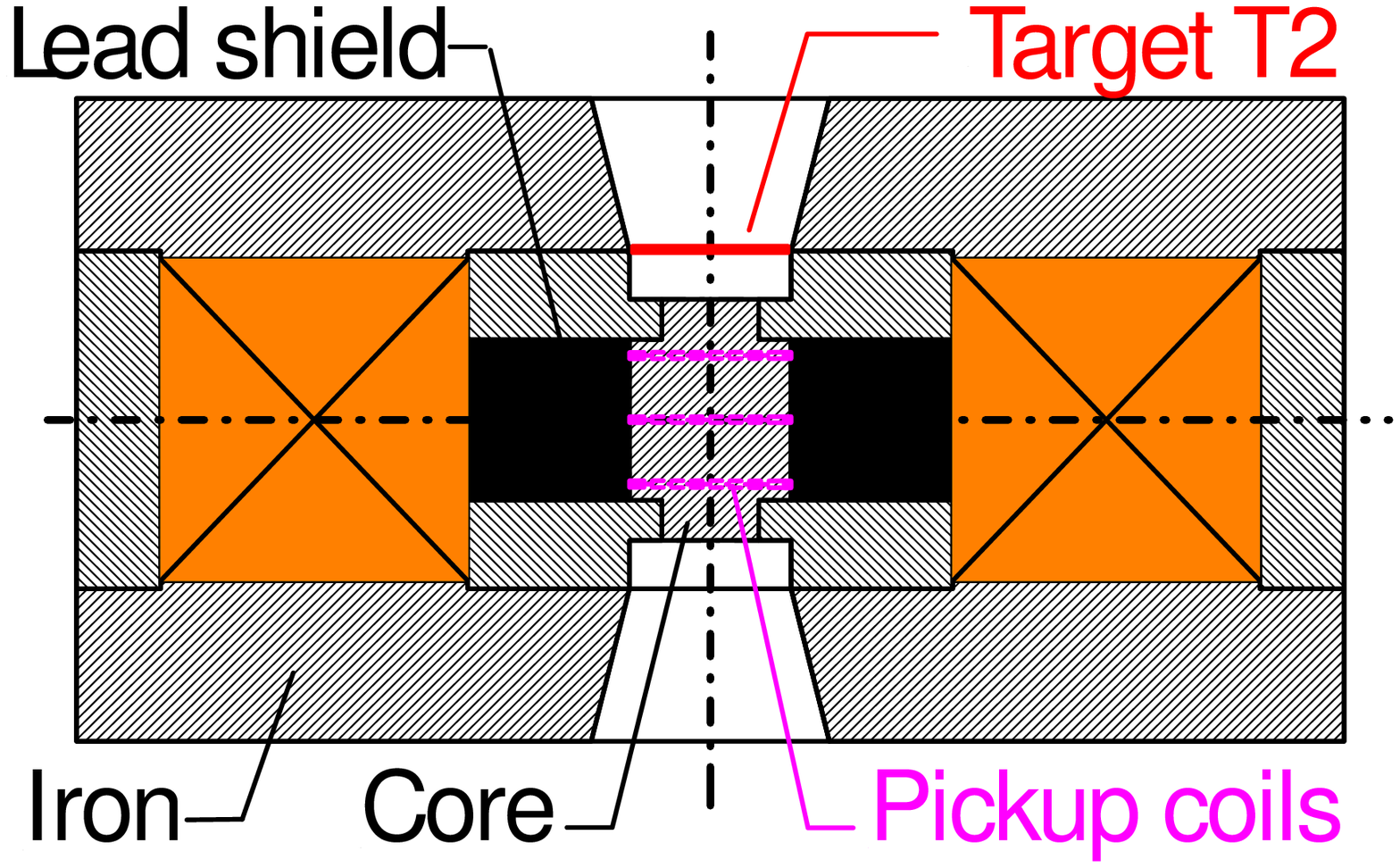} %
  \caption[2d-field model in the $r$-$z$ plane of magnet {\sf TP1}.]{Top: 2d-field model in the $r$-$z$ plane of magnet {\sf TP1}.
Bottom: Horizontal section of magnet {\sf TP1} showing the location of the pickup coils.
The top figure corresponds to the first quadrant of the bottom figure.}
  \label{fig:anal_mag}
\end{figure}

The iron core of the positron analyzer {\sf TP1} was 50\,mm in diameter and 75\,mm in length, as shown in Fig.~\ref{fig:polarimeter_scheme}, while that of the
photon analyzer {\sf TP2} was 50\,mm in diameter and 150\,mm in length. 
These magnets were fabricated at the Efremov Institute \cite{Efremov} based on modeling which showed that the
iron would be saturated over most of the cylindrical core, as shown in upper part of Fig.\:\ref{fig:anal_mag}.

Photons traversing the structure outside the core region were strongly suppressed, as they encountered more iron and an additional lead shield (see Figs.\:\ref{fig:polarimeter_scheme} and \ref{fig:anal_mag}).

\begin{figure}
  \centering
  \includegraphics[width=8cm]{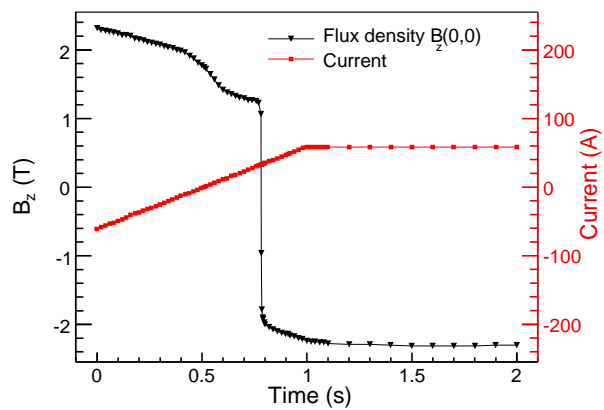} 
  \caption[OPERA-2d calculation of the field in magnet {\sf TP1} for a linear current ramp.]{OPERA-2d \cite{Vector_Fields} calculation of the field in magnet {\sf TP1} for a linear current ramp from $-60$\,A to $+60$\,A over 1\,s.
}
  \label{fig:current_and_bz}
\end{figure}

The main parameters of the polarimeter magnets {\sf TP1} and {\sf TP2} are listed in Table\:\ref{tab_analmag_one}.
\begin{table}
  \centering      
  \caption{Parameters of the polarization analysis magnets {\sf TP1} and {\sf TP2}.
}
  \label{tab_analmag_one}
 \vspace{1mm}
\footnotesize
  \begin{tabular*}{8cm}{l@{\extracolsep\fill}cc}
    \hline
    \hline
    Parameter (Unit)  &  {\sf TP1}  &  {\sf TP2} \\
    \hline
    Overall length (mm)  &    200 &    275 \\
    Overall diameter (mm) &  392 &    320 \\
    Iron-core length (mm)  & 75 & 150 \\
    Iron-core diameter (mm)  & 50 & 50 \\
    Length of internal Pb absorber   &   50 & 125 \\
    Overall mass (kg)   &   175         & 195 \\
    \\  
    Number of coils & 2 & 2 \\ 
    Coil length (mm)  & 49 & 86 \\
    Coil inner diameter (mm) & 152 & 152 \\
    Coil outer diameter (mm)  & 322 & 248 \\
    Number of turns per coil & 160 & 175 \\
    Conductor dimensions (mm) & $4 \times 4$ & $4 \times 4$ \\
    Coolant bore diameter (mm) &  2.5 &  2.5 \\
    Water cooling circuits  &         4 &     4 \\
    Water flow rate (l/min) &         $\sim 2$ &   $\sim 2$   \\
    \\
    Operating current (A) & $\pm 60$ & $\pm 60$\\
    Power (kW) & 1.62 & 1.37 \\
    Current reversal time (s) &     1.0 & 1.0 \\
    Time between reversals (min)  &  5 & 5 \\
    \\
    Field $B_z^{\rm max}$ at center (T) &   2.324 &    2.165\\
    On-axis mean field $\ave{B_z}$ (T) &     2.071 &         2.040\\
    Air field $B_0$ at center (T)  &   0.097 &    0.100\\
    \\
    Number of pickup coils  &         3 & 3 \\
    Pickup-coil diameter (mm) & 48.5 & 48.5 \\
    Number of turns per pickup coil & 160 & 160 \\
    $z$-Position of pickup coils (mm) &    0, $\pm 20$ & 0, $ \pm 57.5$ \\
    \\
    $\ave{P_{e^-}^{\rm Fe}}$ (on axis) & 0.0736       & 0.0723 \\
                             & $\pm 0.0015$ & $\pm 0.0015$ \\
    \hline
    \hline
  \end{tabular*}
\end{table}
Both magnets were energized through fast linear amplifiers\cite{Copley} which permitted rapid polarity reversal. 
For transmission polarimetry, the field direction was automatically reversed every 5 minutes with a linear current ramp over 1 second, as shown in Fig.\:\ref{fig:current_and_bz}.
The nonlinear response was due to eddy effects.

\begin{figure}
  \subfloat[(a)]{\includegraphics[width=8cm]{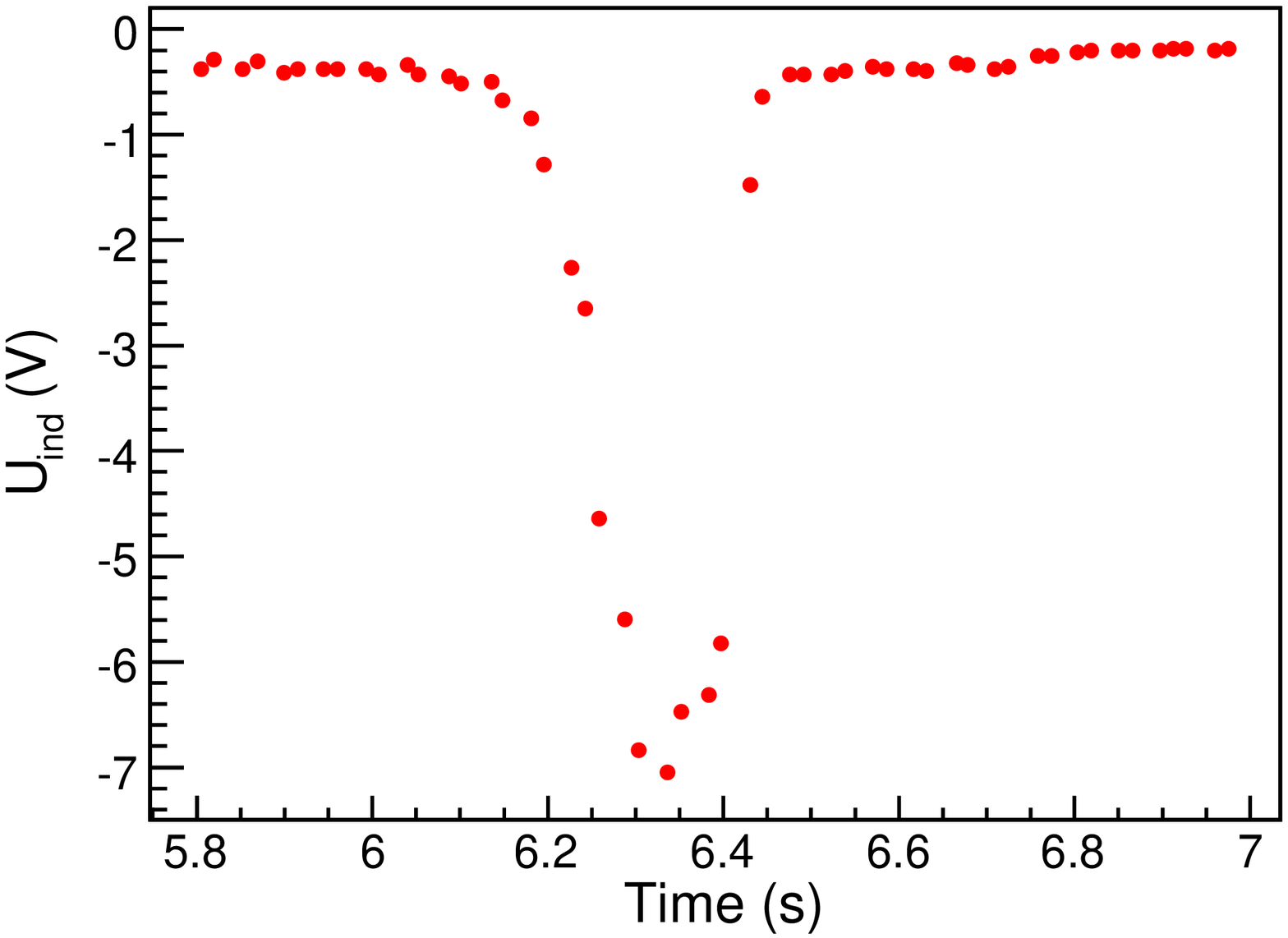}}   
  \hfill
  \subfloat[(b)]{\includegraphics[width=8cm]{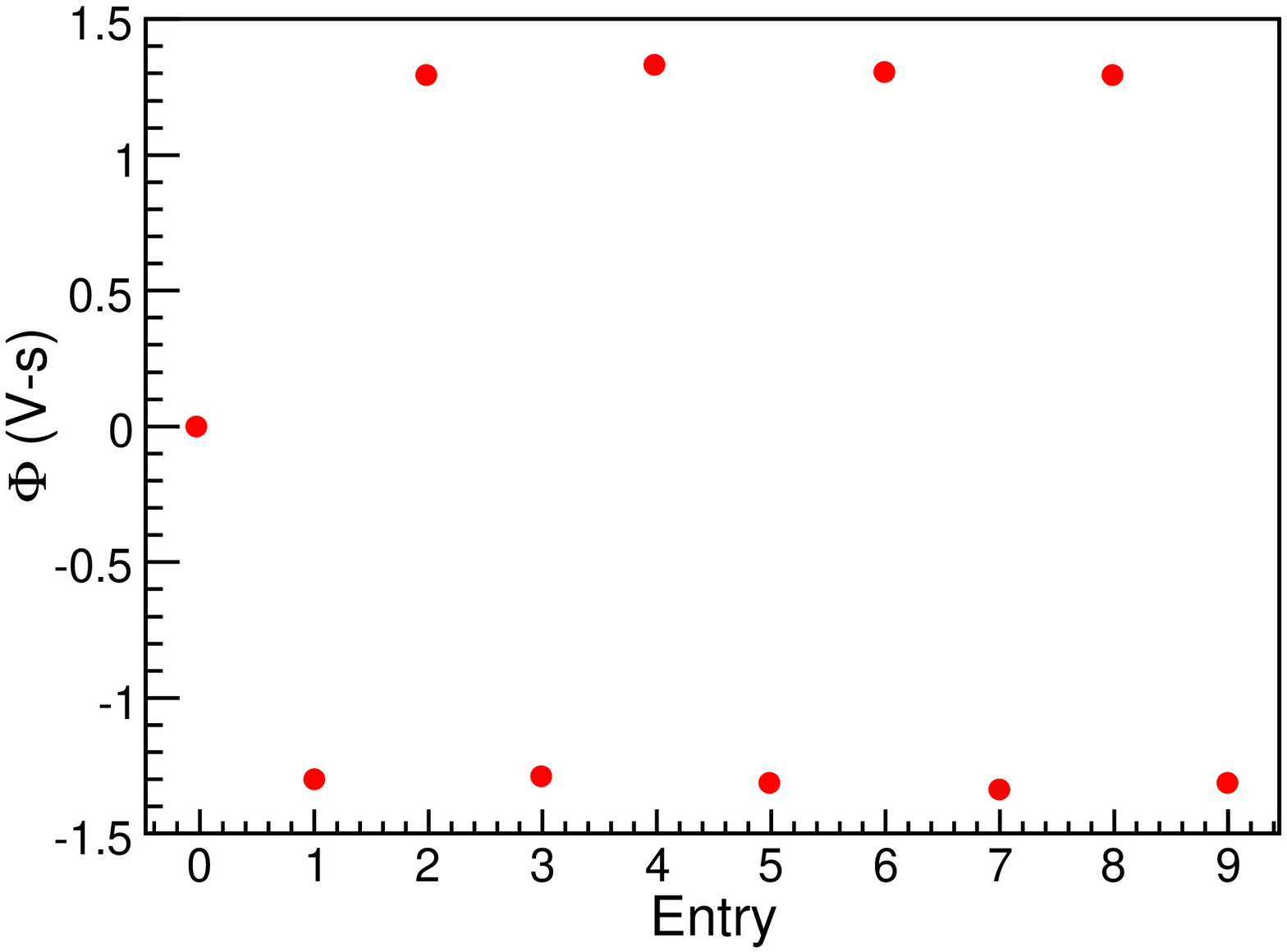}}    
  \caption[Induced-voltage signal upon field reversal in one of
  several pickup coils in magnet {\sf TP1}]{Induced-voltage signal upon field reversal in one of several pickup coils surrounding the iron core of magnet {\sf TP1}: (a) voltage \vs\ time (left); (b) Magnetic flux $\Phi = \int V dt 
= B A$  for several polarity flips (right).}
  \label{fig:pickup_coils}
\end{figure}

\begin{figure}
  \centering
  \includegraphics*[width=8cm]{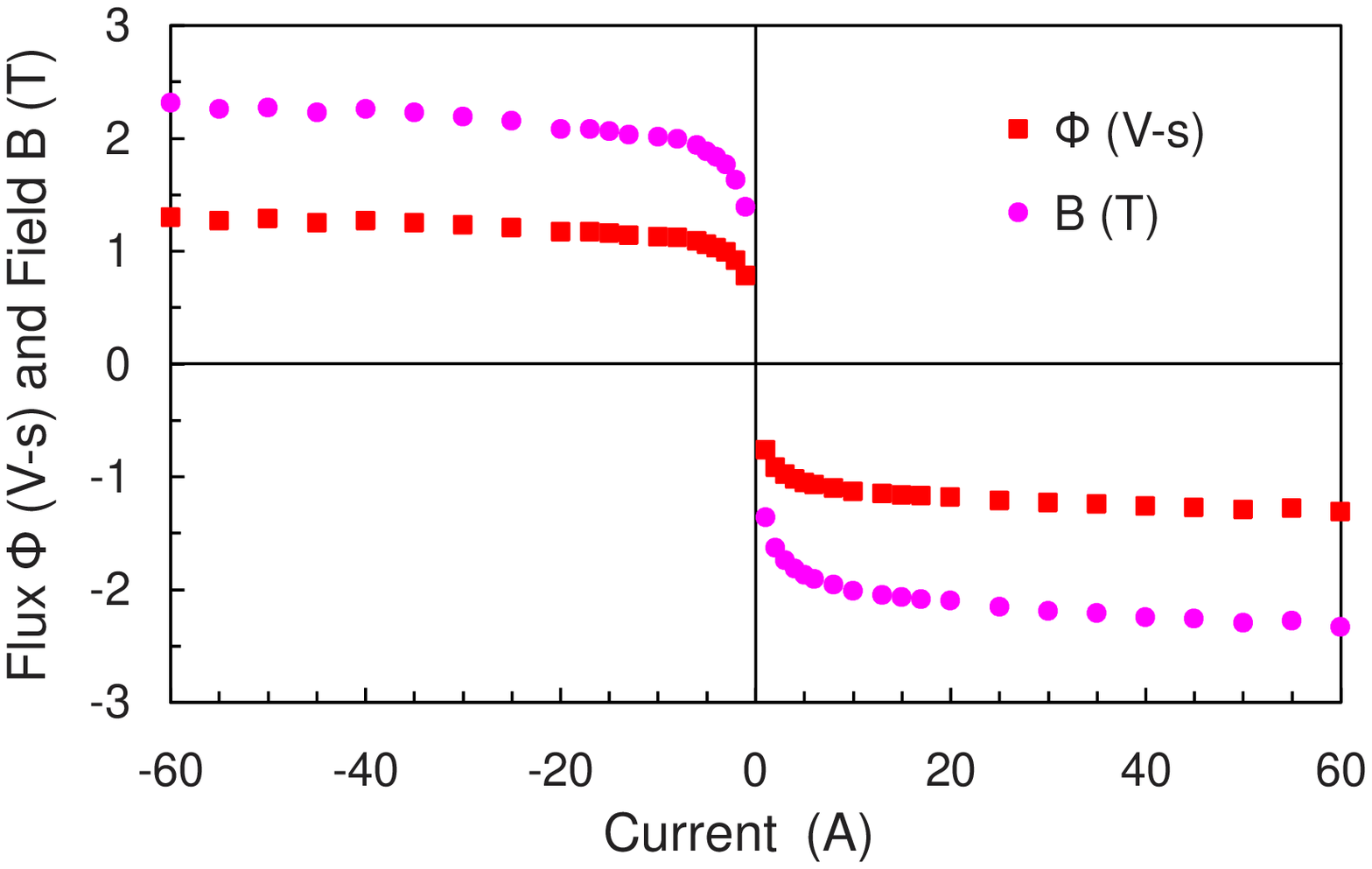} 
  \caption{Flux measurements and derived field values at the center of magnet {\sf TP1} as a function of magnet current.
}
  \label{fig:B_vs_I}
\end{figure}

\paragraph*{Magnetic Field:}
The magnetic field of the iron was measured and monitored with several pickup coils surrounding the core of each magnet, shown for magnet {\sf TP1} in the bottom part of Fig.\:\ref{fig:anal_mag}. 
The induced-voltage signal $V(t) = d \Phi / dt$, where $\Phi$ is the magnetic flux through the pickup coil, upon field reversal was digitized at a readout frequency of 50\,Hz\cite{Esd}. 
An additional pulse with triangular shape from a waveform generator with known repetition frequency was used to calibrate the time scale of these measurements. 
An example of these magnetic flux measurements is shown in Fig.\:\ref{fig:pickup_coils}.
An excitation curve of flux \vs\ current is shown in Fig.\:\ref{fig:B_vs_I}
for the central core region of the positron analyzer. 
The magnetic field $B$ was obtained from the time integral of the
induced voltage, $\int V dt = \Phi = B A$, and the known cross section area $A$ of the pickup coil. 
The central magnetic field at the operating current of $\pm 60$\,A was determined in this way to be 2.324\,T for the positron analyzer, and 2.165\,T for the photon analyzer, with a typical relative measurement error of 1\,\%.

\begin{figure}
  \centering
  \includegraphics[width=8cm]{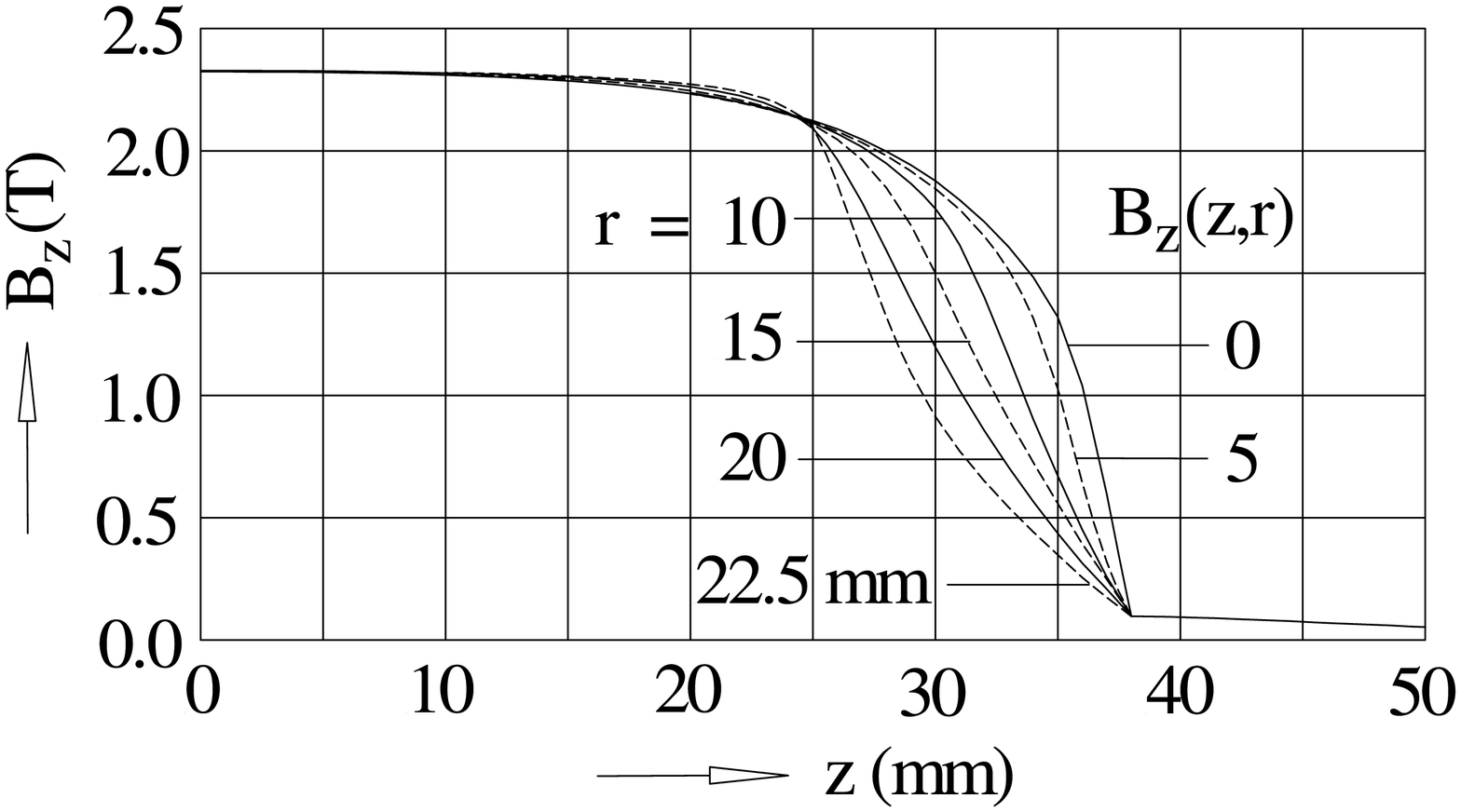} 
  \caption[Longitudinal field component $B_{z}$ in the positron
    analyzer.]{Longitudinal field component $B_{z}$ in the positron
    analyzer (modeled with OPERA-2d \cite{Vector_Fields}) as a
    function of $z$ for different radial  distances from 0 to
    22.5\,mm. 
}
  \label{fig:Bz}
\end{figure}

To obtain $\langle B_z \rangle$ averaged over the entire cylindrical core volume, detailed field modeling was performed with the OPERA-2d code\cite{Vector_Fields}. 
The nominal $B$-$H$ curve for the soft magnet iron (1010 steel) was adjusted until satisfactory agreement with the measured  fields in the pickup coils was obtained. 
Results of the field modeling of the positron analyzer are shown in Fig.\:\ref{fig:Bz},
which plots the longitudinal field component $B_z$ as a function of $z$ for different radii. 
The center of the core is at $z = 0$, and the end faces are located at $\pm 37.5 $\,mm. 
At the operating current of 60\,A, the core remained near its central saturation value $B_z (0,0) = B_z^{\rm max} = 2.324$\,T for much of its length. 
Near the exit face at $z = \pm 37.5$\,mm, however, the longitudinal field component diminished more or less rapidly, depending on the radial distance. 
This general behavior is also apparent in the 2d-field distribution shown in Fig.\:\ref{fig:anal_mag}. 
For photons propagating on axis or with a constant radial offset, the average $\left \langle B_z \right \rangle$ is listed in Table\:\ref{tab_analmag_two}.
\begin{table}
  \centering      
  \caption{Field integrals and average electron polarization of the magnetized iron core. For the positron analyzer the integration covered a cylindrical volume of radius $r$ and the full core length of 75\,mm along $z$. For the photon analyzer only the axial case was evaluated for the full core length of  150\,mm.}
  \label{tab_analmag_two}
 \vspace{1mm}
\footnotesize
  \begin{tabular*}{7.5cm}{c@{\extracolsep\fill}ccc}
    \hline
    \hline
    \rule[-2.5mm]{0mm}{7mm}
    Radius  & $\langle B_z \rangle$ & $\langle B_z - B_0 \rangle$ & $\langle P_{e^-}^{\rm Fe} \rangle$    \\
       {(mm)} & (T) & (T) & \\    
    \hline
    \multicolumn{4}{c}{Positron Analyzer {\sf TP1}}\\
    \hline   
    $r = 0  $         & 2.071 & 1.974 & 0.0736 $\pm$ 0.0015  \\
    $0 < r \leq 5$    & 2.046 & 1.949 & 0.0726 $\pm$ 0.0015  \\           
    $0 < r \leq 10$   & 2.025 & 1.928 & 0.0719 $\pm$ 0.0015  \\           
    $0 < r \leq 15$   & 2.001 & 1.904 & 0.0710 $\pm$ 0.0015  \\           
    $0 < r \leq 20$   & 1.977 & 1.880 & 0.0701 $\pm$ 0.0015  \\           
    $0 < r \leq 22.5$ & 1.963 & 1.866 & 0.0695 $\pm$ 0.0015  \\           
    \hline
    \multicolumn{4}{c}{Photon Analyzer {\sf TP2}}\\
    \hline  
    $r = 0$ &       2.040    &      1.940 &         0.0723 $\pm$ 0.0015     \\
    \hline
    \hline
  \end{tabular*}
\end{table}

\begin{figure}
  \centering
  \includegraphics[width=0.40\textwidth]{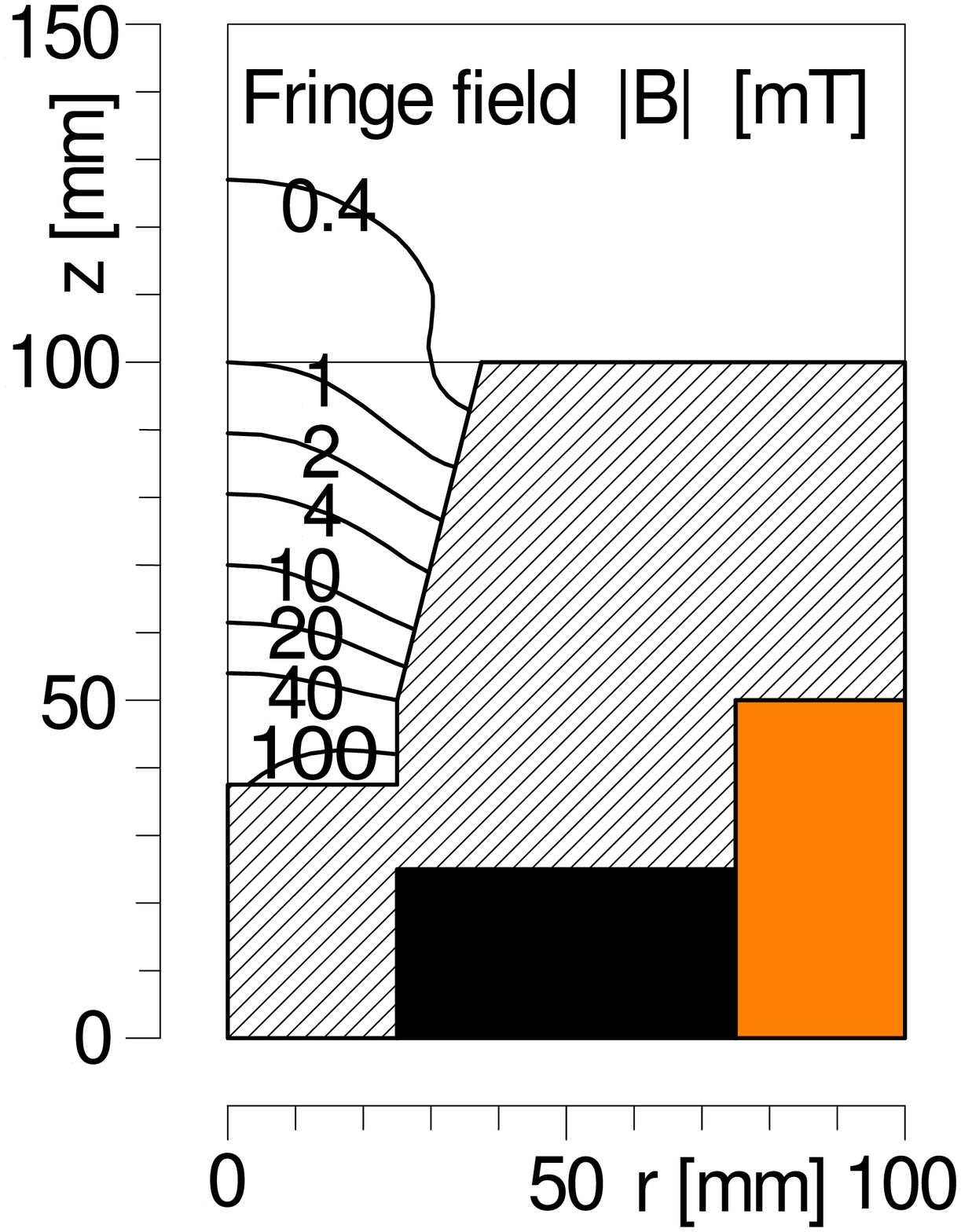} 
  \caption[The fringe-field distribution of the positron analyzer {\sf TP1}, as modeled with OPERA-2d.]{The magnitude $|B|$ of the fringe-field distribution in the first quadrant of the $r$-$z$ plane of the positron analyzer {\sf TP1} (bottom part of Fig.~\ref{fig:anal_mag}), as modeled with OPERA-2d \cite{Vector_Fields}.}
  \label{fig:fringe_field}
\end{figure}

To account for possible systematic effects from fringe fields on positron or electron trajectories, a detailed field map was also generated for the exterior air space surrounding the positron analyzer, which served as input to simulation studies. Figure\:\ref{fig:fringe_field}
shows the magnitude $|B|$ of the fringe field in the $r$-$z$ plane, based on a computation of the axial and radial field components.
At the reconversion target ($z = 52 $\,mm), a fringe field of  44\,mT was determined from the model. Further out, it dropped to 1\,mT  at the start of the trombone-shaped entry throat ($z = 100 $\,mm), which was the location of the positron counter {\sf P1} (Fig.\:\ref{fig:polarimeter_scheme}). 
These field values have been confirmed with Hall-probe measurements.

The electron spin polarization $P_{e^-}^{\rm Fe}$ of the iron is proportional to $\langle B - B_0 \rangle$ according to Eq.~(\ref{eq:p22a}), where $B_0$ is the ``air field", due to the current in the magnet coils, that would exist in the absence of the iron.  
The air field $B_0$ was calculated from the known solenoid parameters shown in Table\:\ref{tab_analmag_one}, and was about 5\,\% of B.
As the air field varied in the core region by less than 11\,\% from its central value, it was sufficient to use a constant air-field correction with a systematic error of less than 0.5\,\% in $P_{e^-}^{\rm Fe}$.

The polarization of atomic electrons in iron was calculated from the measured magnetic fields according to Eq.~(\ref{eq:p22a}) as $P_{e^-}^{\rm Fe} = 0.03727 \langle B\,[{\rm T}] - B_0\,[{\rm T}] \rangle$, with results summarized in Table\:\ref{tab_analmag_two}. 
On axis, the average electron polarization was $0.0736 \pm 0.0015$ for the positron analyzer {\sf TP1}, and $0.0723 \pm 0.0015$ for the photon analyzer {\sf TP2}. 
The errors reflect uncertainties in the field measurements and field modeling. 
Away from the axis there was a slight drop in the average polarization, as detailed in Table\:\ref{tab_analmag_two} for the positron analyzer. 
For the photon analyzer only the on-axis field was relevant, since the beam was narrowly collimated in that case.

Although the average $P_{e^-}^{\rm Fe}$ over the iron core of magnet {\sf TP1} was known to $\pm\,0.0015$ (Table~\ref{tab_analmag_two}), the systematic uncertainty was taken to be $\Delta P_{e^-}^{\rm Fe} = 0.0021$ (\ie, 3\,\%) to account for uncertainties in the knowledge of the spatial distribution of the photons within the cylindrical core.

\subsubsection{The CsI(Tl) Calorimeter}{\label{sec:csi_cal}}

The total energy of each pulse of photons that emerged from the
analyzer magnet {\sf TP1} after the positron reconversion target {\sf
  T2} was measured by a CsI(Tl) calorimeter (shown in
Figs.~\ref{prlfig2}, \ref{fig:E166Scheme}, 
and \ref{fig:polarimeter_scheme}) made of nine crystals arranged in a 3$\times$3 array.

\begin{figure}
  \begin{center}
    \includegraphics[width=8cm]{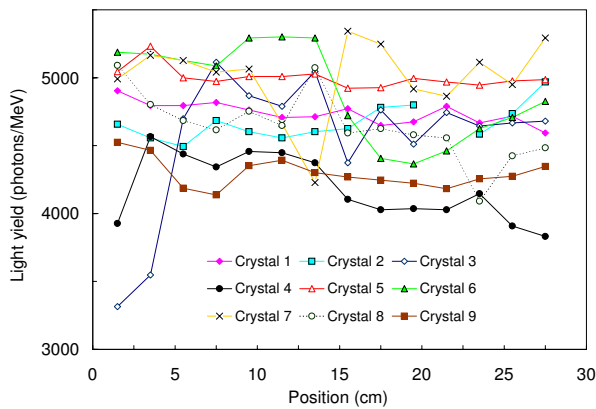}   
    \caption{The photoelectron yield per MeV \vs\ longitudinal position for the nine CsI(Tl) crystals.}
    \label{fig:lightyield}
  \end{center}
\end{figure}

\paragraph*{Crystals:}
The most important properties of the crystals (supplier: Monokristal Institute, Kharkov, Ukraine) are summarized in Table\:\ref{tab:CsI}.
\begin{table}
  \begin{center}
    \caption{Parameters of the CsI(Tl) crystals.}
    \label{tab:CsI} 
 \vspace{1mm}
\footnotesize
    \begin{tabular*}{8cm}{l@{\extracolsep\fill}l}
      \hline 
      \hline 
      Thallium (dopant)        & 0.08\,mol\,\%  \\
      Light yield              & 4000--5500 photons/MeV \\
      Peak emission wavelength & 550\,nm \\
      Length                   & 280\,mm (15.11\,r.l.)\\
      Height                   & 60\,mm        \\
      Width                    & 60\,mm        \\
      \hline
      \hline
    \end{tabular*}
  \end{center}
\end{table}
The homogeneity of the light yield was tested\cite{laihem_thesis} by moving a $^{60}$Co source along the sides of the crystals and analyzing its spectrum. 
From 14 measured points per crystal the homogeneity of each crystal was found to be  within 10\,\% of the average in most cases, with some exceptions (Fig.\:\ref{fig:lightyield}).
The average light yield of the nine crystals was found to vary between 4000 and 5500 photons per MeV of deposited energy.
This variation was accounted for in the calibrations of the individual crystals.

\paragraph*{Mechanical Assembly:}
Each crystal was wrapped with two layers of white Tyvek paper, which provided diffuse reflections at the crystal walls
and increased the scintillation light collection. 
To avoid cross-talk between the crystals, each one was additionally wrapped with a thin copper foil (30\,$\mu$m) which also acted as an electromagnetic shield. 
The wrapping caused about 1.2\,mm of dead space between the crystals. 
The crystals were stacked with the use of plastic spacers inside a brass chamber with 6\,mm wall thickness. 
The front wall, facing the analyzer magnet, was thinned down to 2\,mm in the sensitive area. 
The box was light-tight but not air-tight (a hygroscopic material inside the box kept the humidity low).

Ninety percent of the energy of the reconverted photons was deposited
in the central CsI crystal, so the crystal with the highest light yield
was placed in the center of the array.  The next-best-quality crystals
were used as the four neighbors of the central crystal, and the four
corner crystals had the lowest quality.

\begin{figure*}
  \begin{center}  
    \includegraphics[width=13cm]{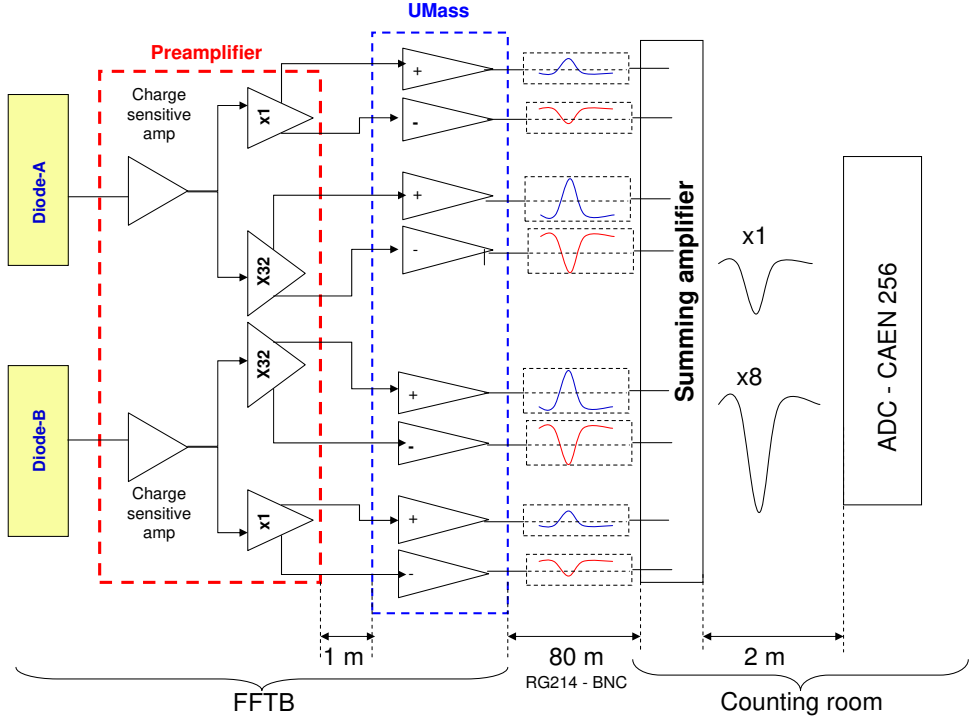} 
    \caption{Scheme of the readout electronic chain:  photodiode, preamplifier, UMass board, summing amplifier and ADC.
}
    \label{fig:electronic_chain}
  \end{center}
\end{figure*}

\paragraph*{Electronic Readout:}
The front end of the electronic readout of the scintillation light (Fig.\:\ref{fig:electronic_chain})
was adapted from the \babar\ experiment\cite{babar_CsI}. 
A module of two photodiodes (called {\sf A} and {\sf B} in Fig.\:\ref{fig:electronic_chain})  with a total active area of 20$\times$20\,mm$^2$ was coupled to the crystal via a thin polystyrene plate using an optical grease. 
The PIN diodes (Hamamatsu model S2744-08) were operated with a reverse bias voltage of $-50$\,V with the cathode grounded. Typical values for the dark current of 3\,nA and a capacitance of 85\,pF for the doublet were reported by the manufacturer. 
The signals from each diode were fed into charge-sensitive preamplifiers, each with two bipolar outputs, one with low gain (LG, gain 1) and the other with high gain (HG, gain 8).

These bipolar signals were fed via flat cables (about 1\,m long) onto the so-called UMass board (adapted from a \babar\ crystal test setup) which drove a total of 72 coaxial cables (one for each polarity of the bipolar signals) over a length of 80\,m from the FFTB tunnel to the counting room. 
There, a summing amplifier (custom made for this experiment) combined the two polarities of each signal by adding the inverse of the positive signals to the negative ones, and by adding the signals with the same gains from diodes {\sf A} and {\sf B} of a crystal.
The resulting 18 pulses were attenuated by 40\,dB and then digitized by three, 8-channel, charge-integrating ADCs (CAEN model  V265).  
The digital output of each ADC channel was two 12-bit words, where the second word was obtained from digitization of the input signal with 8 times higher gain, corresponding to 15-bit resolution over the range of the 12-bit word.  
Thus, digitized signals from each CsI crystal were recorded for four resolutions, designated as LG-LS (Low Gain - 12\,bits), LG-HS (Low Gain - 15\,bits), HG-LS (High Gain - 12\,bits), and HG-HS(High Gain - 15\,bits). 
However, in the final analysis only the 12-bit-resolution data taken with low gain (\ie, LG-LS) were used.    

\paragraph*{Energy Calibration:}
The typical energy deposition in a CsI crystal by photons from reconverted positrons was a few hundred MeV per electron-beam pulse.

\begin{figure} 
  \subfloat[(a)]{\includegraphics[width=8cm]{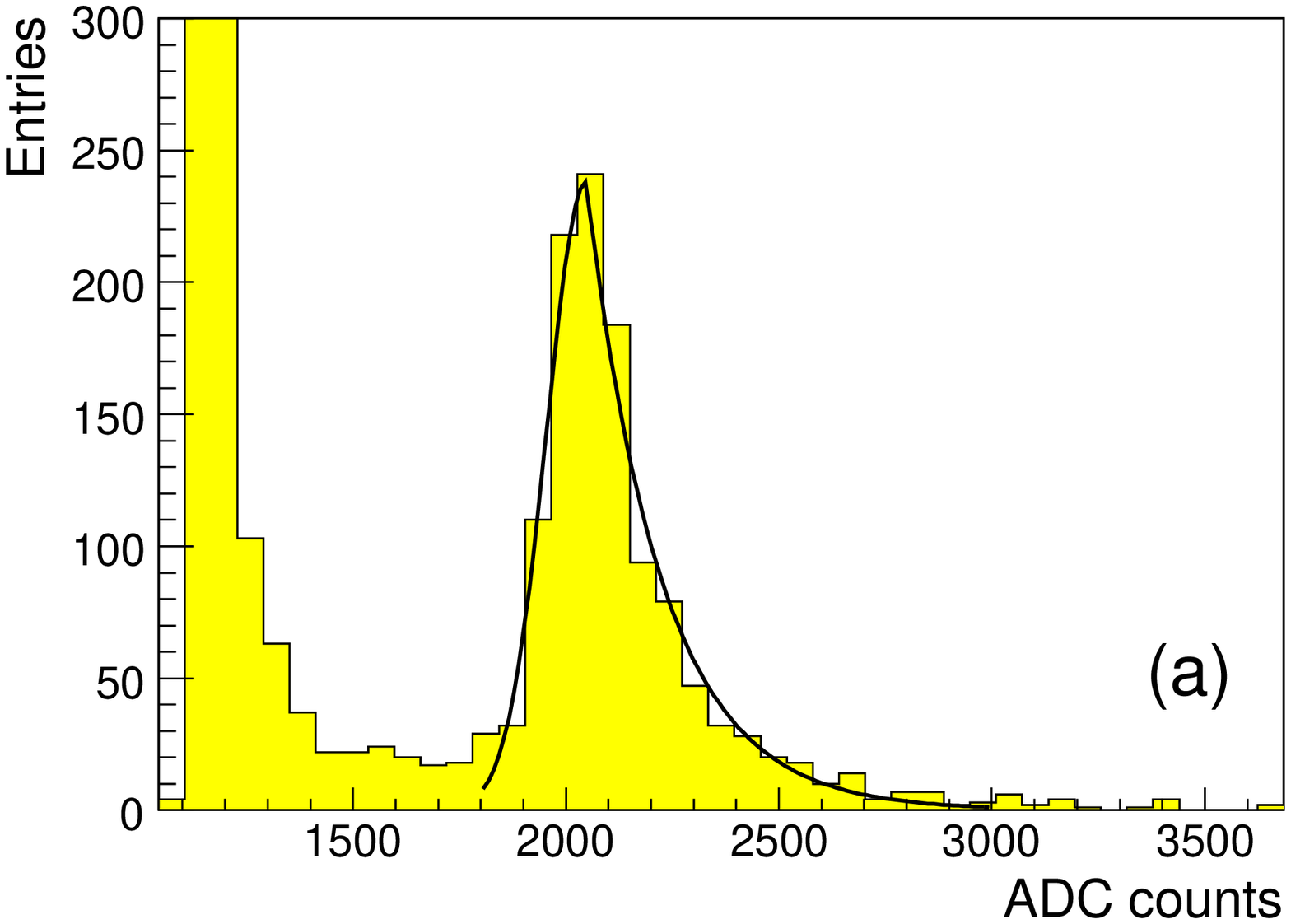}} 
  \hfill  
  \subfloat[(b)]{\includegraphics[width=8cm]{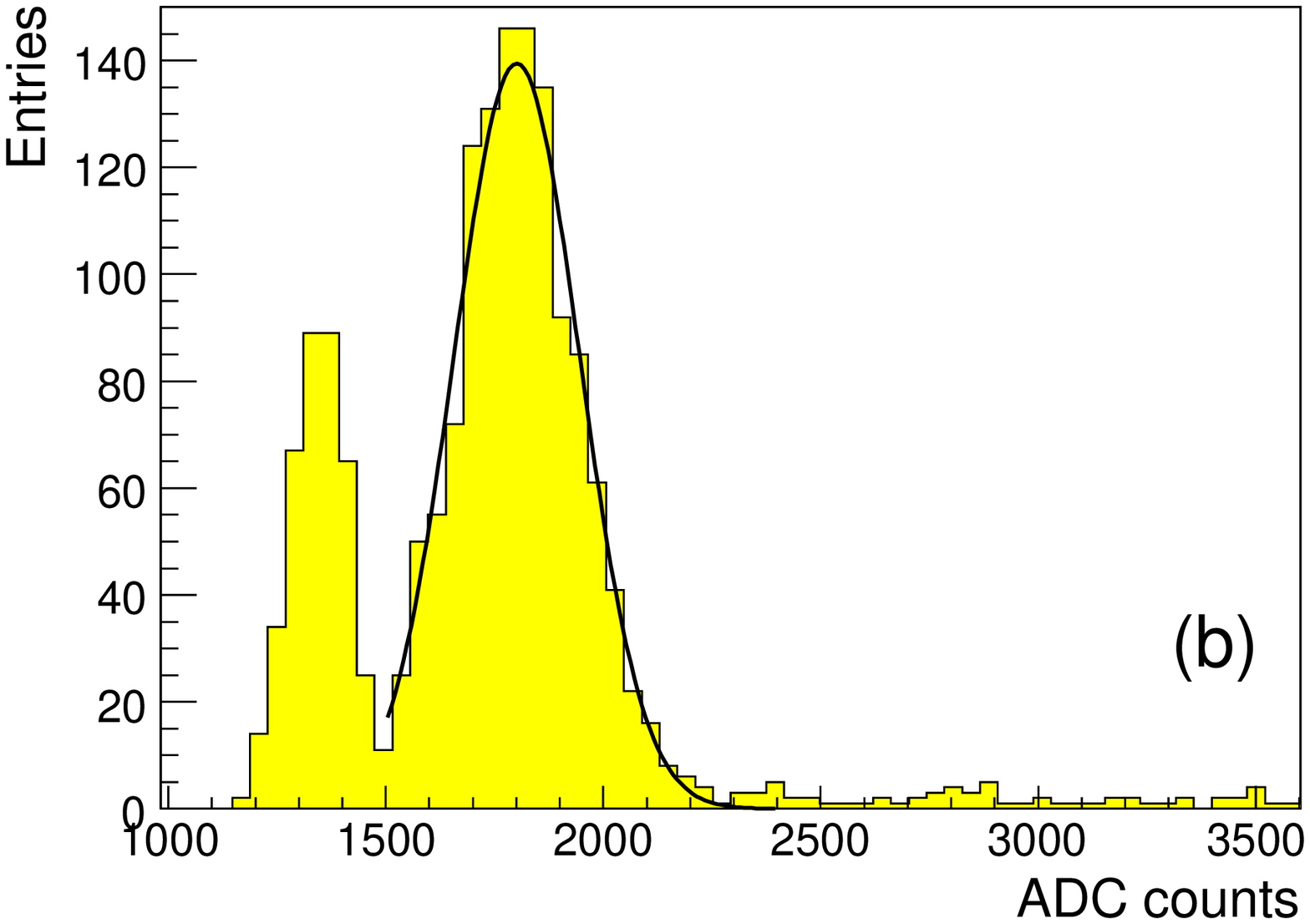}} 
\caption[Examples of calibration spectra from CsI crystal 8.]{Examples of calibration spectra from CsI crystal 8.
(a): Cosmic-muon spectrum (40\,dB attenuation);
(b): $^{228}$Th spectrum (20\,dB attenuation).  The pedestal channel was 1160.
}
\label{fig:csi_csom_spec}
\end{figure}

Prior to data taking with beam, calibration constants for the calorimeter were obtained from high-statistics spectra of cosmic-ray muons using HG-HS resolution, as illustrated in Fig.\:\ref{fig:csi_csom_spec}(a).
They were confirmed for the outer eight crystals via $^{228}$Th decays (2.861\,MeV photo peak) at the same resolution but with only 20\,dB attenuation of the signals (Fig.\:\ref{fig:csi_csom_spec}(b)). 
A more accurate calibration procedure based on high-statistics cosmic runs and pedestals accumulated over nine weeks, as well as measurements using a test beam with energy up to 6\,GeV at DESY, is described in\cite{jans_thesis}. 

The cosmic muons were triggered by a telescope of two scintillation counters. 
The cosmic signal in each crystal was well separated from the pedestal in both low gain and high gain. 
Figure\:\ref{fig:csi_csom_spec}(a) shows the cosmic peak in crystal 8 fitted by the sum of an exponential and a Gaussian function. 
The calibration constants were derived from the individual crystal calibration in the high-sensitivity channels, and then scaled down by a factor of 8 for the low-sensitivity channels. 
Using the thorium calibration, the cosmic peak was determined to correspond to about 40\,MeV. 
This is in good agreement with \geant\ simulations which predicted an energy deposition of 39.7\,MeV for cosmic muons, 
taking into account the angular spectrum of cosmic rays and the acceptance of the trigger telescope.

The calibration constant for the central CsI crystal was 1.74\,MeV per ADC count at LG-LS resolution.

\subsection{Data-Acquisition System}

The data-acquisition-system (DAQ) hardware and software are described in Secs.~\ref{daq_hardware} and \ref{daq_software}.
The data-taking runs are discussed in Sec.~\ref{sec:datasample} and the data-file structure is reviewed
briefly in Sec.~\ref{daq_files}.

\begin{figure}
  \centering
  \includegraphics[width=8cm]{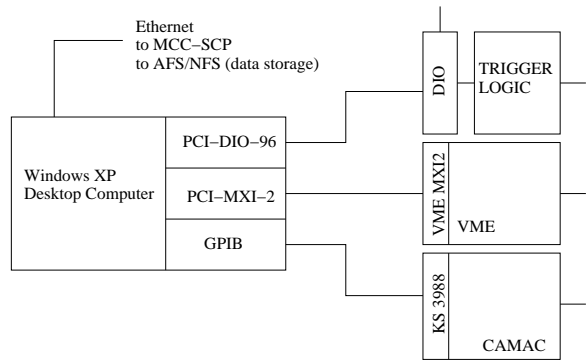} 
  \caption{Layout of the DAQ hardware.}
  \label{fig:DAQ_hw}
\end{figure}

The DAQ was centered around a desktop computer (WinXP) with IO connections via PCI busses to a VME system and an IO register for the trigger logic and a GPIB bus connecting a CAMAC crate (Fig.\:\ref{fig:DAQ_hw}). 
A custom LabVIEW software package monitored and controlled the subsystems of the experiment as well as the data acquisition and storage, and provided an interface to the SLAC accelerator control system. 

\subsubsection{DAQ Hardware}\label{daq_hardware}

The DAQ hardware other than the computer component consisted of three major components: the digital IO register, the trigger logic, and the digitizing electronics.

\begin{figure*}
  \begin{center}
    \includegraphics*[width=0.75\textwidth]{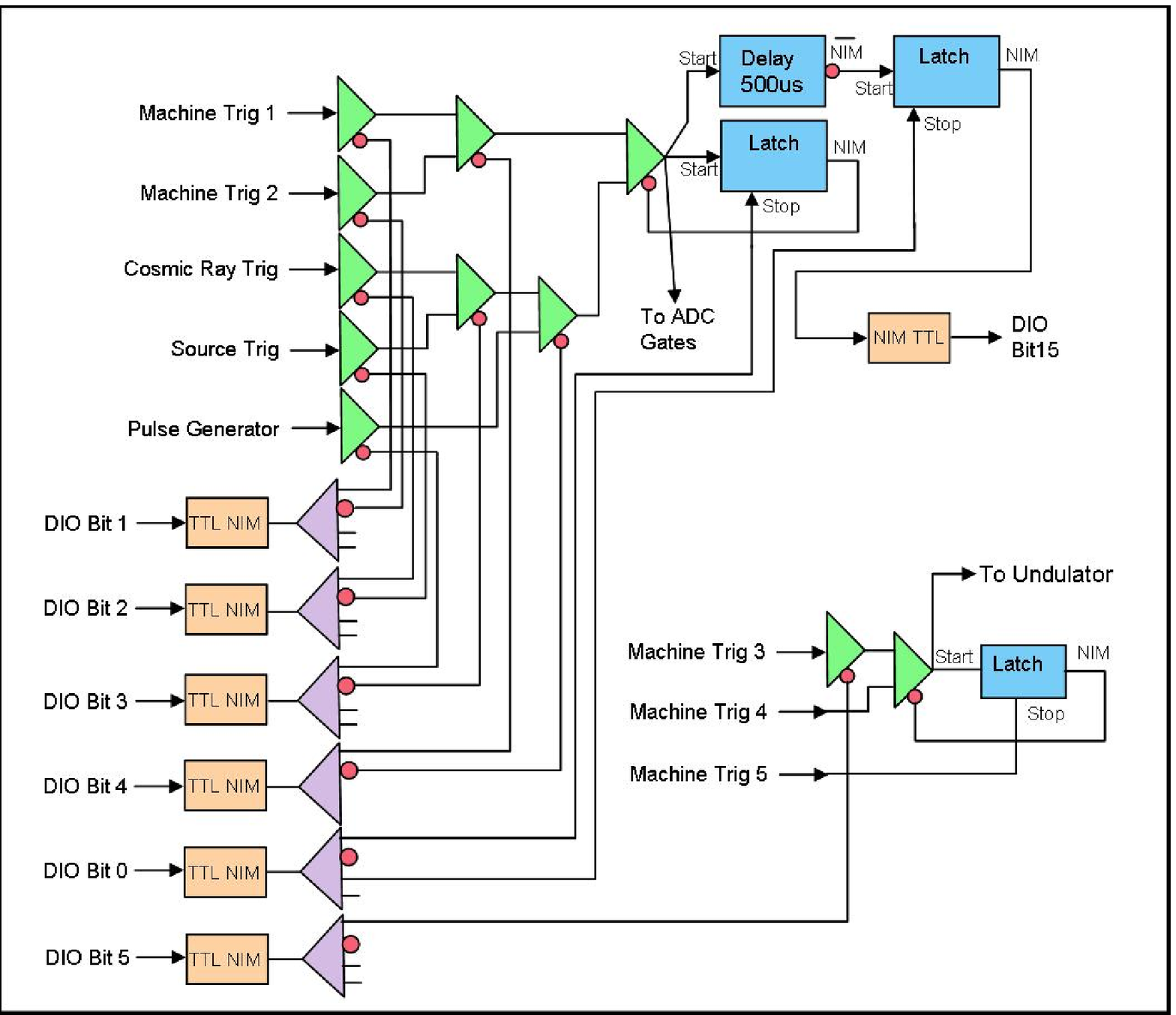} 
  \end{center}
  \caption{Trigger logic}
  \label{fig:trigger_logic}
\end{figure*}

\paragraph*{Digital IO (DIO) Register:} This subsystem consisted of a National Instruments model PCI-DIO-96 interface, connected via ribbon cables to connector blocks (model PXI-6508).  
The connector blocks were integrated into a custom patch panel that had two groups of 16 LEMO connectors each, one for 16 input signals/bits, and the other for 16 output bits.  
Input bit 15 was used to notify the DAQ software when a trigger had been generated in the NIM trigger logic, while six
output bits were used to select options for that logic, as shown in  Fig.\:\ref{fig:trigger_logic}.

\paragraph*{Trigger Logic:} 
This hardware subsystem was implemented using NIM electronic modules, and generated the DAQ software trigger,
the ADC gates, and the undulator trigger associated either with the SLAC electron beam, or with three types of non-beam events,
  \begin{description}
  \item {\sf Cosmic Ray Trigger}, for studying CsI detector performance and calibration.
  \item {\sf Radioactive Source Trigger}, for diagnostics and calibrations of the CsI detector.
  \item {\sf Pulse Generator Trigger}.
  \end{description}
Triggers associated with the electron beam were based on five signals
({\sf Machine Trigger}s) with different timing relative to the arrival of the beam pulse at the experiment,
  \begin{description}
  \item {\sf Machine Trigger 1} generated a DAQ software trigger and ADC gates in time with the electron beam pulse.
  \item {\sf Machine Trigger 2} generated a DAQ software trigger with ADC gates out of time with the electron beam pulse (and was not ordinarily used).
  \item {\sf Machine Trigger 3} came 11\,$\mu$s before the electron beam pulse to trigger the undulator power supply such that its peak current was in time with the electron beam pulse.
  \item {\sf Machine Trigger 4} came 50\,$\mu$s after the electron beam pulse to trigger the undulator power supply such that its peak current was out of time with the electron beam pulse.
  \item {\sf Machine Trigger 5} came 900\,$\mu$s before the electron beam pulse to reset the undulator trigger logic prior to an electron beam pulse.
  \end{description}
DIO output bits 1--4 selected the DAQ software trigger and ADC gates
to be from one of the three non-beam triggers or {\sf Machine Trigger
  1} or {\sf 2}.   
The NIM logic for this was reset by DIO output bit 0 at the end of the data acquisition for each triggered event.
DIO output bit 5 selected the undulator pulse to be in or out of time with respect to the electron beam.

During production data taking the DAQ software trigger and the ADC
gates were derived from {\sf Machine Trigger 1}, while 
the undulator was pulsed in and out of time with respect to the
electron beam on alternate beam pulses by {\sf Machine Triggers 3} and
{\sf 4}.

\paragraph*{Digitizing Electronics:}  Two external bus systems were used:

\begin{description}
\item {\sf VME.}
National Instruments models PCI-MXI-2 and VME-MXI-2 interfaced the VME system to the DAQ computer.  
The VME crate contained three 8-channel V265 ADC modules from CAEN, one VSAM module from BiRa Systems, one VME-CAN2 module from ESD, and one DVME-626 module from DATEL.  
The VME-CAN2 module recorded voltage transients induced in
  the pickup coils upon polarity reversal of the polarimeter magnets {\sf TP1} and {\sf TP2}.
The DVME-626 DAC module set the undulator excitation voltage.
\item {\sf GPIB.}
A National Instruments PCI-GPIB interface was used to drive the GPIB bus.  
A Kinetic Systems model 3988 CAMAC crate controller interfaced the CAMAC crate to the GPIB bus.  
The CAMAC crate contained a LeCroy model 2341A Latch Register, three LeCroy model 2249W 11-bit, charge integrating ADCs, and two LeCroy model 2259B 11-bit, peak-sensing ADCs.
\end{description}

\subsubsection{DAQ Software}\label{daq_software}
 
The DAQ software consisted of several programs written with the use of National Instruments LabVIEW version 7.2. These were executed on an Intel Pentium 4 CPU desktop computer operating under Microsoft Windows XP. 
The activities of these programs were coordinated and they communicated with each other via a set of global variables. Their access to common resources was coordinated using semaphores. 
One program was coded with the C++ programming language and provided a connection to the SLAC accelerator control system. 
The software set consisted of the following:

\begin{description}
\item {\sf Main DAQ}.  
This program initialized parameters prior to data collection, started, paused, resumed (after a pause) and ended data collection.   
During data collection it read out the detector data from the digitizing electronics after a trigger, all of which data was written to a disk file and some of which was monitored via various online displays.
\item {\sf High Voltage}.  
This program controlled and periodically monitored the high voltage
(HV) subsystem, which powered the silicon detectors {\sf GCAL}, {\sf P1}, 
{\sf S1} and {\sf S2}, the {\sf CsI} photodiodes, and photomultiplier tubes of the aerogel 
Cherenkov {\sf A1}, {\sf A2}, and other detectors.
\item 
{\sf Smart Analog Monitor} ({\sf SAM}).  
This program periodically read a 32-channel VME module to which slowly varying signals were attached, and also displayed and recorded these signals.
\item 
{\sf Accelerator Control Data Base Monitor}.  
This program transferred various slowly varying parameters from the accelerator control database for online display and recording in a data file.  
An EPICS-channel-access mechanism implemented retrieval of data from the data base.
\item 
{\sf Magnet Reversing Control}.  
This program, written in C++, issued requests to the SLAC Accelerator Control System to reverse the polarity of the analyzer magnets.  
\item Several other programs  were employed for infrequent control functions and for diagnostic purposes.
\end{description}

\subsection{Data Runs}\label{sec:datasample}


Positron data were collected during two run periods in June and September 2005 at five sets of spectrometer/lens currents (Table~\ref{tab:transport_sim}).
Electron data were collected only at one set of spectrometer/lens currents.
For each of the spectrometer settings a number of {\sf Super} runs was taken. 
One {\sf Super} run consisted of 10 {\sf cycles} with 3000 electron beam pulses (events) each at a 10\,Hz repetition rate, with the undulator pulse in time with only every other beam pulse.
Thus, half of all recorded events were ``signal'' (undulator in time) and half were ``background'' (undulator out of time).
In an automated procedure the polarity of the positron- and photon-polarimeter magnets {\sf TP1} and {\sf TP2} was reversed before each {\sf cycle} by the {\sf Magnet Reversing Control} program. 


The main DAQ program allowed selection of run types from a drop-down menu.
\begin{description}
\item {\sf Patterned:} In this run type the timing between electron beam pulses and undulator pulses was controlled by a software-coded pattern, usually with the undulator pulsed in time with every other electron-beam pulse. 
\item {\sf Pedestal:}  In this type of run {\sf Machine Trigger 1} was used to accumulate a specified number of counts for each ADC channel.  
However, the electron beam was blocked in the SLAC beam switchyard, upstream of the FFTB, to establish the
beam-off state (pedestals) of the data-acquisition system.
At the end of a run, the average value was calculated for each channel.  These values were used in subsequent runs for subtracting ADC pedestal values in the online displays.
\item {\sf Simple:} In this type of run the undulator was pulsed in time with every electron beam pulse.
\item {\sf Special:}  This run type executed the {\sf Magnet Reversing Control} program, and also
recorded the currents in the polarimeter magnets at 50\,Hz via pulse-generator triggers.
\item {\sf Test:}  This run type allowed manual selection of the trigger source for test purposes.
\item {\sf Super:} This run type consisted of a selectable number of
  {\sf cycles}, usually 10, where each {\sf cycle} consisted of a {\sf
    Special} run followed by a {\sf Patterned} run. 
\end{description}

\subsection{Data-File Structure}\label{daq_files}  

Data from each run 
were written to ASCII disk files. 
Following a {\sf Begin Run} record, a string of digitized data from the readout electronics modules
was written for each event.
The bulk of the data in the file came from the electronic modules digitizing the detector data.  
In each event string, the detector data were preceded by five words:
\begin{enumerate}
\item {\sf Data Type Marker} - to identify the source of a data string (detectors, high voltage, accelerator data, pedestals, \etc).
\item {\sf Run Type} - to distinguish among the five run types (other than a {\sf Super} run) listed above.
\item {\sf Event Number} - a sequential number labeling consecutive events. 
This number also identified which {\sf cycle} within a {\sf Super} run the data belonged to.
\item {\sf Relative Time} - the integer time in ms of the event with respect to the run start time. 
For this it could later be determined if any triggers were missing (not processed) due to the DAQ computer being busy.
\item {\sf DIO Register} - a 16-bit pattern used to tell which trigger sources were used for the event.
\end{enumerate}

In addition to the detector data, other data strings such as HV data, SAM data, \etc, were written to
disk asynchronously.
Each type of non-detector data string was identified by a distinct {\sf Data Type Marker}.

\section{Simulation}\label{sec_simulation}

Analysis of the energy dependence of the positron or electron
polarization $P_{e\pm}$ from measurements by polarimeter {\sf TP1},
and predictions for asymmetries in the photon polarimeter {\sf TP2},
required detailed simulations of the experiment. 

The calculation of the analyzing power for positron polarimetry was
rather complex, as it had to include simulations of the undulator, the
collection- and the spectrometer-system, as well as the polarimeter magnet. 
%
%
These simulations were performed using the \geant\ code
\cite{Agostinelli:2002hh,Allison:2006ve}, starting with version 6.2,
which was extended  to include all electromagnetic processes needed
for spin-dependent transport of particles through matter, 
  as discussed in Sec.~\ref{sec:pol_geant}. 
The spin-dependent extensions are available in \geant\ from version 8.2 \cite{g4pol} onwards.

The simulation procedure for positrons (electrons) was performed in
four independent steps, where  each subsequent step used the output of
the previous one as its input:   
\begin{itemize}
\item 
  Generation of undulator photons with the appropriate spectrum and
  polarization, Sec.~\ref{sec:undulator_photons},
\item 
  Conversion of undulator photons to electrons and positrons in the
  production target {\sf T1},  Sec.~\ref{sec:possi_production},
\item 
  Transport of positrons (electrons) through the magnetic field of the
  spectrometer {\sf D2} to the reconversion target {\sf T2}, Sec.~\ref{sec:transport_system}, 
\item 
  Reconversion of positrons (electrons) to photons in the reconversion
  target {\sf T2} and transport of photons (and other particles) through the
  polarimeter magnet {\sf TP1} to the CsI calorimeter, and
  determination of asymmetry $\delta$ of energy deposition with
  respect to the magnet polarity, see Sec.~\ref{section:polarimeter}. 
\end{itemize}
The analyzing powers $A_{e^\pm}$ were determined from the simulated
asymmetries $\delta$ according to the inverse of Eq.\:(\ref{p23}) for
the different spectrometer settings in the experiment.  

Simulations relevant to the photon polarimeter {\sf TP2} were carried
out with a combination of software tools that included semi-analytic
calculations, modified versions of \textsc{Geant3}, and 
parametric modeling.
As the undulator radiation provided a broad distribution of photon
energies, the calculation of asymmetries required a convolution over
energy-dependent detector efficiency, analyzing power, and photon
polarization. 
%
%
However, in contrast to the positron case, the transmitted photon beam was
narrowly collimated. Thus, it was sufficient to consider only
first-generation-scattering events in the iron, which simplified the
task considerably, as described in Sec.~\ref{sec:sim_pl}.

\subsection{Simulation of Polarization in \geant}\label{sec:pol_geant}

The polarization dependence of the following processes has been accounted for in the simulations of the experiment:
\begin{itemize}
\item Compton scattering,
\item Bhabha  and M{\o}ller scattering, 
\item Photoelectric effect,
\item Pair creation and annihilation,
\item Bremsstrahlung.
\end{itemize}

Polarization effects were neglected for low-energy charged particles with less than $200\ \mu$m remaining range.

The polarization states of particles are described in \geant\ by Stokes vectors, and polarization transfer from an initial to a final state is represented by a linear transformation of the corresponding Stokes vectors\cite{Mcmaster}. 
A detailed description of the implementation of the transfer-matrix formalism for all relevant processes can be found in\cite{Geant4:PhysRefMan,laihem_thesis}.

\subsection{Production of Positrons and Electrons in the Target}\label{sec:production}

\subsubsection{Radiation of the Helical Undulator}\label{sec:undulator_photons}

The photon spectrum of undulator radiation has been discussed in Sec.\:\ref{sec:methods}. 
Table\:\ref{tab:undulator_parameters}
\begin{table}
  \begin{center}
    \caption{Parameters for the simulation of the undulator radiation.}
    \label{tab:undulator_parameters}
     \vspace{1mm}
\footnotesize
    \begin{tabular*}{7cm}{l@{\extracolsep\fill}l}
      \hline
      \hline
      Parameter & Value \\
      \hline  
      Electron beam energy $E_e$          & 46.6\,GeV    \\
      Undulator period $\lambda_{\rm u}$  & 2.54\,mm    \\
      $K$-value                           & 0.19    \\
      Energy $E_1$ of first harmonic      & 7.9\,MeV     \\
      Number of harmonics calculated      & 3         \\
      Total number of photons per $e^{-}$ & 0.43\,m$^{-1}$    \\
      Total radiated power                & 1.76\,MeV/m     \\
      \hline
      \hline
    \end{tabular*}
  \end{center}
\end{table}
 summarizes the parameters used for the simulation of the undulator
 radiation.
Only the first three harmonics of the photon spectrum have been included in the simulation. 
The relative contributions of these harmonics to the total undulator radiation are shown in Table\:\ref{tab:harmonic_contribution};
\begin{table}
  \begin{center}
    \caption{The Fraction $N_n / N$ of undulator photons in the first three harmonics and their cutoff energy $E_{n}$ for ${K}=0.19$.}
    \label{tab:harmonic_contribution}
     \vspace{1mm}
\footnotesize
    \begin{tabular*}{6cm}{c@{\extracolsep\fill}cc}
      \hline
      \hline
      Harmonic number $n$ & $N_n / N$ & $E_{n}$\,(MeV)          \\
      \hline
      1        &   0.9584   &   7.9    \\
      2        &   0.0396   &  15.7    \\
      3        &   0.0020   &  23.5    \\
      \hline
      \hline
    \end{tabular*}
  \end{center}
\end{table}
harmonics higher than the third contributed approximately $10^{-4}$ of the total number of photons in this experiment.

In the simulation of the positron polarimeter the $K$-value of the
undulator was assumed to be 10\% higher than its design value
listed in Table \ref{tab:UndPar} of Sec.\:\ref{sec_undulator}.
While this slightly modified 
the predicted positron and
electron rates at high-energy settings of the spectrometer, the impact
on the analyzing power, and consequently on the polarization
results, was negligible. 

\subsubsection{Production of Polarized Positrons and Electrons}\label{sec:possi_production}

The photons generated by the undulator were simulated as striking the production target {\sf T1} uniformly over a circle 3\,mm in diameter, corresponding to the aperture of collimator {\sf C2}.
Although the core of the photon beam was considerably smaller than this (see Sec.~\ref{sec:beamline}, the position
of the beam varied from run to run such that a uniform spatial distribution at collimator {\sf C2} was a good approximation.
Table\:\ref{tab:source} lists the parameters used in the \geant\ simulation of the generation of electrons and positrons in the target. 
\begin{table*}
  \begin{center}
    \caption{Parameters used in the simulation of positron production and transport.}
    \label{tab:source}
 \vspace{1mm}
\footnotesize
    \begin{tabular*}{11cm}{l@{\extracolsep\fill}l} 
      \hline    
      \hline
\multicolumn{2}{c}{Production} \\
\hline 
      Undulator-photon beam shape                & 3-mm-diameter cylinder \\
      Production-target material                 & Tungsten composite W-4Ni-3Cu-3Fe \\ 
      Production-target thickness                & 0.81\,mm (0.2\,r.l.) \\

      \geant\ equiv.\ energy cutoff              & 200\,$\mu$m remaining path length \\
 \hline
\multicolumn{2}{c}{Transport through solenoid {\sf SL} and spectrometer {\sf D2}} \\
\hline 
      Vacuum-pipe diameter (solenoid)            & 36\,mm \\  
      Vacuum (in vacuum chamber)                 & $10^{-3}$\,Torr \\
      Energy-selection slit width                & 30\,mm \\ 
      Solenoid field map                         & SLAC measurement \\ 
      Spectrometer field map                     & MERMAID calculation \\ 
      Magnetic field (implementation method)     & 3-D linear interpolation \\
      Tracking step length (in magnetic field)   & 100\,$\mu$m \\
\hline
\multicolumn{2}{c}{Transport from exit window of spectrometer to target {\sf T2}} \\
\hline 
      Exit-window diameter                       & 48\,mm \\
      Reconversion-target material               & Tungsten \\ 
      Reconversion-target thickness              & 1.75\,mm (0.5\,r.l.) \\  
      Reconversion-target diameter               & 50\,mm\\ 
      \geant\ equiv.\ energy cutoff              &  200\,$\mu$m remaining path length \\
      \hline
      \hline
    \end{tabular*}
  \end{center}
\end{table*}

Positrons were produced in target {\sf T1} by $\gamma$ conversion to
$e^+e^-$, after which they interacted in the target through Bhabha scttering,
annihilation, Bremsstrahlung, ionization, and subsequent cascade
processes.  
Electrons were produced by Compton scattering and the
photoelectric effect in addition to pair production.
For the conditions
of this experiment, the total electron yield at the exit of
the target was a factor 2.2 higher than that of positrons:
$Y_{e^{+}}^{2\pi} = 0.0085$ and $Y_{e^{-}}^{2\pi} = 0.0187$ per
incident photon.
Because of subsequent interactions in the target, these yields were
much smaller than the 10.5\,\% of the primary photons that were
converted into electrons and positrons. 

\begin{figure*}
  \begin{center}
    \includegraphics[width=16cm]{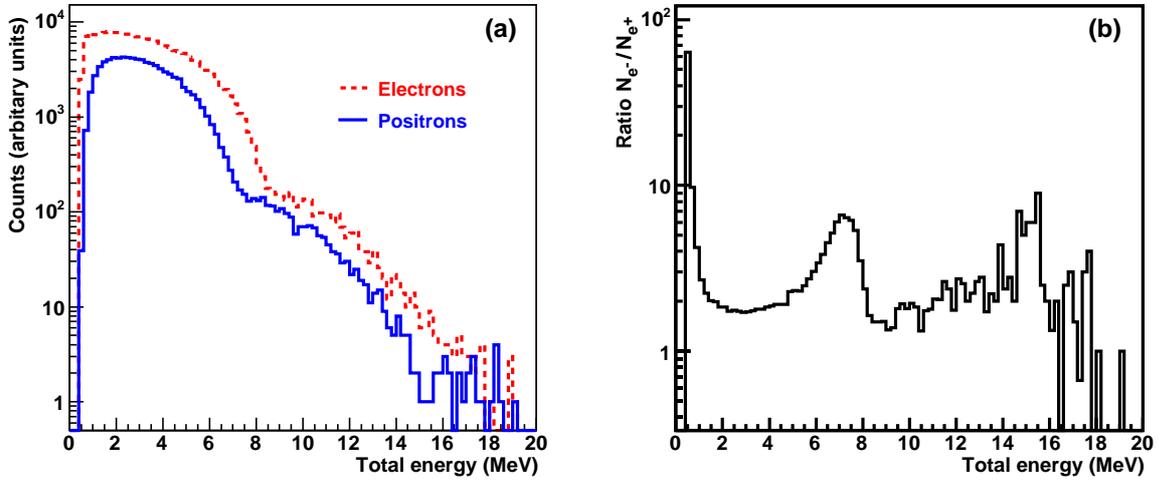} 
    \caption[Electron- and positron-energy spectra and their ratio at
    the exit face of the production target.]{ (a) Electron- and
      positron-energy spectra at the exit face of the production
      target {\sf T1}; (b) ratio of electron and positron spectra plotted in (a).
}
    \label{fig:elect_posit_energy}
  \end{center}
\end{figure*}

\begin{figure}
  \begin{center}
    \includegraphics[width=8cm]{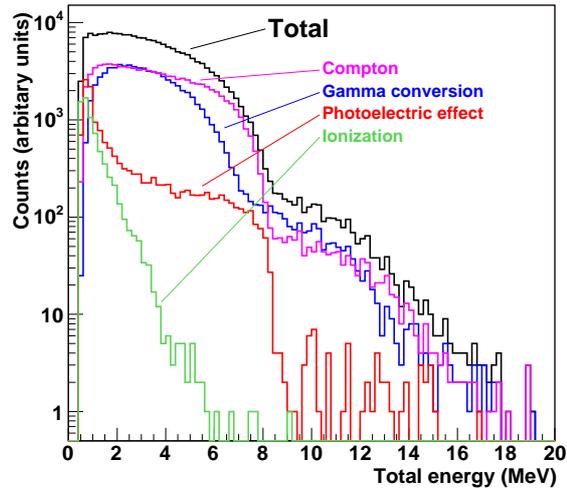} 
    \caption{Relative contributions to the energy distribution of electrons from different processes ($\gamma$-conversion, Compton, photoelectric effect and ionization).
}
    \label{fig:electron_energy_process}
  \end{center}
\end{figure}

Figure\:\ref{fig:elect_posit_energy} shows the simulated energy distributions of electrons and positrons. 
The expected longitudinal polarization of positrons and electrons at
the exit of the production target {\sf T1} was essentially the same as
that for the same set of particles at the reconversion target {\sf T2}
(see Sec.~\ref{section:polarimeter}), and the latter is shown in
Fig.\:\ref{fig:analysing_power_polarization}. 
For production by first-harmonic photons (maximum energy $E_1=7.9$\,MeV), the maximal total energies of the positrons and electrons were
\begin{linenomath}
\begin{align}
  E_{e^{+}}^{\rm max}&= E_{1} - m c^{2} = 7.4\,\mbox{MeV}, \label{eq:positron_Emax}\\
  E_{e^{-}}^{\rm max}&= E_{1} + m c^{2} = 8.4\,\mbox{MeV}, \label{eq:electron_Emax}
\end{align}
\end{linenomath}
noting that the maximal-energy electrons were generated by the photoelectric effect.
Figure\:\ref{fig:electron_energy_process}
shows the energy distribution of the electrons for the  different processes; the fractional  contributions of the electron-production processes in the production target are given in Table\:\ref{tab:process_contribution}.
\begin{table}
  \begin{center}
    \caption{Relative contributions of different processes to the electron production for the parameters in Tables\:\ref{tab:undulator_parameters} and\:\ref{tab:source}.}
    \label{tab:process_contribution}
 \vspace{1mm}
\footnotesize
    \begin{tabular*}{5cm}{l@{\extracolsep\fill}l}
      \hline
      \hline
      Process & Fraction         \\
      \hline
      Compton scattering   & 0.4987  \\
      $\gamma$-Conversion  & 0.3875   \\        
      Photoelectric effect & 0.0773 \\
      Ionization           & 0.0365 \\
      \hline
    \end{tabular*}
  \end{center}
\end{table}

The energy, angular distribution and polarization of the positrons (electrons) at the exit face of the production target were used as input for the next step described in the following section. 

\subsection{Positron/Electron Transport}\label{sec:transport_system} 

\subsubsection{Overview}

The positrons (electrons) generated  in the production target {\sf T1} were guided through the transport system by the solenoid lens {\sf SL} and the spectrometer dipoles {\sf D2} (Sec.~\ref{sec_spectrometer}, Figs.~\ref{prlfig2}, \ref{fig:E166Scheme} and \ref{fig:spectrometerBIS})
while the oppositely charged particles -- the electrons (positrons) -- were dumped. 

The simulation of the transport system and the tracking of the polarized positrons (electrons) yielded the following output for each set of currents $(I_{\rm S},I_{\rm L})$:

\begin{itemize}
\item 
Positron/electron transmission through the spectrometer system,
\item 
Positron/electron energy and spatial distributions at the Si-W detector {\sf P1} and at the reconversion target {\sf T2}.
\end{itemize}
The distributions at the reconversion target {\sf T2} constituted the input for the simulation of measurements by the positron polarimeter described in Section\:\ref{section:polarimeter}.
Depolarization effects due to spin precession in the
  magnetic field \cite{Bargmann} were negligible in the
  present experiment, and have been ignored in the simulation.

\subsubsection{Implementation of the Magnetic Fields}\label{sec:implem_fields}

\begin{figure}
  \includegraphics[width=8cm]{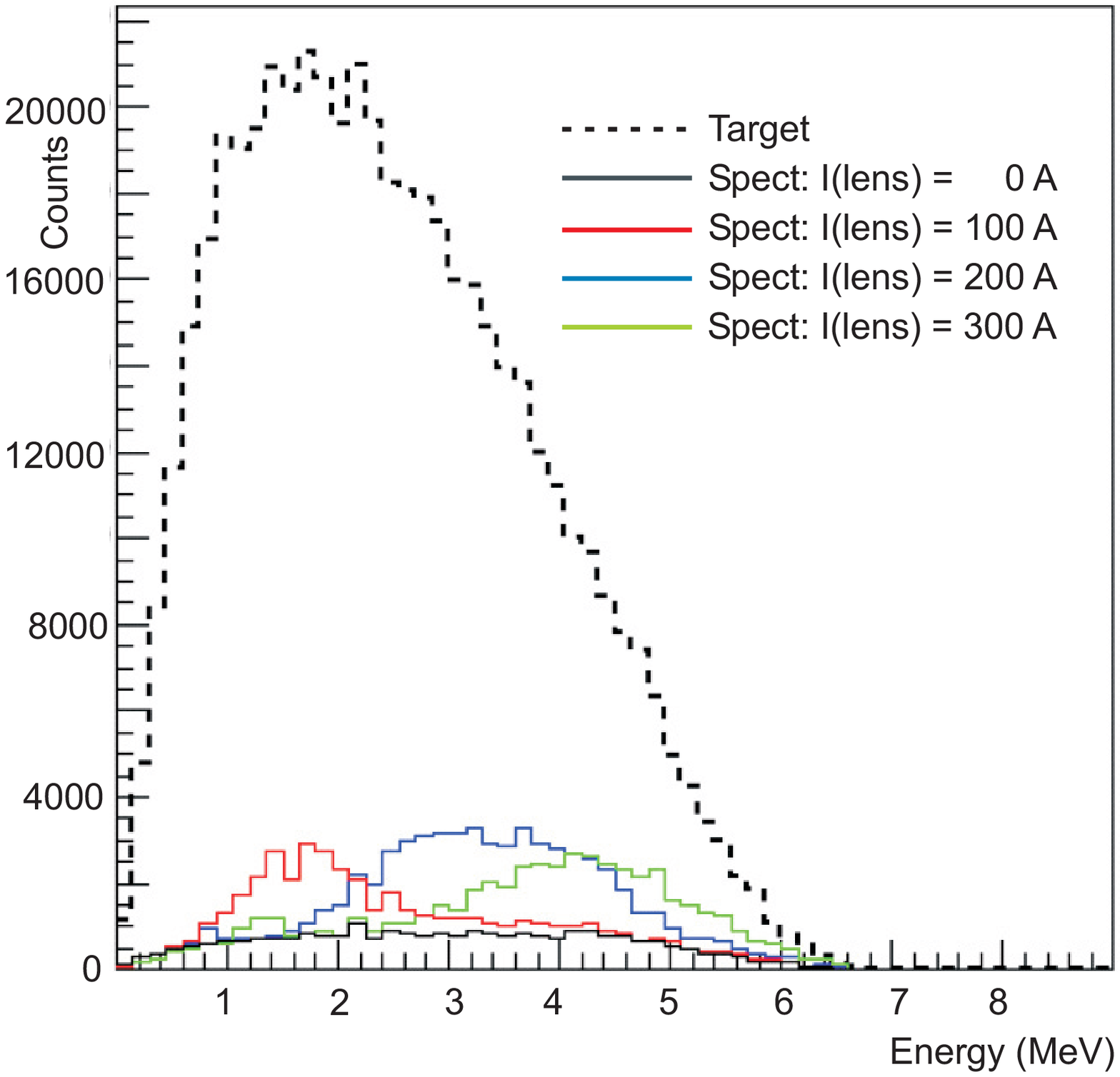} 
  \hfill
  \includegraphics[width=8cm,height=8.3cm]{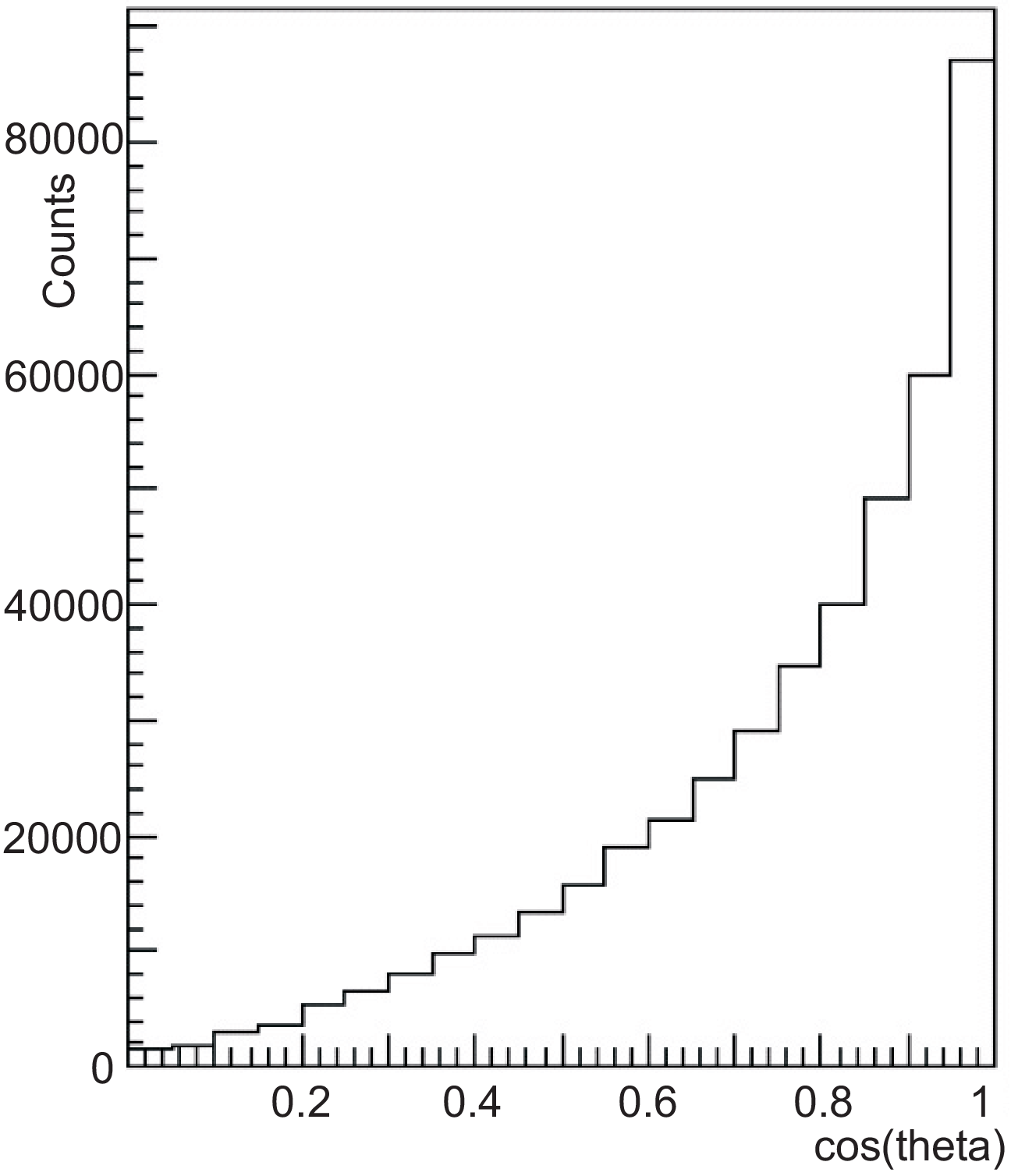} 
  \caption[Simulations of focusing of positrons by the solenoid lens {\sf SL}.]{Simulations of focusing of positrons by the solenoid lens {\sf SL}. 
{\it Left}: The energy spectra of positrons emerging from the production target {\sf T1} and of positrons entering the spectrometer {\sf D2} for different solenoid currents. 
{\it Right}: The polar angle distribution, $\cos\theta$, of positrons at the target exit surface.}
  \label{fig:solenoid_focus_on}
\end{figure}

\paragraph*{Solenoid Magnetic Field:}
The magnetic field of the focusing lens {\sf SL} was implemented in the simulation using the functional forms (\ref{eq:solenoid_Br})--(\ref{eq:solenoid_Bz}) and the measured values of $B_0(z)$.
The focusing effect is illustrated in Fig.\:\ref{fig:solenoid_focus_on}.
For each spectrometer current $I_{\rm S}$ the solenoid current $I_{\rm L}$ used in the simulation was chosen to maximize the number of positrons at the detector {\sf P1}. 
Because the field map used in the simulation differed somewhat from the actual field in the solenoid,
 the currents $I_{\rm L}$ used in the simulation differed slightly from those used in the experiment,
as shown in Table\:\ref{tab:transport_sim}, but the focusing effect of the solenoid lens was thereby well represented in the simulation.

\paragraph*{Spectrometer Magnetic Field:}
In the simulation, the MERMAID field map described in Section\:\ref{sec_spectrometer} was implemented  using a three-dimensional linear interpolation on a cubic grid. 
The magnitude of the field was related to the spectrometer current $I_{\rm S}$ according to Eq.~(\ref{eq_b_is_quadratic}). 

\subsubsection{Energy Selection by the Spectrometer}\label{sec:energy_selection}

\begin{figure*}
  \begin{center}
     \includegraphics*[width=0.3\textwidth]{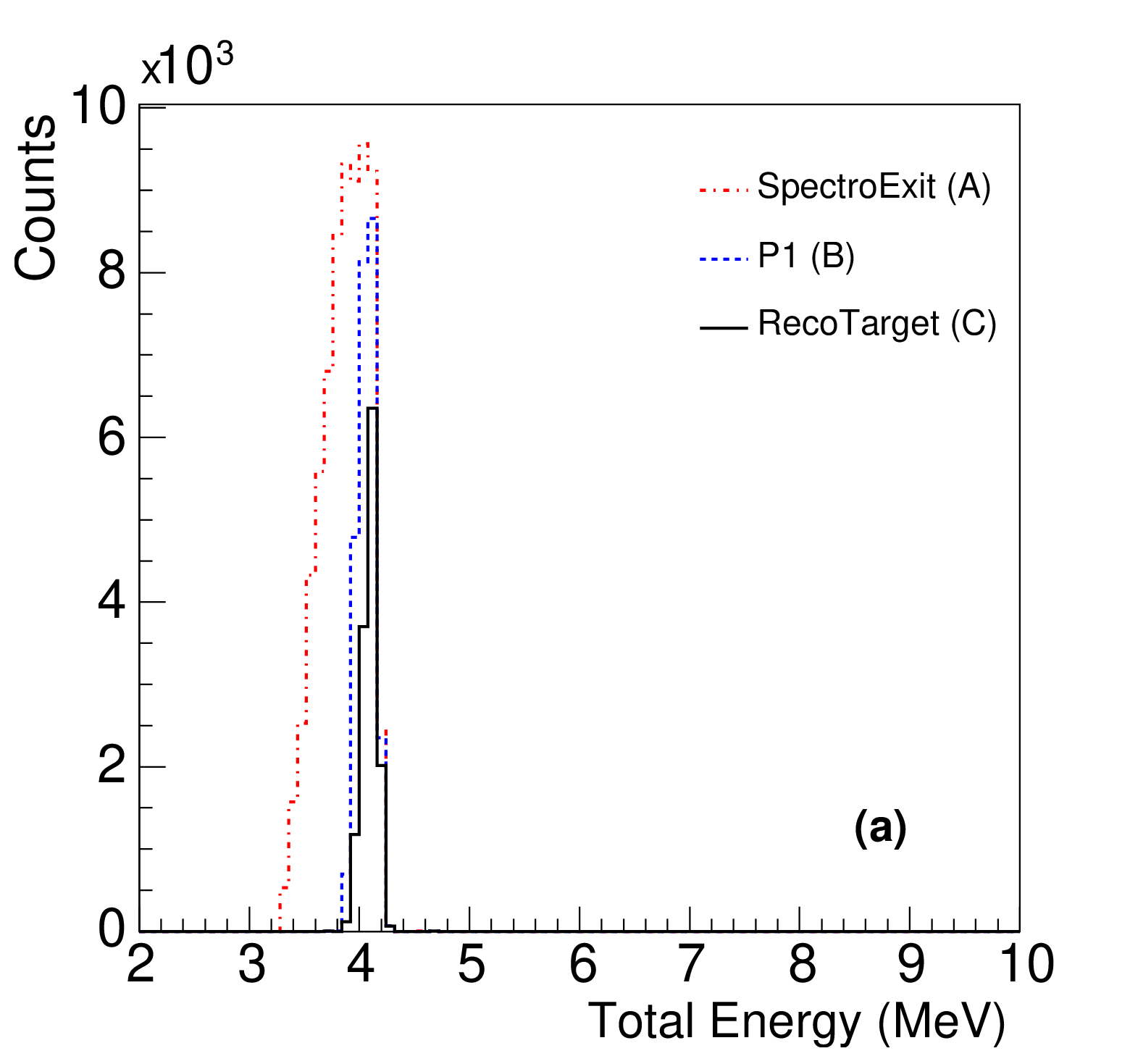} 
     \includegraphics*[width=0.3\textwidth]{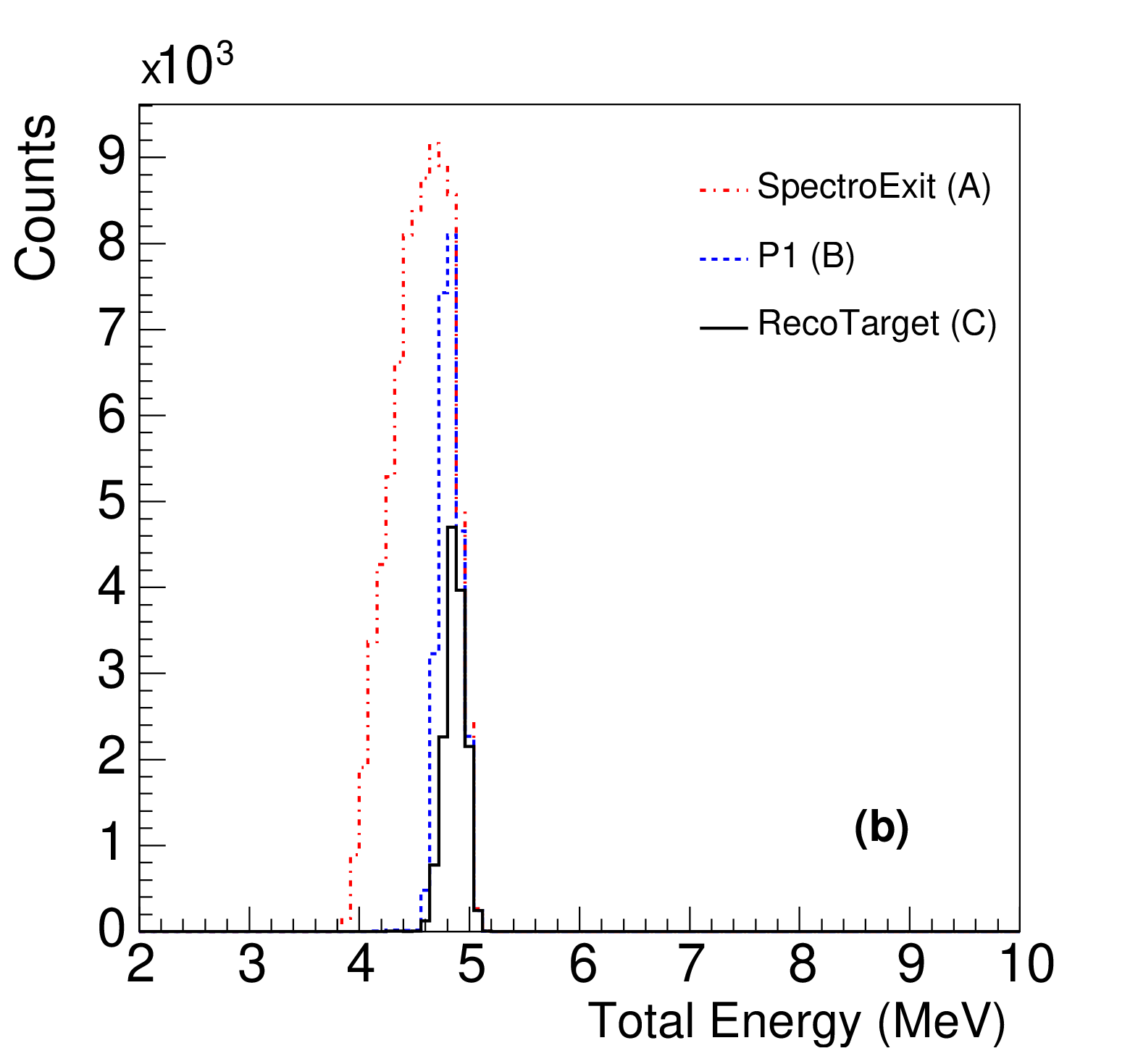} 
     \includegraphics*[width=0.3\textwidth]{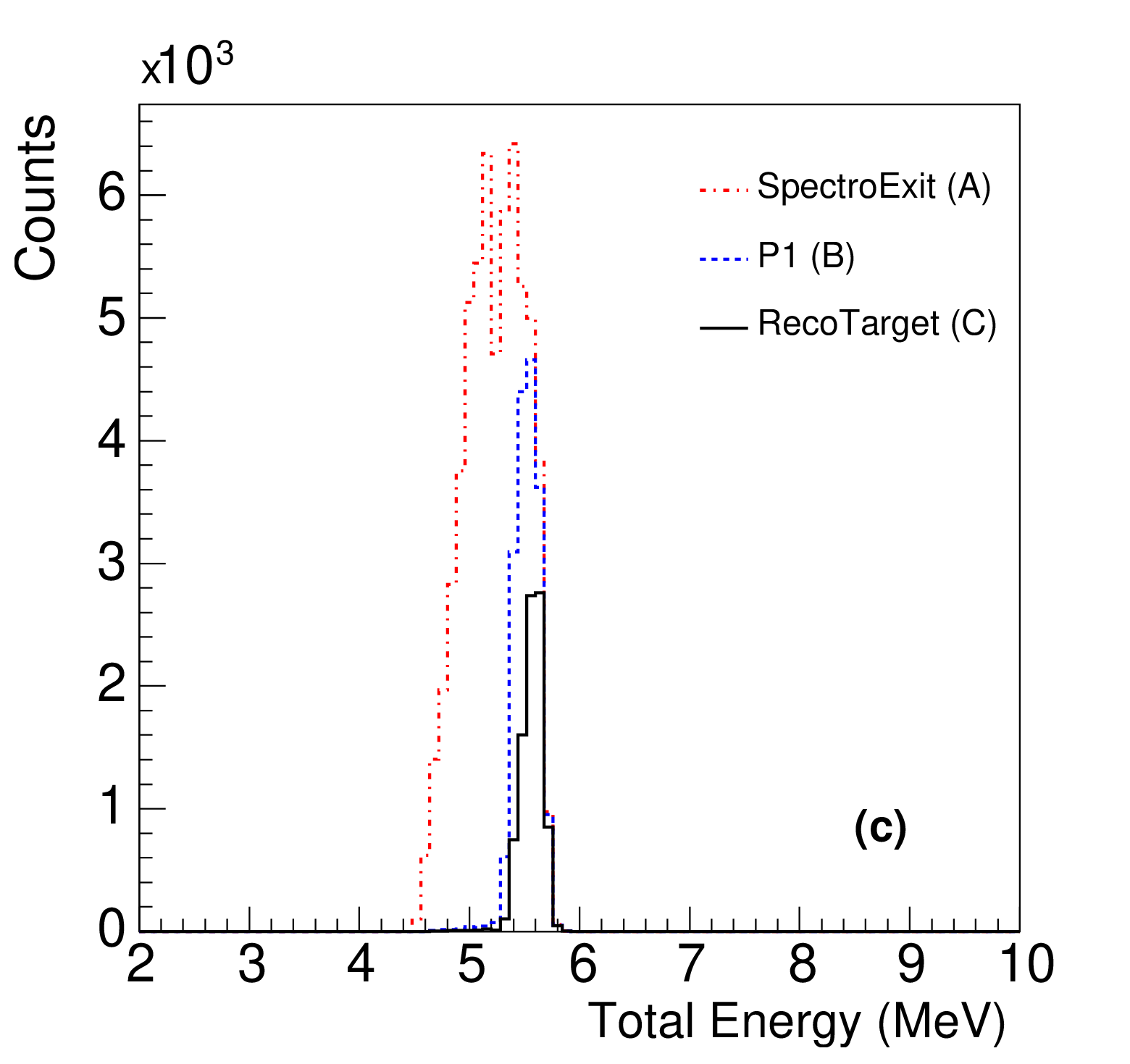} 
     \includegraphics*[width=0.3\textwidth]{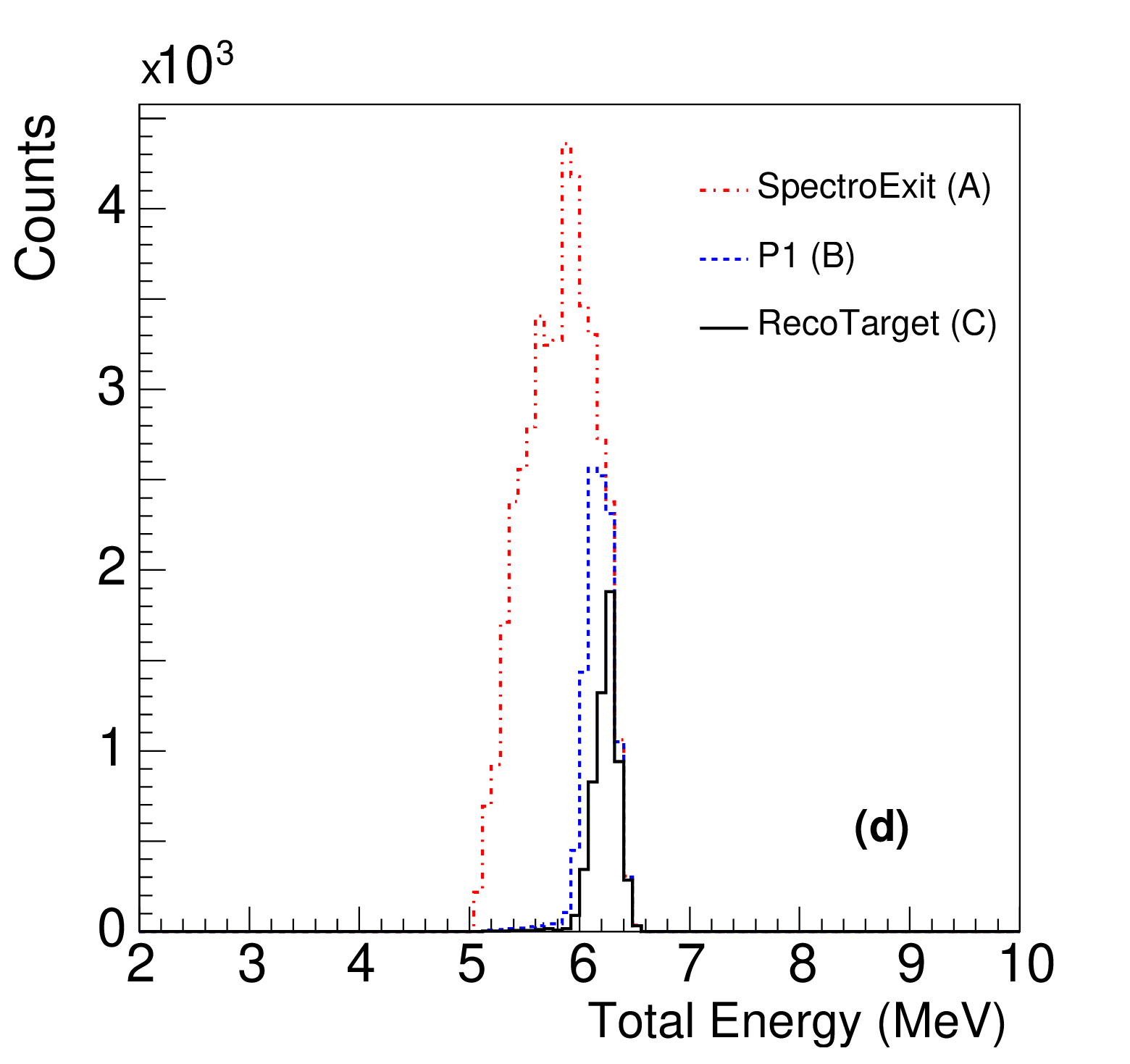} 
     \includegraphics*[width=0.3\textwidth]{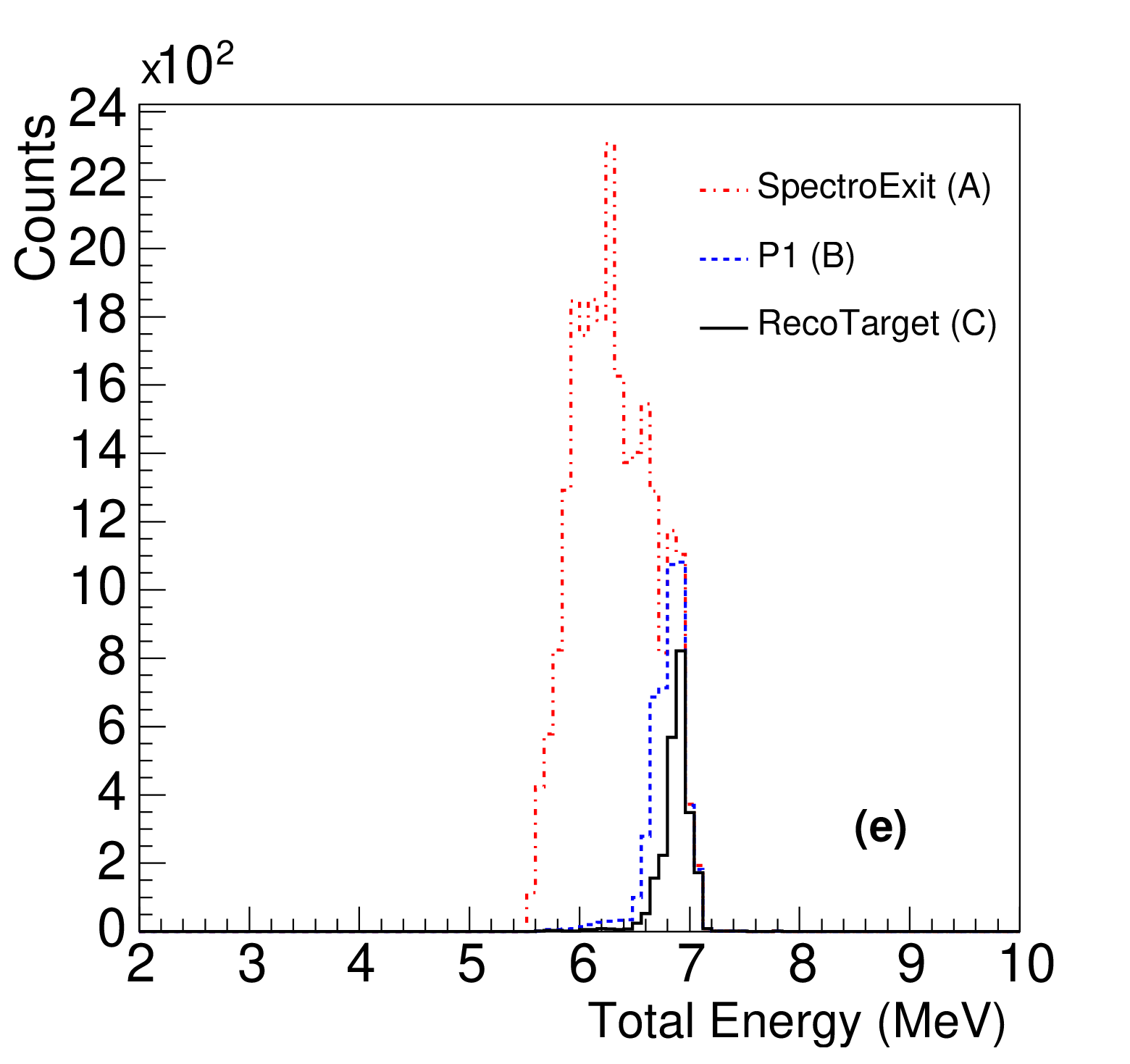} 
  \end{center}
  \caption[Simulated positron-energy spectra at the exit of the spectrometer.]{Simulated positron-energy spectra at the exit of the spectrometer (position A of Fig.\:\ref{fig:spectrometerBIS}), at the detector {\sf P1} (position B), and at the reconversion target {\sf T2} (position C) for the five spectrometer currents, (a): $I_{\rm S} = 100$\, A, (b): $I_{\rm S} = 120$\, A, (c): $I_{\rm S} = 140$\, A, (d): $I_{\rm S} = 160$\, A, (e): $I_{\rm S} = 180$\, A.
}
  \label{fig:possi_rec_energy}
\end{figure*}

The spectra of positron energies at the reconversion target {\sf T2}
as selected by the movable tungsten jaws {\sf J} in spectrometer {\sf D2} was simulated for the five spectrometer currents $I_{\rm S}$, as shown in Fig.\:\ref{fig:possi_rec_energy}. 
While the efficiency of the positron transport was sensitive to the current $I_{\rm L}$ in the solenoid lens,
the shape of the transported energy spectra depended only slightly on this current. 
The resulting dependence on spectrometer current of the central energies of positrons transported to the reconversion target {\sf T2} is listed in Table\:\ref{tab:transport_sim}.

\subsubsection{Positron Transport Near Detector {\sf P1}}\label{sec:sim_veri}

As discussed in Sec.\:\ref{sec_spectrometer} the vacuum chamber was inadvertently constructed with the exit port 6.3\,mm further away from the $\gamma$-line than called for in the magnet design.   
Moreover, the presence of unsuspected magnetism of the tungsten that
faced the iron pole pieces led to a deviation of the positron paths
from their design trajectories. 
In consequence the emergent beam from the vacuum
  chamber was not centered on the axis of the polarimeter magnet. The
  energy selection resulted from a combination of the jaws setting and
  the aperture of the spectrometer exit tube.

\begin{figure*}
  \begin{center}
    {\includegraphics[width=0.8\textwidth]{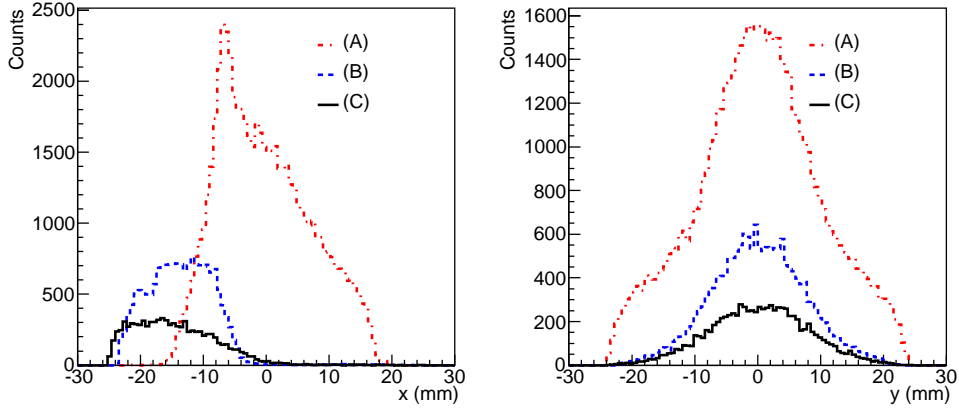}} 
  \end{center}
   \caption[Simulated distributions of positrons along the $x$- and
   $y$-axis at the exit of the spectrometer.]{Simulated distributions
     of positrons along the $x$- and $y$-axis at the exit of the
     spectrometer {\sf D2} (A), at the detector {\sf P1} (B) and at
     the reconversion target {\sf T2} (C) for spectrometer current
     140\,A.  
}
  \label{fig:the_XaY_Rec_100_180}
\end{figure*}

The horizontal asymmetry of the positron distribution at detector {\sf
  P1} (Sec.~\ref{sec_p1}) was confirmed qualitatively by  simulations
shown in Fig.\:\ref{fig:the_XaY_Rec_100_180}, where ideally the
positrons would be centered on $x=0$ at the spectrometer exit (curve
A), at the detector {\sf P1} (curve B) and at the reconversion target
{\sf T2} (curve C); see also \cite{laihem_thesis}.

\subsection{The Positron Polarimeter}\label{section:polarimeter}

Positrons that struck target {\sf T2} were reconverted into photons, some of which passed through the iron-core polarimeter magnet {\sf TP1} and were absorbed in the CsI(Tl) calorimeter, as described in Sec.\:\ref{sec:pos_diag}.  
The flux of photons in the calorimeter depended on the polarization of the
positrons (as well as that of the electrons in the magnet iron).
The spatial, momentum and polarization distributions of positrons at the reconversion target from the preceding simulations were the input for the simulation of the positron polarimeter.

\subsubsection{The Analyzing Power}\label{sec:pos_power}

\begin{figure*}
  \begin{center}
    \includegraphics[width=\textwidth]{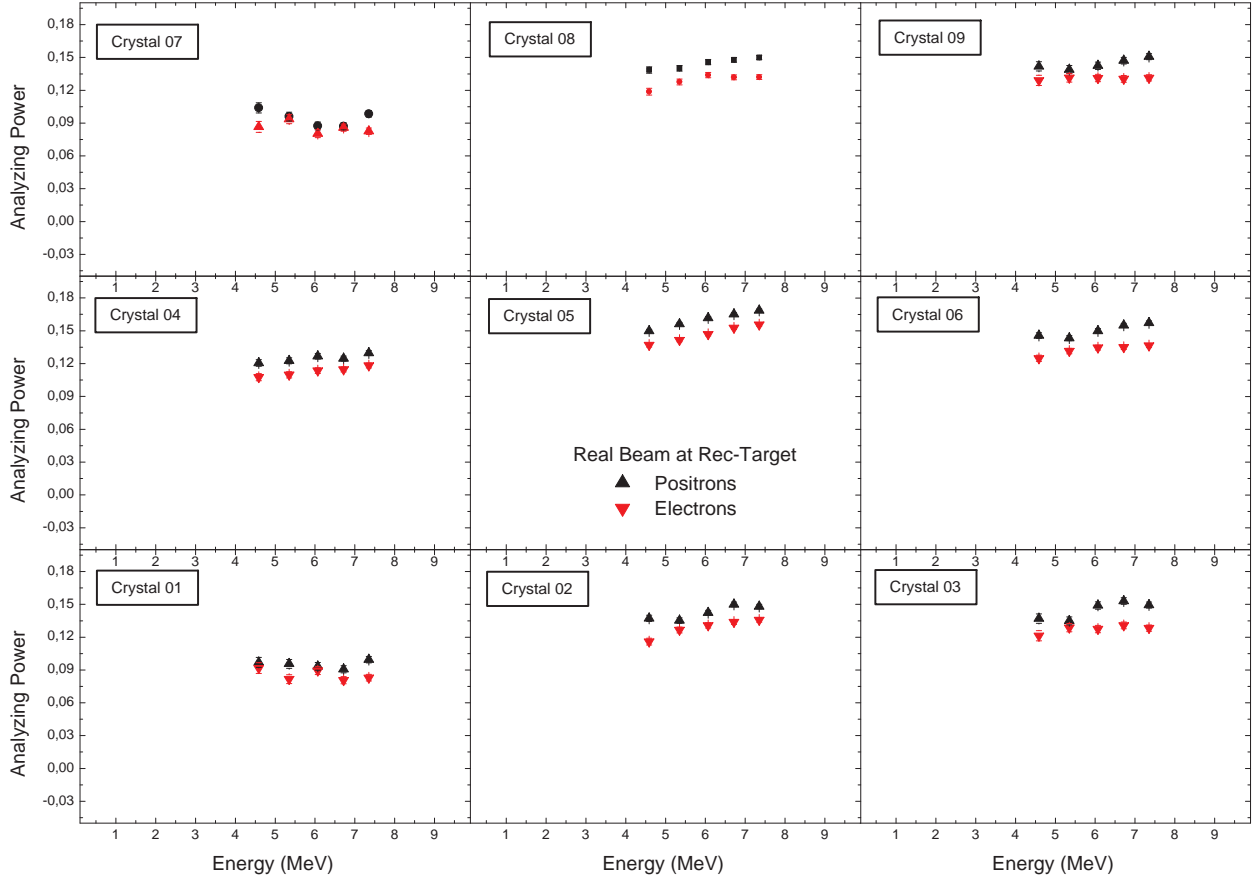}  
  \end{center}
  \caption{The simulated analyzing power $A_{e^\pm}$ of each of the nine CsI crystals for positron and electron polarization. }
  \label{fig:realistic_electron_positron}
\end{figure*}

By reversing the polarity of magnet {\sf TP1} an asymmetry $\delta$, defined in Eq.~(\ref{eq:analpower}), in the energies observed in the nine CsI crystals was obtained. 
The analyzing power $A_{e^\pm}$ for positron (electron) polarization is given by 
\begin{linenomath}
\begin{equation}
  A_{e^\pm} = \frac{\delta}{P_{e^\pm}P_{e^-}^{\rm Fe}} ~~,
\end{equation}
\end{linenomath}
where $P_{e^\pm}$ and $P_{e^-}^{\rm Fe}\ (=0.069 \pm 0.002)$ are the positron (electron) longitudinal polarization and the polarization of the electrons in iron, respectively. 
To improve the statistical significance of the simulation, the positron polarization as well as the absolute polarization of the iron absorber were set to 100\,\% and the results scaled to the actual polarization values.
The results are shown in Fig.\:\ref{fig:realistic_electron_positron}
as a function of electron and positron energy. 
The analyzing power differed slightly for the individual crystals in
the CsI(Tl) calorimeter, 
due to their different geometrical acceptance. 
The analyzing power for the central crystal is listed in Table\:\ref{tab:analysing_power_centr} for the various spectrometer current settings; the uncertainties were due to limited statistics in the simulation.

\begin{table}[t]
 \centering
  \caption{The analyzing powers $A_{e^{\pm}}$ for electrons and positrons in the central CsI crystal at different spectrometer currents $I_{\rm S}$.}
  \label{tab:analysing_power_centr}
 \vspace{1mm}
\footnotesize
    \begin{tabular*}{8cm}{c@{\extracolsep\fill}ccc}
      \hline
      \hline
      $I_{\rm S}$ (A) &  E$_{e^{\pm}}$ (MeV) &$A_{e^{+}} \pm \Delta A_{e^{+}}$   &$A_{e^{-}} \pm \Delta A_{e^{-}}$  \\
      \hline
      100        &  4.59      & $0.1498\pm 0.0016$ &  $0.1371\pm 0.0018$ \\
      120        &  5.36      & $0.1563\pm 0.0015$ &  $0.1417\pm 0.0016$ \\
      140        &  6.07      & $0.1616\pm 0.0014$ &  $0.1469\pm 0.0015$ \\
      160        &  6.72      & $0.1651\pm 0.0013$ &  $0.1528\pm 0.0014$ \\
      180        &  7.35      & $0.1686\pm 0.0013$ &  $0.1557\pm 0.0014$ \\
      \hline
      \hline
    \end{tabular*}
\end{table}

The systematic uncertainty on the simulation of the analyzing power was estimated to 7\,\% by varying the \geant\ physics parameters within their allowed range.

\subsubsection{The Asymmetries $\delta$ and the Polarizations $P$}\label{sec:delta_sim}

\begin{figure}[t]
  \begin{center}
    \includegraphics*[width=8cm]{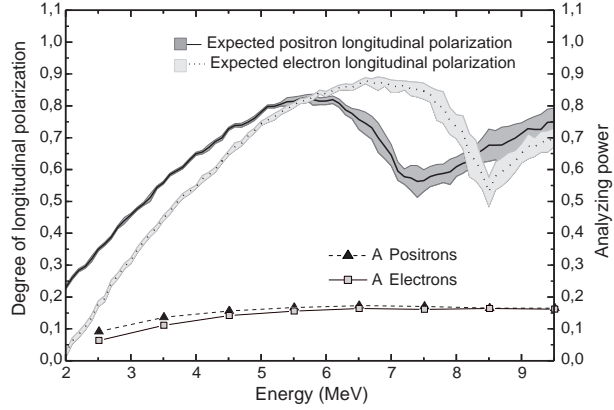} 
  \end{center}
  \caption{The simulated polarization $P_{e^\pm}$ of positrons/electrons at target {\sf T2} and the respective analyzing powers $A_{e^\pm}$ in the central CsI crystal as a function of particle energy.}
  \label{fig:analysing_power_polarization}
\end{figure}

\begin{figure}
  \begin{center}
    \includegraphics*[width=8cm]{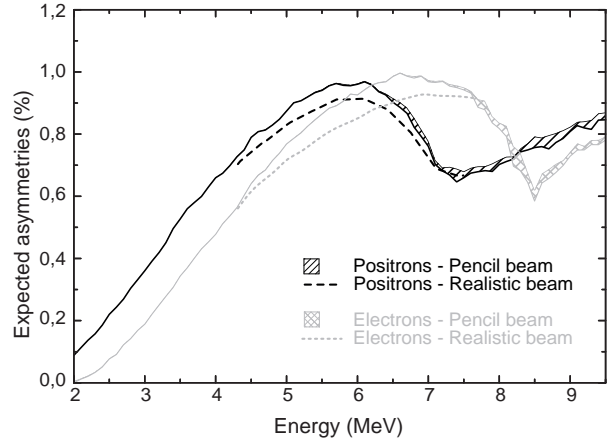} 
  \end{center}
  \caption[Simulations of the asymmetries of the energies observed in
  the central CsI crystal.]{Simulations of the asymmetries $\delta$,
    Eq.~(\ref{eq:analpower}), of the energies observed in the central CsI crystal  
on reversal of the polarity of magnet {\sf TP1} as a function of positron/electron energy for the experimental beam and for a pencil beam.}
  \label{fig:expected_asymmetry}
\end{figure}

The asymmetries $\delta$ in the energies observed in the crystals of the CsI calorimeter on reversal of the polarity of magnet {\sf TP1} was predicted from the simulated polarization $P_{e^\pm}$ of positrons (electrons) after the production target (see Sec.\:\ref{sec:production}) and the analyzing power $A_{e^\pm}$, 
\begin{linenomath}
  \begin{equation}
    \delta = A_{e^\pm} P_{e^\pm} P_{e^-}^{\rm Fe} ~~.
  \end{equation}
\end{linenomath}
Figure\:\ref{fig:analysing_power_polarization}
shows the simulated polarization of positrons (electrons) after the production target, together with the analyzing power in the energy range 2.0--9.5\,MeV. 
The asymmetries $\delta$ expected in the central CsI crystal are shown in Fig.\:\ref{fig:expected_asymmetry}, where the results of a simulation of an ideal pencil-like beam are also presented.
%


\subsection{The $\gamma$-Line}\label{sec:sim_pl}

\subsubsection{Photon Spectra}\label{sec:photon_spectra}

The simulation of the $\gamma$-line was carried out with two special simulation tools based on {\sc Geant}3, as modified to include polarization-dependent absorption cross sections. 
The first of these generated the undulator-photon spectrum through
third order in the undulator-strength-parameter $K$,
and corrected the spectrum for absorption in the production target
{\sf T1}, the incident-flux detectors {\sf A1} and {\sf S1}, the
magnetized-iron polarimeter {\sf TP2}, and the transmitted-flux detectors {\sf A2}, {\sf S2}, and {\sf GCAL}. 
The code allowed separate examination of contributions from Compton scattering, pair production, and the photoelectric effect.  
It also incorporated the response functions discussed in Sec.~\ref{sec:gamma_response} to calculate the signal generated in each detector by the passage of the photon beam.
The program did not treat effects of beam divergence or collimator acceptance.

\begin{figure}
  \centering
  \includegraphics[width=8cm]{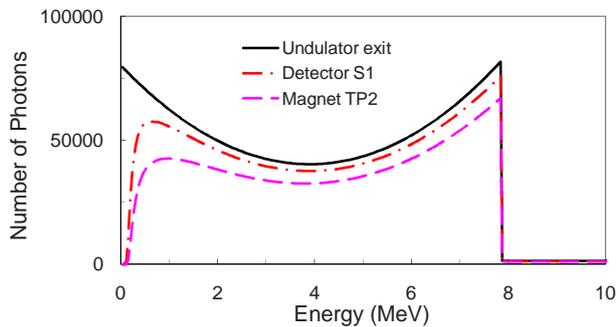}   
  \caption{Simulated photon spectra $N_\gamma(E)$ at the exit of the undulator, at detector {\sf S1}, and incident on
polarimeter magnet {\sf TP2}.}
  \label{fig:51}
\end{figure}

\begin{figure}
  \centering
  \includegraphics[width=8cm]{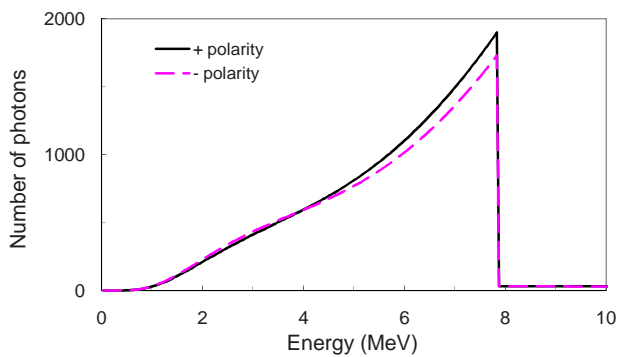}    
  \caption{Simulated spectra $N_\gamma(E)$ of photons at detector {\sf S2} for both polarities of magnet {\sf TP2}.}
  \label{fig:52}
\end{figure}

Figure\:\ref{fig:51}
shows photon spectra $N_\gamma(E)$ as simulated by this program at the exit of the undulator, at detector {\sf S1}, and incident on the polarimeter magnet {\sf TP2}, assuming a first-order cutoff energy of $E_1 \approx$ 7.9\,MeV ($K \approx 0.17$) and that the production target {\sf T1} was 0.81\,mm of tungsten alloy. 
The loss of photons at very low energies (and large angles) in the first simulation was due to absorption in the production target {\sf T1}, which also removed 10\,\% of photons of all higher energies. 
Another 16\,\% of the photons were absorbed in the 0.5-mm-thick tungsten convertor of detector {\sf S1} and in 1\,cm of aluminum in detector {\sf A1} before the beam reached magnet {\sf TP2}.
Figure\:\ref{fig:52}
shows the predicted photon spectrum at counter {\sf S2} behind polarimeter magnet {\sf TP2} for both states of polarization of the iron, assuming electron polarization $P^{\rm Fe}_{e^-} = 7.23$\,\% (Table~\ref{tab_analmag_one}). 
The magnet iron reduced the photon flux by a factor of $\approx 50$.
The relative number of photons and the relative total energy in the photon beam at various components along the $\gamma$-line are summarized in Table\,\ref{tab:16}; not included are possible losses in collimators {\sf C2-C4}. 
\begin{table}
  \caption{Results of a {\sc Geant3} simulation of the relative number of photons, their relative total energy, and the average energy per photon at the upstream face (except as noted) of various $\gamma$-line components.
In this simulation the photon beam incident on target {\sf T1} was the same as that at the exit of the undulator.
Beyond {\sf TP2} results are reported for both polarities of that magnet.
}
  \label{tab:16}
  \centering
 \vspace{1mm}
\footnotesize
  \begin{tabular*}{8 cm}{l@{\extracolsep\fill}ccc}
    \hline
    \hline
    Component  	& Photon 	&	Total	&	$\langle$Energy$\rangle$ per	\\
			& number	&	energy	& 	photon (MeV)	\\
    \hline
Target {\sf T1}			&	1.000	&	1.000	&	4.116	\\
Exit of {\sf T1}		&	0.895	&	0.939	&	4.316	\\
Detector {\sf S1}		&	0.881	&	0.927	&	4.332	\\
Detector {\sf A1}		&	0.835	&	0.888	&	4.380	\\
Magnet {\sf TP2} 		&	0.739	&	0.807	&	4.494	\\
Detector {\sf A2} $(+)$		&	0.01406	&	0.02010	&	5.885	\\
{\sf A2} $(-)$			&	0.01334	&	0.01880	&	5.802	\\
Detector {\sf S2} $(+)$		&	0.01280	&	0.01840	&	5.912	\\
{\sf S2} $(-)$			&	0.01214	&	0.01720	&	5.831	\\
Detector {\sf GCAL} $(+)$	&	0.01229	&	0.01765	&	5.909	\\
{\sf GCAL} $(-)$		&	0.01165	&	0.01650	&	5.828	\\
    \hline
  \end{tabular*}
\end{table}

\begin{figure}
  \centering
  \includegraphics[width=8cm]{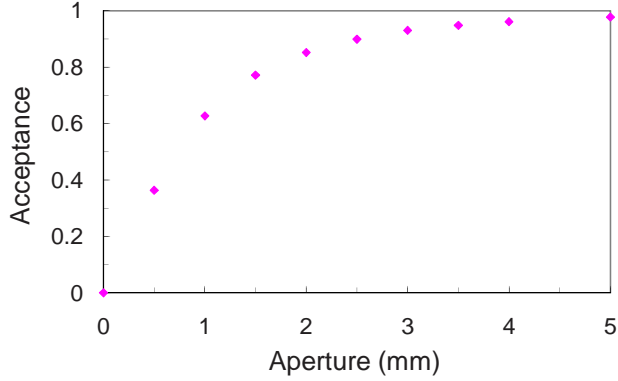} 
  \caption{Simulated acceptance of the collimator {\sf C2} as a function of the collimator aperture, assuming that the photon beam axis and the collimator axis coincided.}
  \label{fig:col_accept}
\end{figure}

The second simulation tool folded the beam emittance with the undulator spectrum to examine the geometric transmission of the photon beam through the aperture of collimator {\sf C2}. 
It could independently vary the size of the {\sf C2} aperture, the
lateral displacement, and the angle of the collimator axis with respect to the axis of the photon beam. 
For example, Fig.~\ref{fig:col_accept} shows the {\sf C2} acceptance of undulator photons ($N_{\rm transmitted} / N_{\rm incident}$) as a function of collimator aperture when the beam axis and collimator axis coincide.
The nominal diameter of that collimator was 3\,mm, so that 93\,\% of the photons, and 98.9\,\% of their energy, passed through the collimator if the beam was properly aligned.   
Because of the correlation (\ref{p3}) between
photon energy and angle, the photons that were absorbed by the collimator had very low energy.
Lateral displacement of the photon beam had a relatively small effect on the {\sf C2} acceptance if the collimator axis was parallel to the beam axis; a 1\,mm displacement resulted in a loss of acceptance of 10\,\%. 
The simulation also showed that for a 5\,mrad tilt of the C2 axis with
respect to the beam axis the loss of acceptance would be only 3\,\%,
but a combination of the tilt plus a 1\,mm displacement of the axes
could cause reduction in acceptance of up to 48\,\%, in which case
{\sf S1} and {\sf PRT} would only cover a corresponding fraction of
the full undulator-photon flux.

\subsubsection{$\gamma$-Detector Response}\label{sec:gamma_response}

\begin{figure}
  \centering
  \includegraphics[width=8cm]{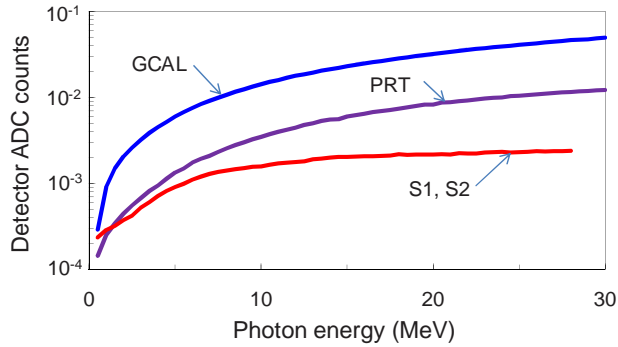}    
  \caption{{\sc Geant3} simulation of the single-photon response $R(E)$ of the $\gamma$-detectors {\sf S1/S2}, {\sf GCAL}, and {\sf PRT}.}
  \label{fig:Si_responses}
\end{figure}

Figure\:\ref{fig:Si_responses} shows the results of the {\sc Geant3}
simulation 
of energy deposition $R(E)$
in various Si-W detectors as a function of the energy $E$ of single
incident photons. 
Estimates of the detector sensitivities (in ADC counts, Sec.~\ref{sec:detectors}) were obtained by convoluting the single-photon response $R(E)$ of the detectors with the corresponding spectrum $N_\gamma(E)$ of incident photons (Sec.~\ref{sec:photon_spectra}).

Without attenuation, the average sensitivity of detector {\sf S1} for the photon spectrum labeled {\sf S1} in Fig.~\ref{fig:51} would have been approximately 1250 photons per ADC count, corresponding to 1420 photons at the exit of the undulator. 
It was therefore necessary for normal data acquisition to attenuate the signal by a factor of 199.5 (46\,dB), such that one ADC count for {\sf S1} corresponded to 283,000 photons.

Although detectors {\sf S1} and {\sf S2} were identical in construction, the energy spectra of incident particles were quite different and the unattenuated sensitivity of {\sf S2} was about 950 photons/ADC count; and that of {\sf GCAL} about 135 photons/ADC count. 
In the experiment the {\sf S2} signal was attenuated by a factor of 10 (20\,dB) and that of {\sf GCAL} by 100 (40\,dB). 
For the {\sf PRT} detectors, assuming that the incident spectrum was an undisturbed undulator spectrum, the sensitivity was 860 photons/ADC count and the attenuation used was 316.2 (50\,dB).
For the {\sf PCAL} detectors, which monitored positrons of GeV energy, the unattenuated sensitivity was about 600 MeV/ADC count. 
The attenuation was 40, 50, and 70\,dB for {\sf PCAL}c, {\sf PCAL}d, and {\sf PCAL}e, respectively. 
The {\sf A1} and {\sf A2} Cherenkov counter sensitivities were obtained by comparison with the {\sf S1} and {\sf S2} signals shown in Fig.~\ref{fig:und_phot1_and_trans1}.

\subsubsection{Transmission Ratios and Asymmetries}\label{sec:gamma_trans}

The first {\sc Geant3} simulation (Sec.~\ref{sec:photon_spectra}), which included the response functions $R(E)$ for detectors {\sf S1}, {\sf S2}, and {\sf GCAL} and the corresponding spectra $N_\gamma(E)$ of incident photons, was used to calculate the signal transmission ratios {\sf S2}/{\sf S1} and {\sf GCAL}/{\sf S1} for both polarities of polarimeter magnet {\sf TP2}, and the transmission asymmetries $\delta$ of Eq.~(\ref{eq:analpower}), as shown in Table\:\ref{tab:17}.  
\begin{table}
  \caption{Simulations of the transmission of photons from the undulator though detectors {\sf S2} and {\sf GCAL}, normalized to the photon flux at detector {\sf S1}, and of the signal asymmetries with respect to the polarity of magnet {\sf TP2} in detectors {\sf A2}, {\sf S2}, and {\sf GCAL}.}
  \label{tab:17}
  \centering
 \vspace{1mm}
\footnotesize
  \begin{tabular*}{4cm}{l@{\extracolsep\fill}c}
    \hline
    \hline
    Detector			&	Transmission 	\\
    \hline
    {\sf S2}$(+)$/{\sf S1} 	&	0.0202	\\
    {\sf S2}$(-)$/{\sf S1}	&	0.0189	\\
    Average			&	0.0195	\\
    \hline
    {\sf GCAL}$(+)$/{\sf S1}	&	0.1366 \\
    {\sf GCAL}$(-)$/{\sf S1}	&	0.1274 \\
    Average			&	0.1320 \\
   \hline
   \hline
		&	Asymmetry	\\
   \hline
    {\sf A2}    &	0.0363	\\
    {\sf S2} 	&	0.0339	\\
    {\sf GCAL}	&	0.0347	\\
    \hline
    \hline
  \end{tabular*}
\end{table}
A reliable calibration was obtained for the absolute response of the silicon detectors, but not for the Cherenkov detectors, so simulation of transmission ratios was possible only for {\sf S2} and {\sf GCAL} relative to {\sf S1}. 
For detector {\sf A2} only the transmission asymmetry $\delta$ was simulated, assuming a Cherenkov cutoff of 3.8\,MeV and a response proportional to $1-1/n^2\beta^2$ above threshold.

\section{Background Studies}\label{sec:backgrounds}

Two classes of backgrounds were studied: those associated with the primary electron beam; and those associated with energizing the undulator.

\subsection{Backgrounds Related to the Primary Electron Beam}\label{sec:back}

Backgrounds associated with the primary electron beam were due to secondary particles that reached various detectors from electromagnetic showers of particles in the tails of the primary beam, as well as due to backsplash from the interaction of the primary beam with the dump at the end of the FFTB.   
These backgrounds were mitigated by protection collimators prior to the last ``dogleg'' bend in the FFTB, and by extensive lead shielding between the experimental detectors and undulator, and also between the detectors and the beam dump.

The major source of electron-beam-related backgrounds was interaction of the tails of the beam
with the protection collimator {\sf C1} (0.71\,mm in aperture and
6.35\,cm long; see Sec.~\ref{sec:beamline}) 
and with undulator bore tube (0.9\,mm in aperture and 1\,m long).  
This was confirmed by comparison of rates in normal data runs with those in occasional runs with collimator {\sf C1} and the undulator moved out of the electron beam and a 2.5-cm-diameter bypass vacuum pipe inserted in the beam using an articulated mover, sketched in Fig.~\ref{fig:E166Scheme}.
Table\:\ref{tab:undulator_backgr} compares the responses of various detectors, normalized to a beam intensity of $3 \cdot 10^9\,e^-$/pulse, for two such runs.

\begin{table}
  \centering
  \caption{Comparison of rates in ADC counts for the signals (after correction for attenuation) in various detectors for two runs, No.~1702 with undulator out, and No.~2803-5 with the undulator inserted but energized out of time with the electron beam.}
  \label{tab:undulator_backgr}
 \vspace{1mm}
\footnotesize
  \begin{tabular*}{6cm}{l@{\extracolsep\fill}cc}
    \hline
    \hline
    Detector & Undulator &  Undulator in      \\
             & removed   & but out of time   \\
    \hline
    {\sf A1} 		& 0.7 	& 81.6 \\
    {\sf A2} 		& 0.2 	& 61.2 \\
    {\sf CsI} 		& 11.6 	& 71.3 \\
    {\sf GCAL} 		& 17.9 	& 16725.2 \\
    {\sf P1} 		& 0.1 	& -0.2 \\
    {\sf PCAL}c 	& -0.1 	& 7287.3 \\
    {\sf PCAL}d 	& -0.4 	& 17590.2 \\
    {\sf PCAL}e 	& 0.0 	& 24740.4 \\
    {\sf PRT}left 	& 1.2 	& 23171.0 \\
    {\sf PRT}right 	& -10.4 & 3308.4 \\
    {\sf PRT}top 	& 47.5 	& 7807.4 \\  
    {\sf PRT}bottom 	& -5.5 	& 3962.2 \\ 
    {\sf S1} 		& 35.5 	& 1178.6 \\
    {\sf S2} 		& 0.1 	& 425.1 \\
    \hline
    \hline
  \end{tabular*}
\end{table}

In the absence of the undulator, the rates were small and we were unable to ascertain the 
relative contribution from interactions of the electron beam with material upstream and downstream of
the location of the undulator.
In particular, the level of background due to backsplash from the electron beam dump was very small and no correction was made for it in the data analysis.

When the undulator was in place, but not energized in time with the electron beam, the rates increased slightly in the two detectors of the positron polarimeter, {\sf CsI} and {\sf P1}, which were displaced by 464\,mm from the $\gamma$-line.
In contrast, the rates in detectors along the $\gamma$-line (including the {\sf PCAL}) increased by factors of several hundred when the undulator was introduced.

Secondary photons from interactions of the tails of the electron beam with the undulator followed a path similar to undulator-generated photons and were the main source of background rates in the various $\gamma$-line detectors.  
Secondary electrons from interactions in the undulator (or material upstream thereof) were deflected downward by the dump magnet string {\sf D1} (Fig.~\ref{fig:explayc2k}) and were largely absorbed in the magnet yokes or in the lead shielding (shown in Fig.~\ref{fig:E166Scheme}) upstream of the detectors.
Secondary positrons were deflected upwards and could have created minor backgrounds in the $\gamma$-line detectors via tertiary particles from interactions with the ceiling of the FFTB tunnel; this may have been the major source of
background in the {\sf CsI} detector.

The large rates in detectors {\sf S2} and {\sf GCAL} downstream of polarimeter magnet {\sf TP2} were due to low-energy particles from showers of high-energy background photons in the magnet iron.   These low-energy particles were not observed in Cherenkov counter {\sf A2} which threshold was about 4\,MeV.

The energy observed in the {\sf PCAL} detectors was largely due to high-energy positrons from interactions of primary electrons with collimator {\sf C1} and the undulator bore tube. 
From this energy it was estimated that only about one beam particle in $10^5$ so interacted.
 
During normal data collection, backgrounds associated with the tails of the primary electron beam were separated from  signals due to undulator photons by energizing the undulator in- and out-of-time with respect to the electron beam on alternate pulses.
Table\:\ref{tab:ImportanceOfBackground} shows the average background and undulator-photon signal (undulator-on minus undulator-off) in various detectors for a sample of 211 runs during September 2005.
\begin{table}
  \centering
  \caption{Background and signal ADC counts in various detectors for a sample of September 2005 runs. 
The background is from undulator-off data and the signal is undulator-on minus undulator-off.}
  \label{tab:ImportanceOfBackground}
 \vspace{1mm}
\footnotesize
  \begin{tabular*}{8cm}{l@{\extracolsep\fill}ccc}
    \hline
    \hline
    Detector &      Background &            Signal &                Signal/Background \\
    \hline
    {\sf A1}  		& 242.3 & 543.7 &  2.4 \\
    {\sf A2}  		& 210.7 &  13.7 &  0.1 \\
    {\sf CsI} 		& 136.0 &  29.0 &  0.2 \\
    {\sf GCAL} 		& 580.5 &  91.0 &  0.2 \\
    {\sf P1} 		& 9.3   & 509.2 & 54.9 \\
    {\sf PCAL}e 	& 30.9  &  -1.4 & -0.1 \\
    {\sf PRT}top 	& 59.8  & 259.8 &  4.3 \\
    {\sf PRT}right 	& 27.5  & 229.1 &  8.3 \\
    {\sf PRT}bottom 	& 29.9  & 292.3 &  9.8 \\
    {\sf PRT}left 	& 159.9 & 598.2 &  3.7 \\
    {\sf S1}		& 5.8   & 164.4 & 28.5 \\
    {\sf S2} 		& 169.5 & 95.5  &  0.6 \\
    \hline
    \hline
  \end{tabular*}
\end{table}

The signal-to-background ratios were different for the central crystal (CC), the four crystals touching the central one along a side (NC = nearest crystal), and the four crystals in the corners (DC = distant crystal). 
The undulator-off signal levels were typically about 66\,\%, 84\,\%,  and 91\,\% of those for CC, NC, and DC in undulator-on events, respectively. 
The average energy deposition per beam pulse for undulator-off events in the central crystal, for example, varied between 130 and 280\,MeV, such that the signal (undulator-on $-$ undulator off) to background (undulator-off) ratio varied between 0.2 and 1.2. 

Details of the background subtraction in the positron analysis will be presented in Sec.~\ref{sec:background}.
 
\subsection{Undulator-On Backgrounds}\label{sec_und_back}

Background in the {\sf CsI} detector associated with energizing the undulator principally were due to secondary particles from interactions of undulator photons with collimators, beam windows, and target {\sf T1} which reached the detector by other than the nominal path through the positron spectrometer {\sf D2}.  
In addition, the small deflection of the primary electron beam by the undulator, discussed in Sec.~\ref{sec:steering}, might have changed the level of backgrounds associated with interaction of tails of the primary electron beam with the undulator body.

\begin{figure}
  \centering
  \includegraphics[width=8cm]{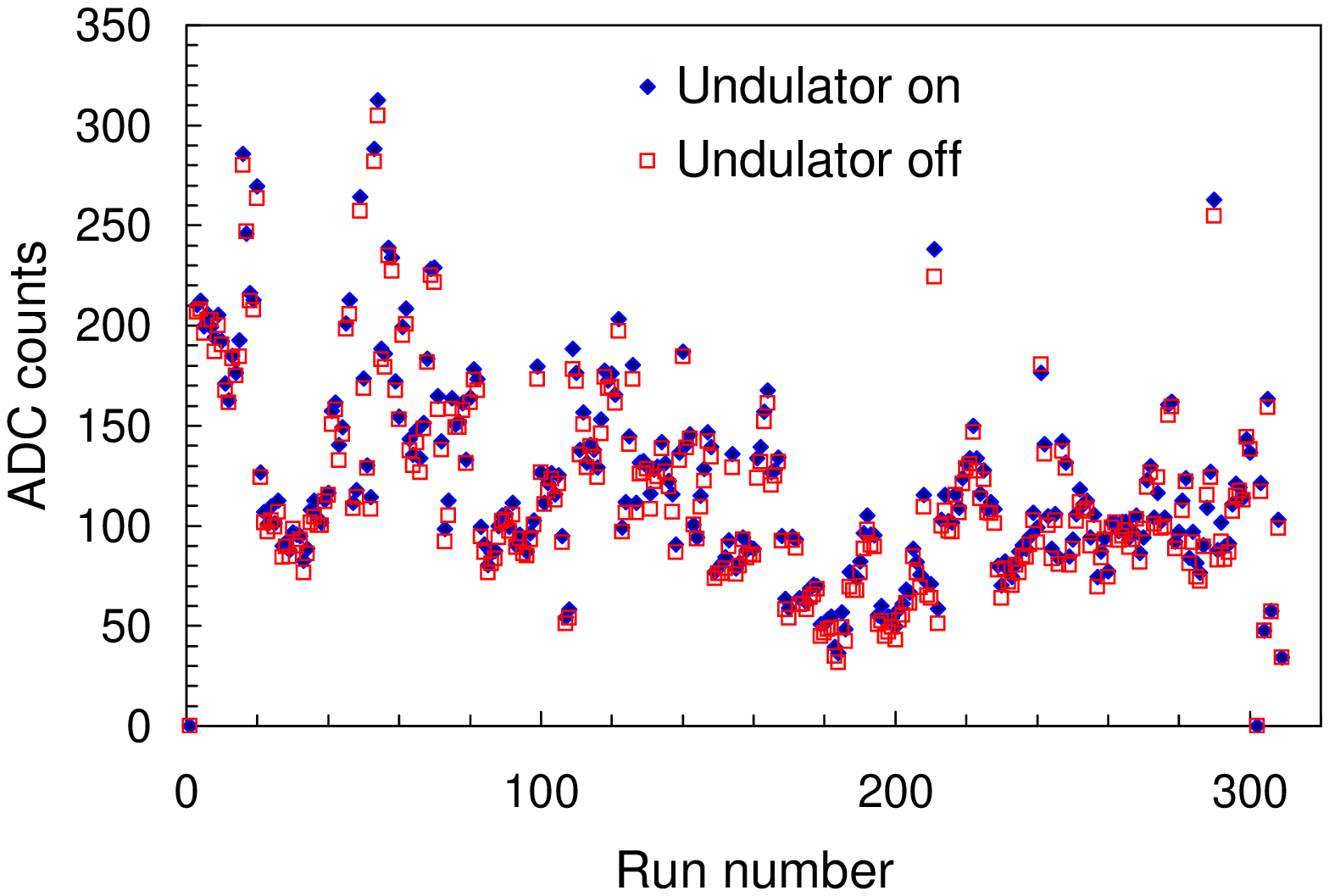} 
  \caption{CsI-central-crystal signal in ADC counts for all September 2005 runs with positron target {\sf T1} removed and
undulator on (circle) or off (rhombus).}
  \label{fig:CsI_on_off.}
\end{figure}

A measure of the undulator-induced backgrounds in the central CsI crystal was obtained in a set of runs during September 2005 in which target {\sf T1} was removed, and data collected, as usual, with the undulator both on (in time with the electron beam) and off (out of time), as shown in Fig.~\ref{fig:CsI_on_off.}.
The average target-out, undulator on-off difference was 4.2\,ADC counts out of the typical background level of 100\,ADC counts due to effects of the primary electron beam.

\begin{figure}
  \centering
  \includegraphics[width=8cm]{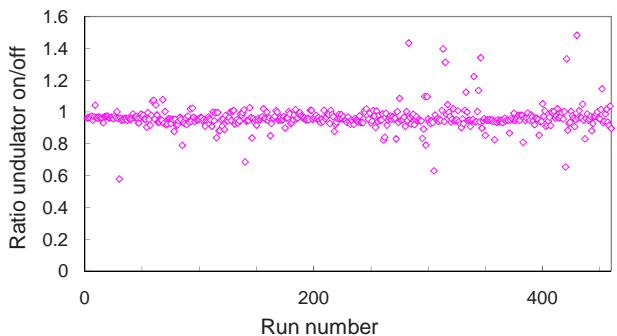}   
  \caption{Ratio of energies for undulator-on/off measured by detector {\sf PCAL}e for a series of runs.}
  \label{fig:Pcale_on_off}
\end{figure}

Additional evidence for a small change of the interaction of the tails of the primary electron beam with the undulator due to energizing the undulator was obtained from the {\sf PCAL} detectors.
For example, {\sf PCAL}c and {\sf PCAL}e consistently recorded an undulator-on/off ratio of $(95.5\pm0.5)\,\%$, as shown in Fig.\:\ref{fig:Pcale_on_off}, while {\sf PCAL}d gave a ratio of $(100.0\pm0.5)\,\%$. 

No correction was made for this small difference between undulator-on and undulator-off background, but it was included as a term in the estimate of the systematic uncertainty of the positron polarization (Sec.~\ref{sec:syserr}).

\section{Photon Analysis}\label{sec_phot_anal}

\subsection{Undulator Performance}\label{sec_undulator_performance}

\subsubsection{Photon Beam Intensity}\label{sec:und_phot_int}

Based on the design parameters of the undulator, as described in Secs.\:\ref{sec_undulator_physics} and \ref{sec_undulator}, the undulator generated 0.35 photons per beam electron with angle-energy distribution (\ref{p3}).
Thus, half of the photons were emitted at angles larger than $1.1\,\mu$rad to the direction of the electron beam,
such that the flux observed in the $\gamma$-detector {\sf S1} was strongly influenced by the alignment of the primary electron beam.  
During the experiment, the electron beam was steered to minimize backgrounds generated by interactions with collimator {\sf C1} upstream of the undulator, rather than to maintain optimal alignment of the $\gamma$-beam.
To include a measure of the phtoton flux outside
  collimator {\sf C1}, photon intensity measurements were based on the
  sum of signals 
in detector {\sf S1} and in the quadrant detector {\sf PRT} that
observed photons outside the aperture of collimator {\sf C2}. 

\begin{figure}
  \centering
 \includegraphics[width=8cm]{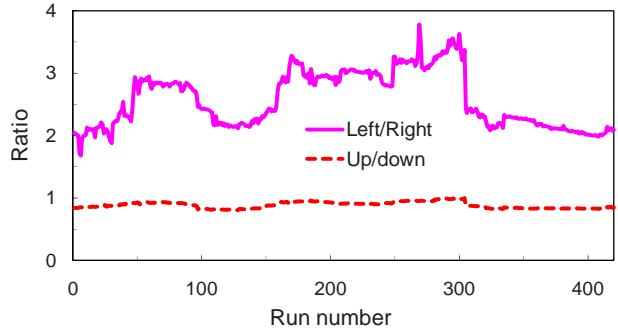} 
  \caption{The left/right and the up/down ratios measured in the {\sf PRT} quadrant detectors during 420 runs.}
  \label{fig:prt_asym}
\end{figure}

\begin{figure}
  \centering
  \includegraphics[width=8cm]{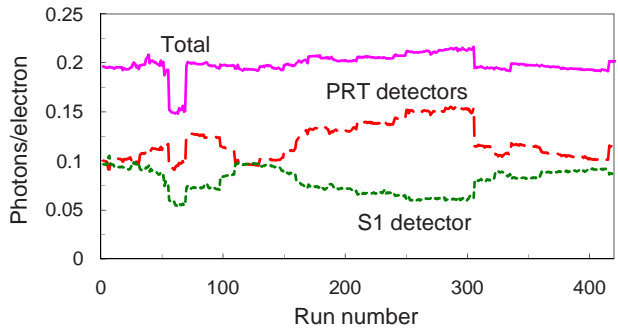}   
  \caption[Undulator photons per beam electron in the {\sf PRT} and {\sf S1} detectors.]{Undulator photons per beam electron in the {\sf PRT} and {\sf S1} detectors for the same 420 runs as in Fig.\:\ref{fig:prt_asym}.}
  \label{fig:phot_prt_s1}
\end{figure}

Figure\:\ref{fig:prt_asym}
 shows the left/right and up/down ratios of rates in the four {\sf PRT} $\gamma$-detectors for 420 runs at spectrometer current $I_{\rm S} = 100$\,A. 
The beam displacement at collimator {\sf C2} was primarily in the $x$ (horizontal) direction, and the observed
left/right ratios corresponded to displacements from 500 to 1100\,$\mu$m according to simulations. 
The average $y$ (vertical) displacement was less than 100\,$\mu$m. 
The anticorrelation of the signals in detectors {\sf PRT} and {\sf S1} is shown in Fig.\:\ref{fig:phot_prt_s1} for the same set of runs as in Fig.\:\ref{fig:prt_asym}.
The sum of the signals in the {\sf PRT} and {\sf S1} detectors  remained fairly constant at about 0.2 photons per electron.  
Averaged over the entire experiment, the rate of undulator photons observed in the {\sf PRT} and {\sf S1} detectors
per beam electron was  0.199$\pm$0.008, about 60\,\% of the design performance.

\subsubsection{Intensity and $K$-Value \vs\ Undulator Current}

\begin{figure}
  \centering
  \includegraphics[width=8cm]{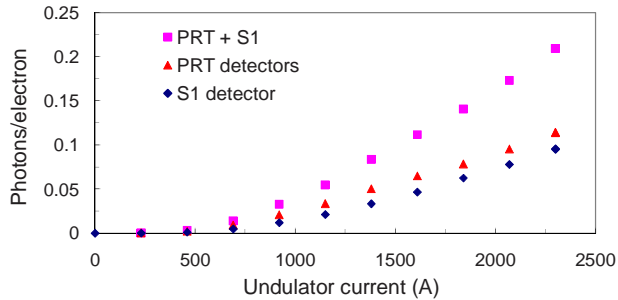} 
  \caption{Photons per beam electron in the {\sf PRT} and {\sf S1} detectors \vs\ undulator current.}
  \label{fig:current_scan}
\end{figure}

The undulator $K$-value (\ref{eq_u2}) depended linearly on the excitation current of the undulator, while the photon intensity (\ref{eq_Ngamma}) varied as the square of $K$-value, and hence as the square of the current.
Several scans of photon intensity \vs\ undulator current were conducted during the experiment.
Results from one of these scans, shown in Fig.\:\ref{fig:current_scan},
\begin{figure}
  \centering
  \includegraphics[width=8cm]{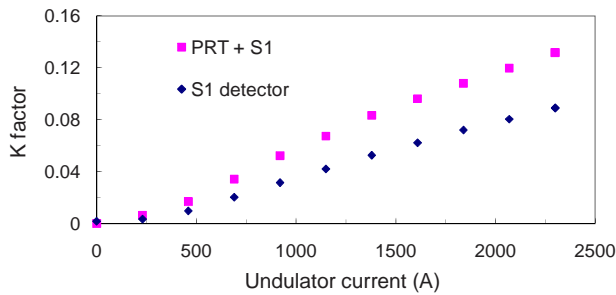} 
  \caption{Dependence of the $K$-value on the undulator current as determined from the
number of photons per beam electron observed in the {\sf PRT} and {\sf S1} detectors.}
  \label{fig:Kvalue}
\end{figure}
exhibit the expected quadratic dependence of intensity on current. 
The observed photon intensities in detectors {\sf PRT} and {\sf S1} were converted to estimates of the $K$-value via Eq.~(\ref{eq_Ngamma}), with results shown in Fig.~\ref{fig:Kvalue}.
The estimated $K$-value varied approximately linearly with current, but was below the design value of 0.17 at
2300\,A. 
The discrepancy can be 
explained by the observed misalignment between collimator {\sf C2} and the 
photon beam, as discussed in Sec.~\ref{sec:photon_spectra}.

\subsubsection{Effect of Ferrofluid in the Undulator}\label{sec:ferro}

\begin{figure}[ht]
  \centering
  \includegraphics[width=8cm]{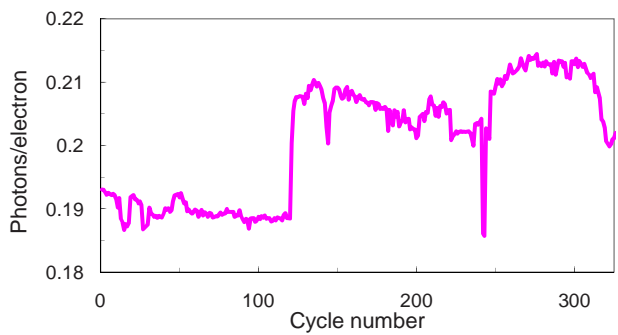}   
  \caption{Photons observed in the {\sf PRT} and {\sf S1} detectors per beam electron during operation with spectrometer current $I_{\rm S} = 180$\,A. 
Ferrofluid was used in the undulator beginning with cycle 132.}
  \label{fig:photon_per_electron}
\end{figure}

Near the end of the experiment the cooling oil in the undulator was replaced by Ferrofluid EMG 900, as discussed in Section\:\ref{sec_undulator}. 
The increase in undulator magnetic field resulted in an increase of the photon flux by 10--12\,\%, as shown in Fig.\:\ref{fig:photon_per_electron}.

\subsection{Photon Transmission through Magnet {\sf TP2}}

Measurements of the transmission of undulator photons through the magnetized iron of the polarimeter magnet {\sf TP2} were made by comparing rates in detectors {\sf GCAL} and {\sf S2} to those in detector {\sf S1}. 
Detector {\sf A2} was not used for this measurement as its signal normalization was not well determined.

\begin{figure}
  \centering
   \includegraphics[width=7.8cm]{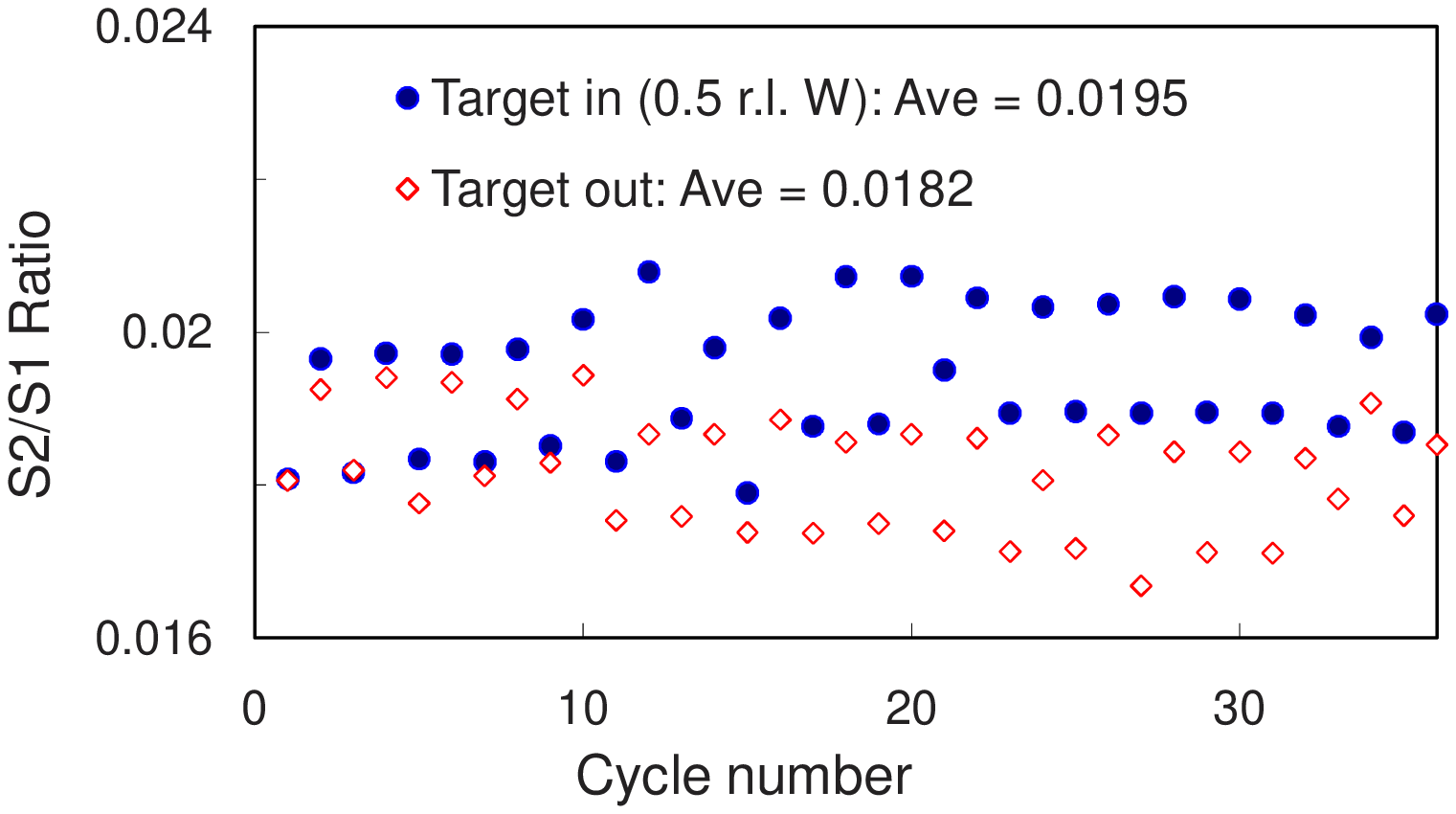}  
    \hfill
   \includegraphics[width=7.8cm]{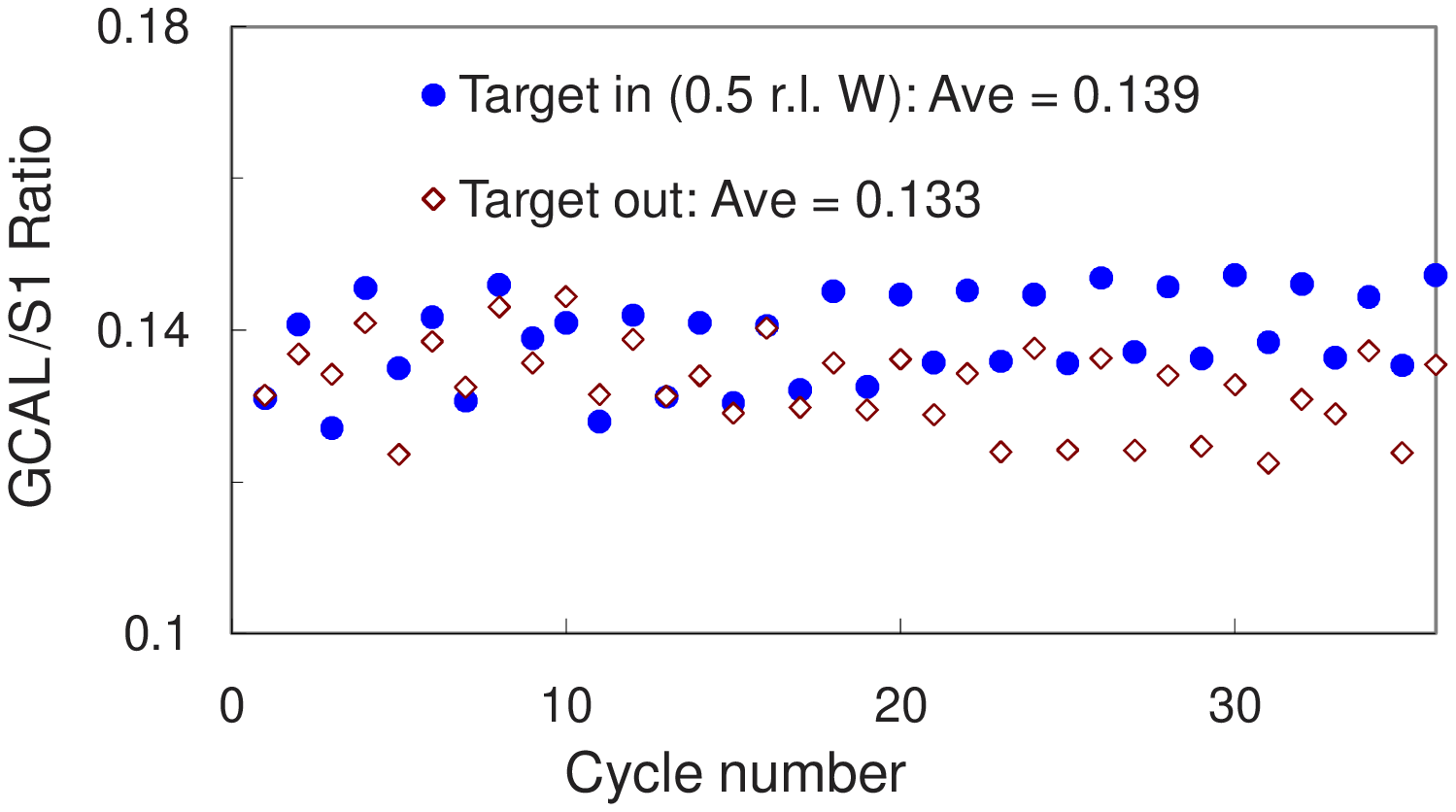}  
  \caption{Transmission of undulator photons through magnet {\sf TP2} as measured with detectors {\sf S2} (left) and {\sf GCAL} (right), both normalized to detector {\sf S1}.
}
  \label{fig:trans_s2GCAL}
\end{figure}

Two rather small samples of data from the June 2005 provided the most reliable direct transmission measurements, as shown in Fig.\:\ref{fig:trans_s2GCAL}.
One set of data was taken with the target removed and the other with the 0.5\,r.l.\ tungsten target.   The
variation in rate with alternating cycles was due to the reversal of polarity of magnet {\sf TP2}.
The transmission ratios {\sf S2/S1} and {\sf GCAL/S1} are given in Table\:\ref{tab:phot_trans1}
\begin{table}
  \caption{The transmission ratios {\sf S2/S1} and {\sf GCAL/S1} averaged over the run cycles shown in Fig.\:\ref{fig:trans_s2GCAL}, and a comparison with a \geant\ simulation.}
  \label{tab:phot_trans1}
  \centering
 \vspace{1mm}
\footnotesize
  \begin{tabular*}{8cm}{l@{\extracolsep\fill}ccc}
    \hline
    \hline
    & Target out & 0.5\,r.l.\ Target & 0.2\,r.l.\ Target \\
    &             &                  & (simulation)   \\
    \hline  
    {\sf S2/S1}   & 0.0182 & 0.0194 & 0.0195 \\	
    {\sf GCAL/S1} & 0.133  & 0.139  & 0.132\\
    \hline
    \hline
  \end{tabular*}
\end{table}
along with a comparison with a simulations described in Sec.\:\ref{sec:gamma_trans}. 
The agreement with predictions is reasonable given the intrinsic uncertainty in unmeasured undulator parameters and beam conditions. 
Qualitatively, the measured transmission is greater with the target in place as the target hardens the spectrum resulting in increased penetration of the iron absorber. 

During data collection in September 2005 collimator {\sf C4}
downstream of magnet {\sf TP2} was inadvertently displaced so as to
intercept a portion of the transmitted photon beam, resulting in a loss of about half of the transmitted signal. 
While this did not interfere appreciably with the measurement of photon polarization (Sec.~\ref{sec:gamma_asym}), it resulted in unreliable transmission measurements. 
 
\begin{figure}
  \centering
  \includegraphics[width=8cm]{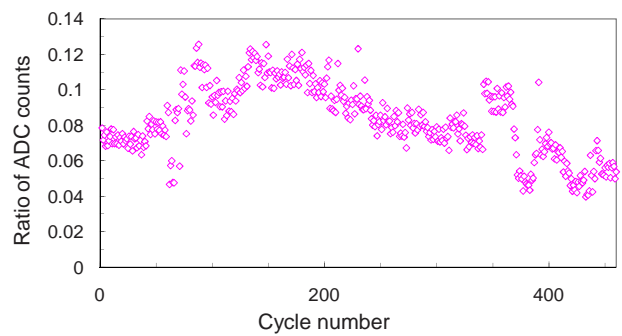}   
  \caption{Ratio of the signal in detector {\sf S2} to that in {\sf S1} for a set of cycles with spectrometer current $I_{\rm S} = 140\,A$.}
  \label{fig:phot_trans1}
\end{figure}

Gradual drifts with time of the transmissions measured by both {\sf S2} and {\sf GCAL} were observed,
as shown in Fig.\:\ref{fig:phot_trans1}.
The source of this drift with run conditions has not been clearly identified, but it may have been due to
changes in beam alignment relative to collimators {\sf C2}--{\sf C4} that altered the energy spectrum of the photons in the $\gamma$-line. 

\subsection{Photon Transmission Asymmetry}\label{sec:gamma_asym}

The transmission asymmetry $\delta$, defined in Eq.~(\ref{eq:analpower}), was measured with each of the three
detectors {\sf S2}, {\sf A2}, and {\sf GCAL} (in sequence along the $\gamma$-line).  

\begin{figure}
  \begin{center}
    \includegraphics[width=8cm]{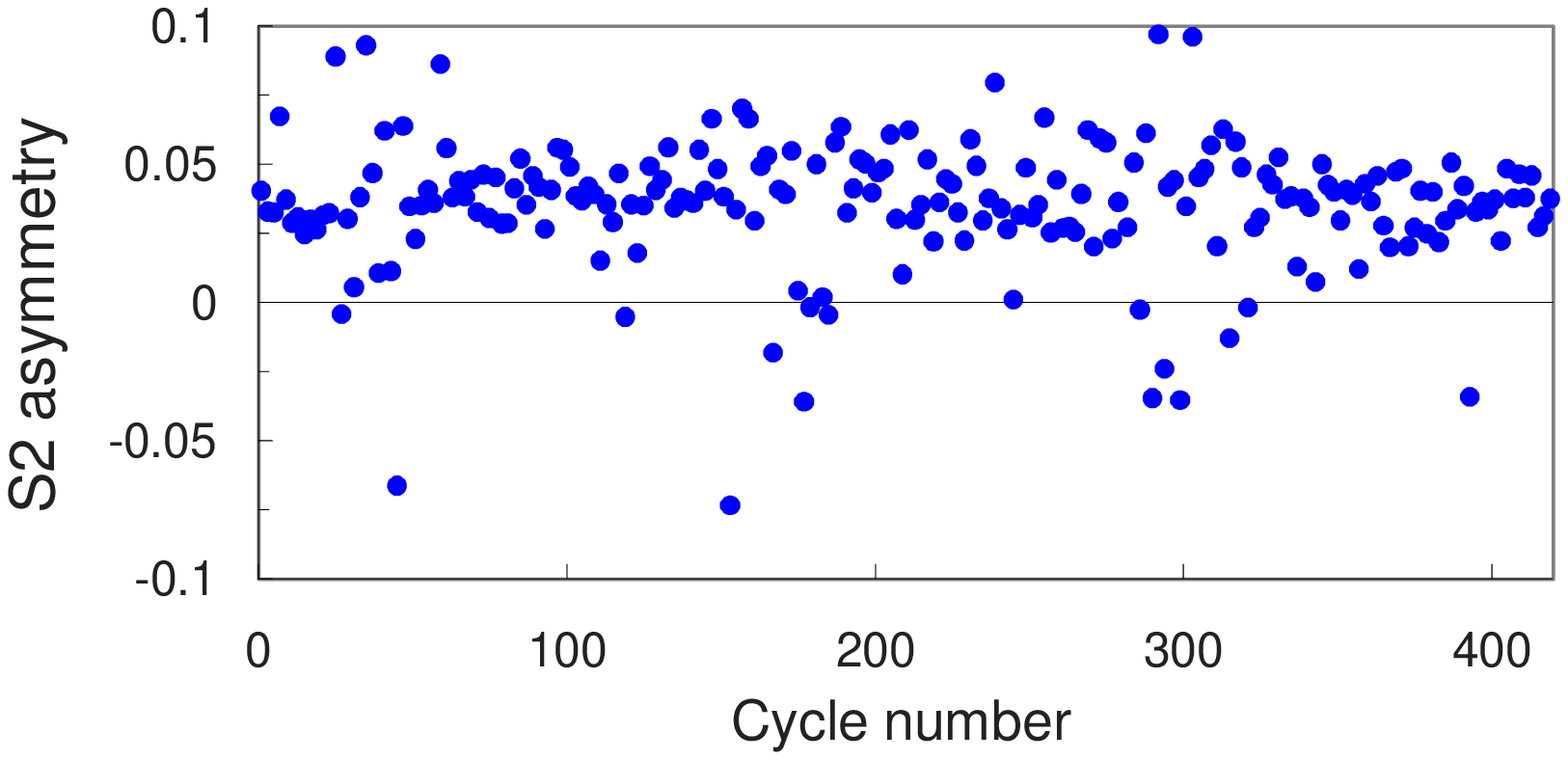}  

\vspace{0.1in}
    \includegraphics[width=8cm]{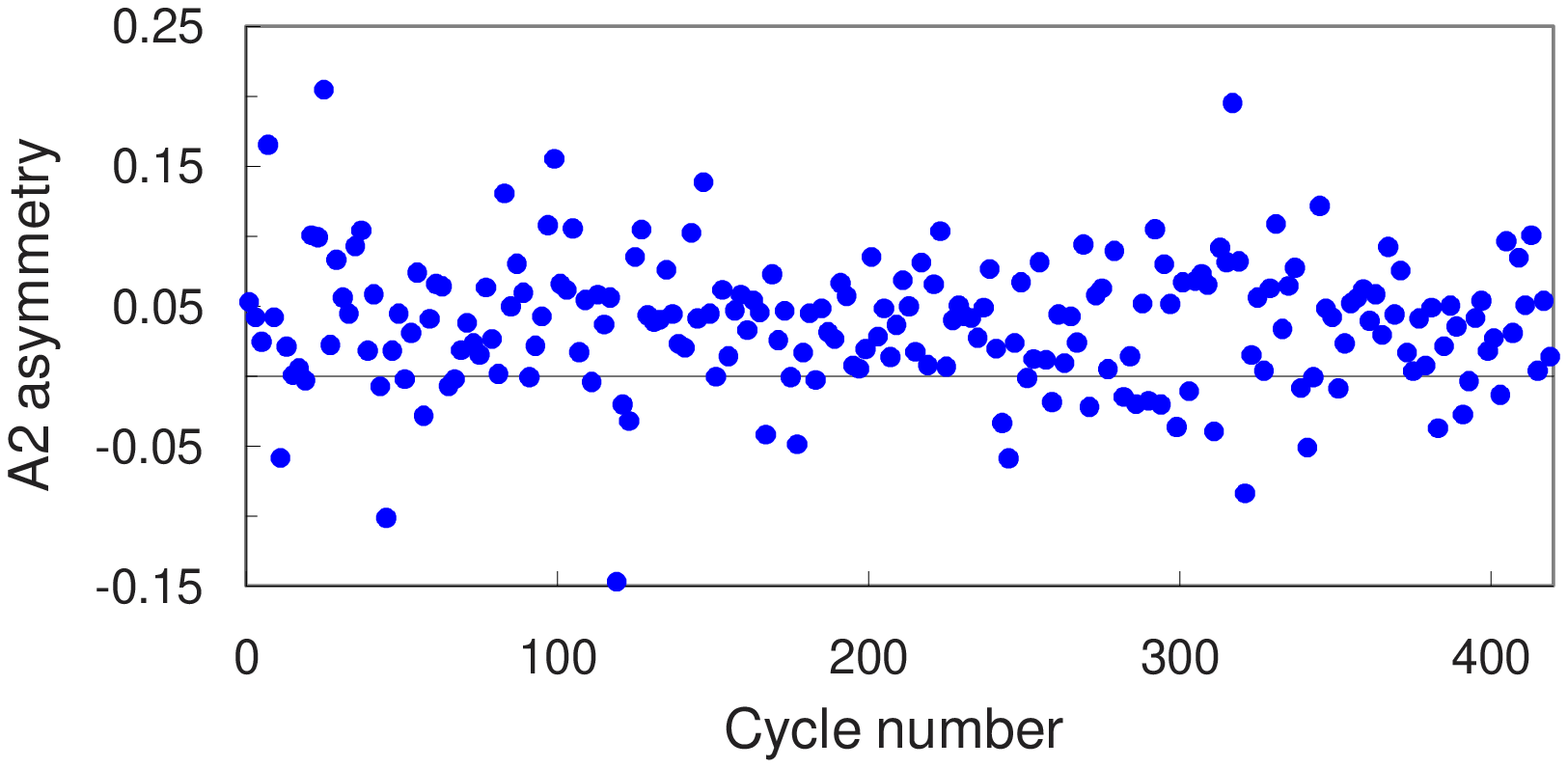}   

\vspace{0.1in}
    \includegraphics[width=8cm]{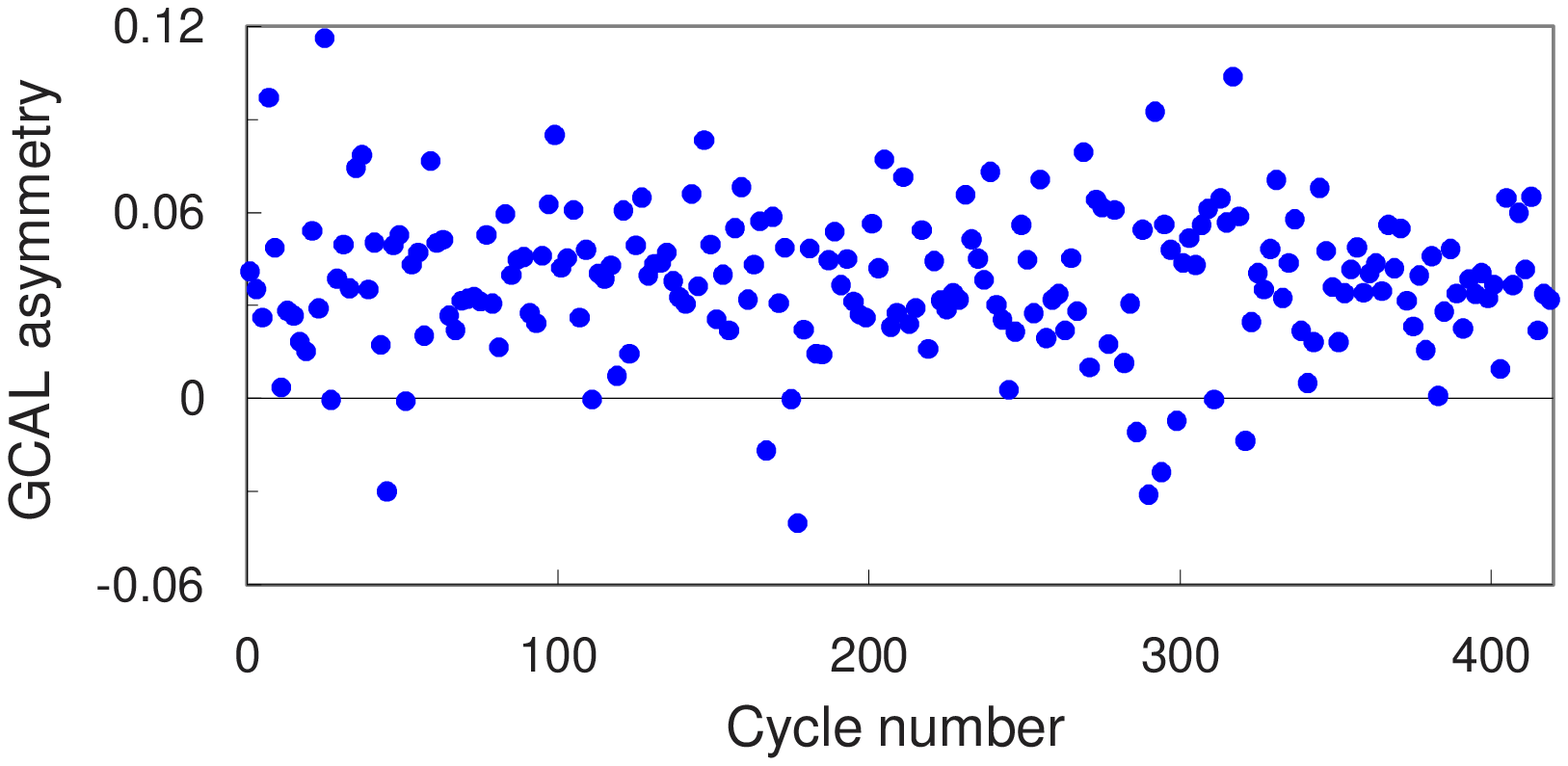}   
  \end{center}
  \caption{The asymmetries $\delta$ observed between adjacent cycles in detectors {\sf S2}, {\sf A2}, and {\sf GCAL}  during 420 cycles with spectrometer current $I_{\rm S} = 100\,A$.}
  \label{fig:as_s2GCALa2}
\end{figure}

The data shown in Fig.\:\ref{fig:trans_s2GCAL} illustrate the dependence of the rate of transmission of
photons through the iron-core magnet {\sf TP2} on the polarity of its field.
The transmission asymmetries observed in detectors  {\sf A2}, {\sf
  GCAL}, and {\sf S2} between adjacent cycles (with reversal of the
polarity of magnet {\sf TP2} on alternate cycles) during a sequence of
420 cycles during 42 {\sf Super} runs (Sec.~\ref{sec:datasample}) are shown in Fig.\:\ref{fig:as_s2GCALa2}.

 Average backgrounds were subtracted on a run-by-run basis. 
Runs with unstable beam conditions were eliminated by visual inspection of a standard set of plots, and noisy single events were removed by cuts on background detectors. 
Data for individual runs were normalized to incident photon intensity observed in detector {\sf S1}. 
This normalization was superior to that based on the electron beam current as the former included effects of undulator fluctuations and of steering of the photon beam through collimator {\sf C2}. 
Normalization to aerogel detector {\sf A1} gave similar results. 

\begin{figure}
  \centering
  \includegraphics[width=8cm]{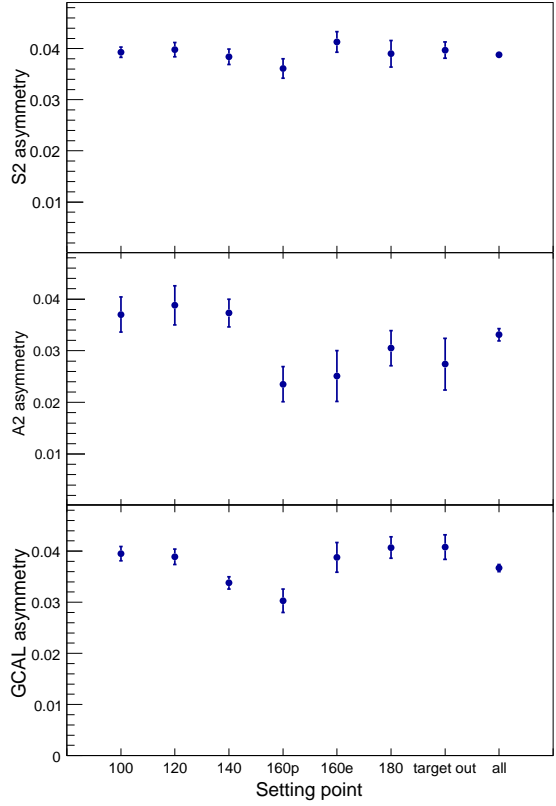}  
  \caption[The average asymmetry $\delta$ observed in the photon detectors {\sf S1}, {\sf A1}, and {\sf GCAL}.]{The average asymmetry $\delta$ observed in the photon detectors {\sf S1}, {\sf A1}, and {\sf GCAL} in the seven data segments (Table~\ref{tab:transport_sim}) of the experiment, as well as the asymmetries averaged over the entire experiment.}
  \label{fig:Photonanalysistext_images_3}
\end{figure}

Figure\:\ref{fig:Photonanalysistext_images_3}
shows the averaged asymmetries observed in the three detectors during seven positron-data segments (Table~\ref{tab:transport_sim}), as well as the overall asymmetries. 
Typically, a single data segment consisted of a few hundred cycles
from 30--50 {\sf Super} runs. 
The smallest uncertainties were obtained with detectors {\sf S2} and {\sf GCAL}; this was not due to an intrinsic limitation of the aerogel detector {\sf A2} but rather to insufficient attention to its PMT gain during data collection. 
The variation of the asymmetries observed in {\sf S2}, {\sf A2}, and {\sf GCAL} with spectrometer current was not predicted and is not understood.

Table\:\ref{tab:phot_asyms}
\begin{table}
\caption{Measured and predicted asymmetries $\delta$ of signals in the $\gamma$-detectors on reversal of the polarity of magnet {\sf TP2}, averaged over the entire experiment.
The uncertainties are statistical.
}
\label{tab:phot_asyms}
  \begin{center}
 \vspace{1mm}
\footnotesize
    \begin{tabular*}{6cm}{l@{\extracolsep\fill}cc}
      \hline
      \hline
      Detector & \multicolumn{2}{c}{Asymmetry}\\
      &  Measured  &  Predicted \\
      \hline
      {\sf A2} & 0.0331 $\pm$ 0.0012 & 0.0363 \\
      {\sf GCAL} & 0.0367 $\pm$ 0.0007 & 0.0347 \\
      {\sf S2} & 0.0388 $\pm$ 0.0006 & 0.0339 \\
      \hline
      \hline
    \end{tabular*}
  \end{center}
\end{table}
shows the average measured and predicted asymmetry for each detector for the entire run. 
The uncertainties listed are purely statistical.
The good agreement between the measured and simulated asymmetries indicates that the undulator produced the expected spectrum of photon polarization.

\section{Positron Polarization Analysis}\label{pos_anal}

The longitudinal positron polarization $P_{e^+}$ was derived from the asymmetry $\delta$, introduced in Secs.~\ref{sec_phot_pol_meth}--\ref{sec:posipol}, of the energy deposition $E^{\pm}$ in the CsI calorimeter with respect to the two polarization states (+ and $-$) of the iron-core magnet {\sf TP1},
\begin{linenomath}
\begin{equation}
  \delta = \frac{E^- - E^+}{E^- + E^+}
         = A_{e^+} P_{e^-}^{\rm Fe} P_{e^+},
\label{eq:delta_positron}
\end{equation}
\end{linenomath}
where the (positive) analyzing power $A_{e^+}$ was obtained from the
simulations described in Sec.~\ref{sec:pos_power}, and the average
longitudinal polarization $P_{e^-}^{\rm Fe}$ of atomic electrons in
the magnet iron was determined to be $0.0695 \pm 0.0015$ for $r \leq
22.5$\,mm (Table~\ref{tab_analmag_two}). 
The CsI energy signals $E^{\pm}$ were corrected for background and
normalized to the number of positrons observed in detector {\sf P1} at
the exit of the spectrometer {\sf D2}, as discussed below. 

The best results, in terms of highest signal-to-background ratio and least systematic uncertainty, were obtained using only the central CsI crystal, and only the analysis of the data from that crystal is described here. 
See \cite{Ralph} for details of the analysis of the other nine crystals.

\subsection{General Analysis Procedure}\label{sec:pos_ana_general}

The steps in the positron polarization analysis were:
\begin{itemize}
\item 
Energy calibration of the signals in the CsI crystals (Sec.~\ref{sec:csi_cal}).
\item 
Selection of events for each positron-energy setting that were recorded under stable beam and background conditions (Sec.~\ref{sec:ana_data_sel}).
The data structure has been described in Sec.\:\ref{sec:datasample}.
\item 
Subtraction of the background for every undulator-on event in each CsI crystal, and normalization of the signals in the calorimeter for each event to the number of positrons observed in detector {\sf P1} (Sec.~\ref{sec:background}).
\item 
Determination of the asymmetry at each positron energy from a fit to the asymmetries between pairs of adjacent cycles (Sec.~\ref{sec:asym_det}).
\item
Evaluation of systematic uncertainties in the asymmetry/polarization measurements (Sec.~\ref{sec:syserr}).
\item
Comparison of the asymmetries with simulation (Sec.~\ref{sec:asym_result}). 
\item 
Calculation of the positron (electron) polarizations from the asymmetries observed in the central CsI crystal (Sec.~\ref{sec:posi_pol}).

\end{itemize}

\subsection{Data Selection}\label{sec:ana_data_sel}

There were six spectrometer/lens settings for positron data (see Table\:\ref{tab:runlist}), and a test was made with the undulator filled with ferrofluid at the highest-current setting.
Electron data were taken at only one spectrometer setting, $I_{\rm S} = 160$\,A.
The two data sets at $I_{\rm S} = 140$\,A were taken at the beginning and at the end of the second run period in September 2005, respectively. 
The data set with $I_{\rm S} = 150$\,A was recorded during June 2005 and, unlike the other data sets, its cycles were not combined into runs.
Instead, every cycle was recorded individually and the polarity of magnet {\sf TP1} was reversed manually in between. Since detector {\sf P1} did not exist in the first running period, these data could not be normalized to the positron flux and do not enter in the final results. 
\begin{table*}
  \begin{center}
    \caption{Event samples (data segments) at various spectrometer
      settings, together with the corresponding numbers of events
      before and after the beam-current cuts described in the text, and the fraction $\eta$ of events which were used to determine the average asymmetries of the crystal signals. 
$I_{\rm S}$ is the current in spectrometer magnet {\sf D2} which defined the positron (electron) energy. 
The label ``ff" indicates the data set for which the undulator was run with ferrofluid.} 
    \label{tab:runlist}
 \vspace{1mm}
\footnotesize
    \begin{tabular*}{\textwidth}{c@{\extracolsep\fill}ccccccc}
      \hline
\hline
      I$_{S}$ &  $E_{e^\pm}$ & Particle & Number of & \multicolumn{3}{c}{Number of Events ($\cdot 10^{3}$)} & $\eta$ \\ 
      
      ~(A) & (MeV) & type & cycle pairs &  Before cuts & After cuts & Used in fits &      \\
      \hline
      100    & 4.59 & e$^{+}$ &       207 &1240 &       1202 &       958  & 0.800     \\
      120    & 5.36 & e$^{+}$ &       187 &1119 &       1082 &       867  & 0.804     \\
      140    & 6.07 & e$^{+}$(1)&    283 &1422 &       1368 &       1057 & 0.787     \\
      140    & 6.07 & e$^{+}$(2)&    220 &1320 &       1280 &       1015 & 0.795     \\
      150    & 6.41 & e$^{+}$ &      \;51 &224  &       215  &       169  & 0.791     \\
      160    & 6.72 & e$^{+}$ &       169 &1014 &       983  &       767  & 0.781     \\
      160    & 6.72 & e$^{-}$ &       145 &870  &       846  &       676  & 0.801     \\
      180    & 7.35 & e$^{+}$ &      \;65 &390  &       380  &       294  & 0.776     \\
      180    & 7.35 & e$^{+}$(ff) &  104 &624  &       601  &       481  & 0.803     \\
      \hline
      \multicolumn{3}{r}{$\sum $} &\multicolumn{1}{c}{1431} &\multicolumn{1}{c}{8223} &\multicolumn{1}{c}{7957} & \multicolumn{1}{c}{6284} & \multicolumn{1}{c}{}\\
      \hline
      \hline
    \end{tabular*}
  \end{center}
\end{table*}

\begin{figure}
  \centering
  \includegraphics[width=8cm]{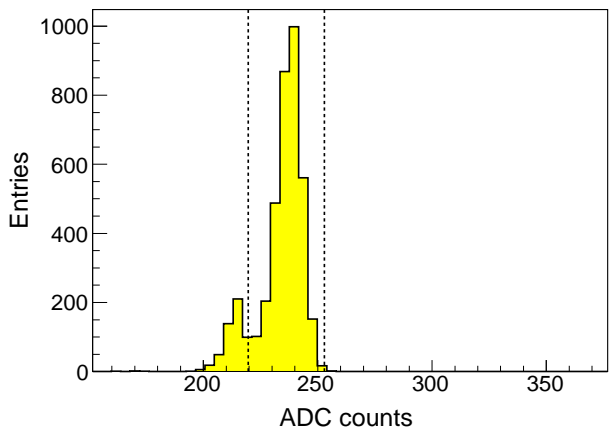}  
  \caption[Example of a distribution of the beam current in toroid {\sf BT}.]{Example of a distribution of the beam current in toroid {\sf BT} (see Fig.~\ref{fig:explayc2k}) that shows a fluctuation to low current.
Only events with current within the $\pm 3 \sigma$ cuts about the mean were used in the subsequent analysis. 
}
  \label{fig:toroid}
\end{figure}

In all steps of the analysis adjacent cycle pairs were treated in parallel. 
For example, both cycles of a pair should have similar and stable beam and background conditions. 
To assure stable beam conditions, only those events from both cycles
were selected that had a beam-current value, measured by toroid {\sf BT} (see Fig.~\ref{fig:explayc2k}), within $\pm 3 \sigma$ of the mean of that value during both cycles, as illustrated in Fig.\:\ref{fig:toroid}.
That example is not typical; in most cases the beam-current distribution was Gaussian with a $\sigma$ of 2-3\,\% about the mean.

\subsection{Background Subtraction, Normalization, and Energy Deposition in the Crystal} \label{sec:background}

\begin{figure}
  \centering
    \includegraphics[width=8cm]{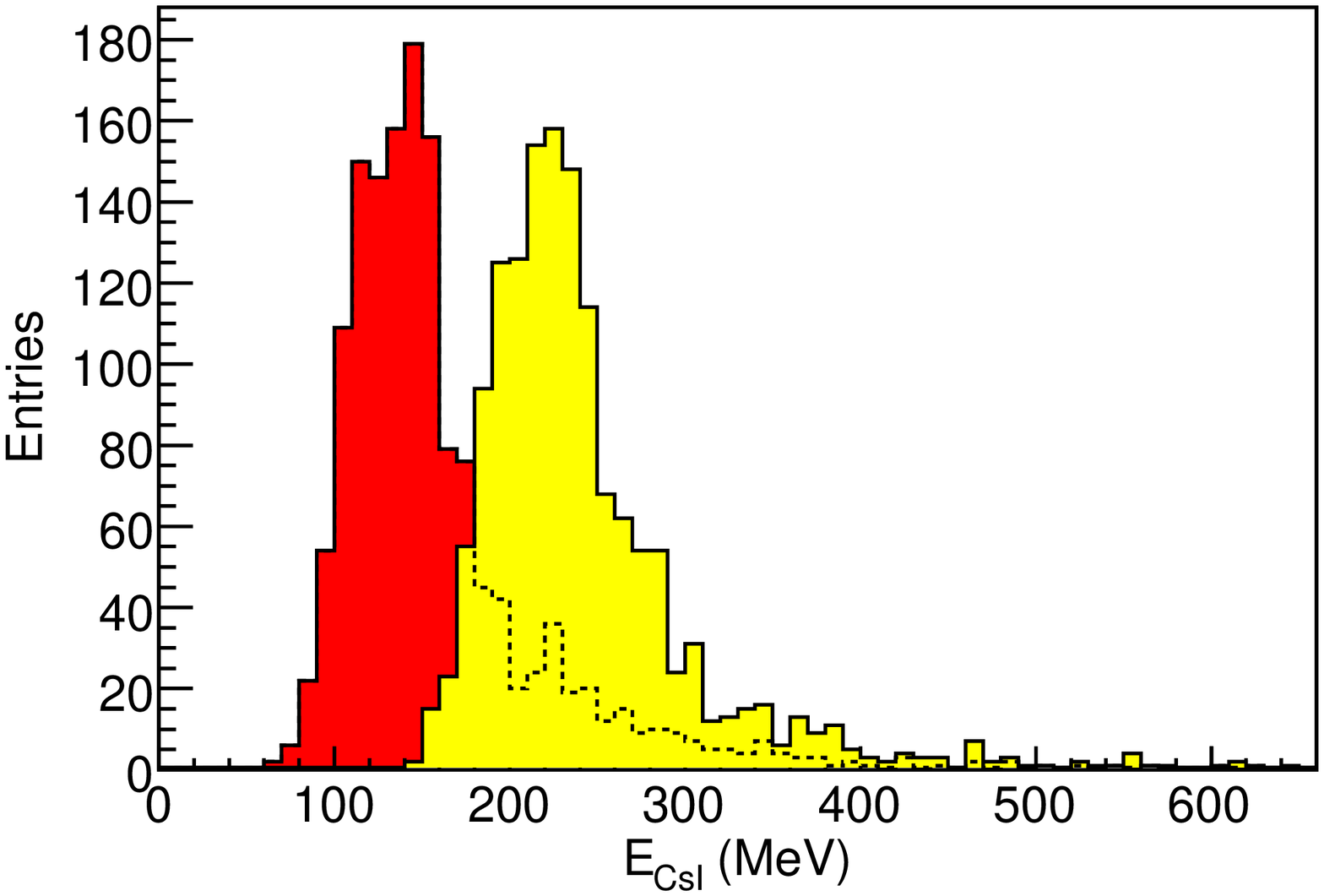}   
    \hfill
    \includegraphics[width=8cm]{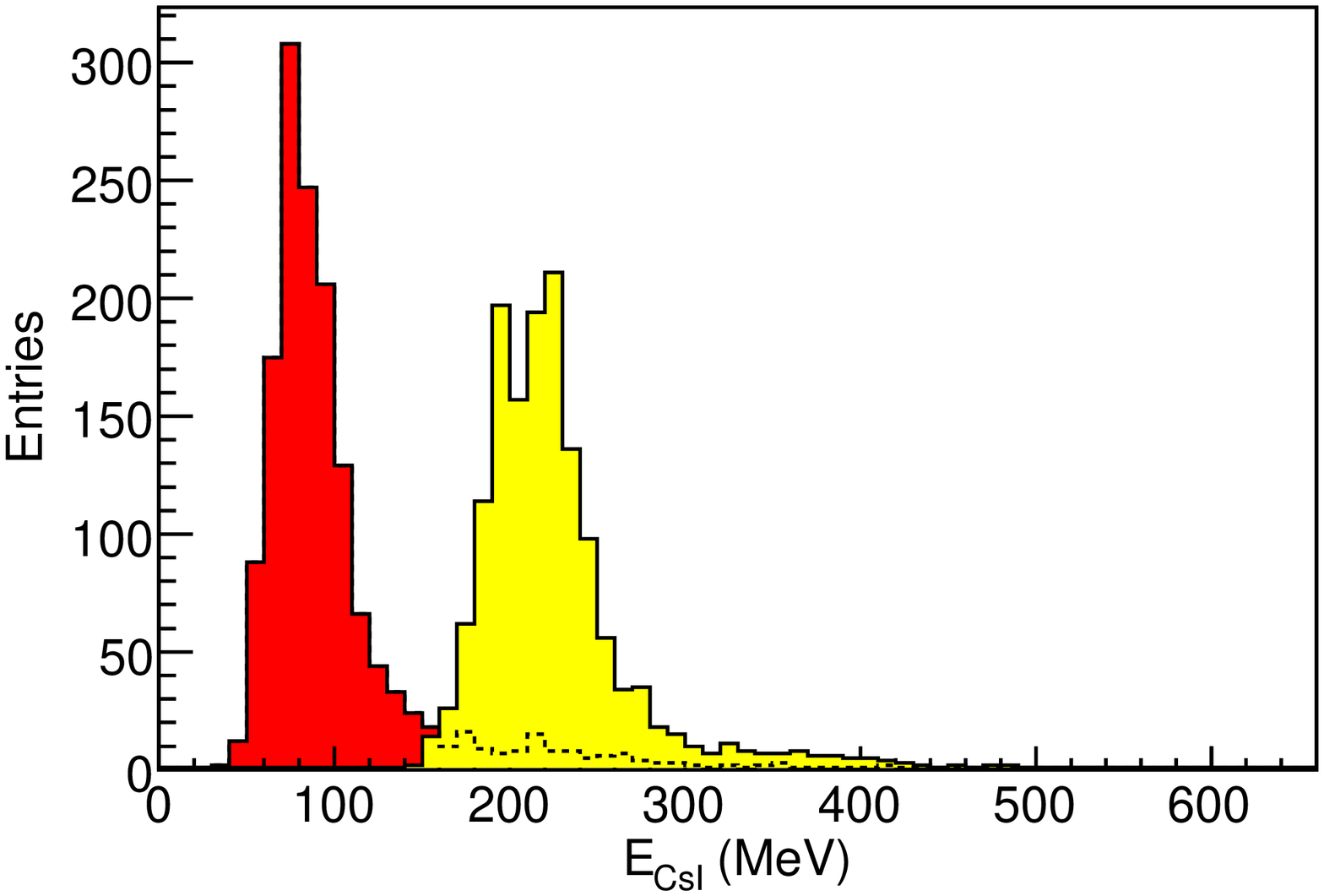}  
  \caption[Two examples of energy distributions for events in the central CsI crystal.]{Two examples of energy distributions for events in the central CsI crystal, taken with spectrometer current $I_{\rm S} = 160$\,A; (left) positrons, (right) electrons. 
Within each plot the left histogram is for undulator off (out of time) and the right histogram is for undulator on.}
  \label{fig:energy_s_b}
\end{figure}

As discussed in Sec.~\ref{sec:back}, backgrounds in the CsI crystals due to interactions of primary beam electrons were monitored by operation of the undulator out of time with the electron beam (``undulator off'') during alternate beam pulses. 
Figure\:\ref{fig:energy_s_b} gives examples of energies observed in the central CsI crystal during undulator-on and -off events.
The variation in the relative amount of background in different CsI crystals required determination of the asymmetries and polarizations, and their corresponding uncertainties,  separately for each crystal. 
No correlations of fluctuations in neighboring undulator-on and -off events were observed. 
Hence, there was no advantage to a background subtraction based only on neighboring events. 
Rather,
the energy deposition $E_{j}^{\rm off}$ in every undulator-off event of a cycle was subtracted from energy $E_{i}^{\rm on}$ of every undulator-on event of that cycle.  
The signal $I$ in the beam toroid {\sf BT} was used to correct for variation in the intensity of the primary electron beam from event to event.
This resulted in $n_{\rm on} n_{\rm off}$ background-subtracted energies of the form 
\begin{linenomath}
\begin{equation}
E_{i}^{\rm on}  - E_{j}^{\rm off} \frac{I_{i}^{\rm on}}{ I_{j}^{\rm off}}\, , 
\label{eq:subtraction2}
\end{equation}
\end{linenomath}
from a cycle of $n_{\rm on}$ undulator-on events and $n_{\rm off} \approx n_{\rm on}$ undulator-off events.

\begin{figure}
  \begin{center}
    \includegraphics[width=8cm]{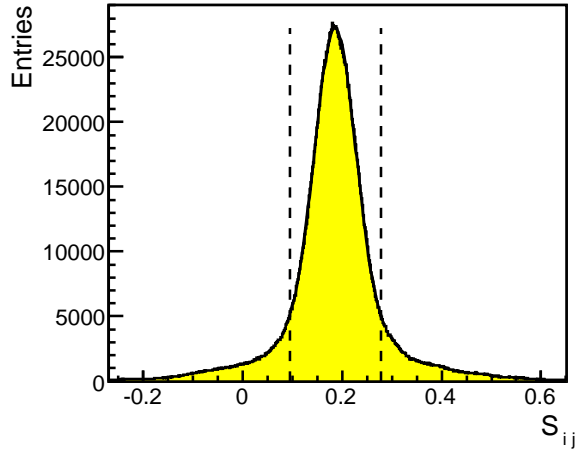}  
    \caption[The distribution of renormalized, background-subtracted energies $S_{ij}$ in the central CsI crystal for a typical data cycle.]{The distribution of renormalized, background-subtracted energies $S_{ij}$, defined in Eq.~(\ref{eq:subtraction}), in the central CsI crystal for a typical data cycle. 
The dotted lines indicate the $\pm 2\, \sigma$ fit range described in the text.}
    \label{fig:Efit3}
  \end{center}
\end{figure}

Signals in the silicon detector {\sf P1} (Sec.\:\ref{sec_reconversion}), just upstream of reconversion target {\sf T2}, showed a small asymmetry, $\delta^{\sf P1}$, with respect to the polarity of magnet {\sf TP1} (see Table\:\ref{tab:Asymmetries}).
This was probably caused by  the fringe field of magnet {\sf TP2} in conjunction with the misalignment of the positron beam (Sec.\:\ref{sec:sim_veri}). 
The asymmetries $\delta^{\sf P1}$ varied from $-3$ to $-16$\,\% of the asymmetries $\delta$ measured in the central CsI crystal for positrons, and was $+40$\,\% of the asymmetry for electrons. 
Correction for this effect was made by renormalizing the subtracted energy depositions (\ref{eq:subtraction2}) in the CsI crystal to the (subtracted and normalized by the beam toroid) signal in detector {\sf  P1}, defining renormalized, background-subtracted energies $S_{ij}$ according to
\begin{linenomath}
\begin{equation}
S_{ij} = \frac{E_{i}^{\rm on} - E_{j}^{\rm off} \frac{I_{i}^{\rm on}}{I_{j}^{\rm off}}}{{\sf P1}_{i}^{\rm on} - {\sf P1}_{j}^{\rm off} \frac{I_{i}^{\rm on}}{I_{j}^{\rm off}}}\, . 
\label{eq:subtraction}
\end{equation}
\end{linenomath}
An example of a distribution of renormalized, background-subtracted energies is shown in Fig.\:\ref{fig:Efit3}.
The multiple use of each event was accounted for in the computation of statistical uncertainties (Sec.~\ref{sec:asym_det}).

The distributions of the $S_{ij}$ had a roughly Gaussian core and relatively long tails to both sides. 
These tails were excluded by an iterative procedure in which each distribution of the $S_{ij}$  was fit to the sum of two Gaussian functions with common mean but different variances; entries were then cut outside a $\pm 2\, \sigma$ range defined by the narrower Gaussian, after which the mean of the remaining entries was determined by a new fit to two Gaussians.

Alternatively, the truncated mean over a cycle of the renormalized energy $S$ (per positron) in a CsI crystal could be defined as
 \begin{linenomath}
   \begin{equation}
  \ave{S}  = \ave{S_{\rm on} - S_{\rm off}}
   =  \frac{1}{n_{\rm on}n_{\rm off}} \sum_{i=1}^{n_{\rm on}}  \sum_{j=1}^{n_{\rm off}} S_{ij},
  \label{eq_SBmean}
 \end{equation}
 \end{linenomath}
where the sum is over only those $S_{ij}$ that pass the cut on the tails of their distribution.

If instead the entries in a distribution of background-subtracted events were the result of subtraction of undulator-off events only from the preceding undulator-on event, a cut on the tails of the distribution would exclude both the undulator-on and the undulator-off event if either one of them had a large fluctuation. 
The advantage of a procedure based on
Eqs.~(\ref{eq:subtraction})--(\ref{eq_SBmean}) compared to use of a
neighboring event subtraction was the ability to cut on large fluctuations in energies while retaining good statistical power from events with behavior near the mean.
Note that without cuts on fluctuations, the mean $\ave{S_{\rm on} - S_{\rm off}}$ in (\ref{eq_SBmean}) would be equal to $\ave{S_{\rm on}} - \ave{S_{\rm off}}$, and then would not be dependent on any signal-background ordering.

The effectiveness of this method and, in particular, the assignment of uncertainties to the  asymmetries derived therefrom, has been verified by Monte Carlo simulations. 
The mean of a truncated distribution as given by Eq.~(\ref{eq_SBmean}) was found to be a more stable measure of the average signal of a cycle. 

The statistical uncertainties on the means $\ave{S}$ of Eq.~(\ref{eq_SBmean}) were calculated from the full width at half maximum of the distribution of the $S_{ij}$ and the numbers of signal events, $n'_{\rm on}$, and of background events, $n'_{\rm off}$, that contributed to the truncated mean:
\begin{linenomath}
   \begin{equation}
  \Delta \ave{S} = \frac{\sigma_{\ave{S}}}{\sqrt{2}} \cdot \sqrt{\frac{1}{n'_{\rm on}} + \frac{1}{n'_{\rm off}}} \approx \frac{\sigma_{\ave{S}}}{\sqrt{n'}} \, ,
  \label{eq:Eerr1}
\end{equation}
\end{linenomath}
where $n' = n'_{\rm on} \approx n'_{\rm off}$ and $\sigma_{\ave{S}} = {\rm FWHM} / 2.35$ is the width parameter of the approximately Gaussian central distribution of the $S_{ij}$. 
Taking $\eta$ to be the fraction of entries in the distribution (\ref{eq:subtraction}) that survived the cuts
(see Fig.\:\ref{fig:Efit3}), the quantities $n'_{\rm on}$ and $n'_{\rm off}$ were defined as:
\begin{linenomath}
  \begin{equation}
    n'_{\rm on} = \eta\cdot n_{\rm on}\, , \qquad  n'_{\rm off} = \eta\cdot n_{\rm off}\, ,
  \end{equation}
\end{linenomath}
where $n_{\rm on}$ and $n_{\rm off}$ are the numbers of signal and background events of a cycle. For the central CsI crystal, $\eta$ was between 77 and $80\,\%$ (see Table\:\ref{tab:runlist}).

The systematic uncertainties were small, as confirmed by repeating the background subtraction with different methods and independent analyses. 
For example, an analysis was performed by first averaging (for each crystal) all signal and all background data independently, before the background was subtracted. 
The results obtained with this alternative procedure were fully consistent with the results presented here. 

\subsection{Asymmetry Determination}\label{sec:asym_det}

\begin{figure}[htbp]
  \centering
  \includegraphics[width=8cm]{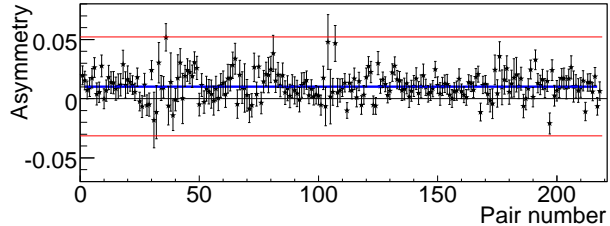} 
  \caption[The asymmetries $\delta_{k}$ observed for the 218 cycles pairs in the positron polarimeter for $I_{\rm S} = 140$\,A.]{The asymmetries $\delta_{k}$ observed for the 218 cycles pairs in the positron polarimeter for $I_{\rm S} = 140$\,A during the second running period.
 Also shown is the best-fit $\delta = \ave{\delta_k}$ (horizontal blue line) and the $\pm 3\sigma$ band used for
the cut on outliers.}
  \label{fig:140d}
\end{figure}

The asymmetries
\begin{linenomath}
  \begin{equation}
    \delta_{k} = \frac{S_{k}^{-} - S_{k}^{+}} {S_{k}^{-} + S_{k}^{+}}
    \;,
    \label{eq:asym1}
  \end{equation}
\end{linenomath}
were obtained from the renormalized, background-subtracted energy depositions $S^{-}_k$ and $S^{+}_k$ of Eq.~(\ref{eq_SBmean}) in the CsI calorimeter for $-$ and $+$ polarity of magnet {\sf TP1} during cycle pair $k$, and the corresponding uncertainties $\Delta \delta_{k}$ evaluated as
\begin{linenomath}
  \begin{equation}
    \Delta \delta_{k} =  \frac{2
    \sqrt{\left( S_{k}^{+} \;\Delta S_{k}^{-}\right)^2 + \left(
        S_{k}^{-} \;\Delta S_{k}^{+}\right)^2}}{ \left(S_{k}^{-} + S_{k}^{+}\right)^2} \;,
    \label{eq:asymerr1}
  \end{equation}
\end{linenomath}
where $\Delta S_{k}^{-}$ and $\Delta S_{k}^{+}$ were determined according to\:(\ref{eq:Eerr1}). 
As an example, the asymmetries obtained for all cycle pairs of one spectrometer setting are shown in Fig.\:\ref{fig:140d}.

The best-fit value $\delta$ for the asymmetry for each positron energy (setting of spectrometer {\sf D2}) was determined by fitting a constant to the $\delta_{k}$, taking their uncertainties into account. 
To exclude possible outliers, only asymmetries within a $\pm 3\sigma$ range were used in the fit. 
The effect of outliers was small, and was included in the estimate of the systematic uncertainties (Sec.~\ref{sec:syserr}). 

The best-fit asymmetries $\delta$ with uncertainties $\Delta \delta$, and the $\chi^{2}$ per degree of freedom of the fits are listed in Table\:\ref{tab:Asymmetries}.  
The uncertainty $\Delta \delta$ is smaller than the standard deviation $\sigma$ of the asymmetries $\delta_k$ prior
to the exclusion of outliers.
Also listed are the asymmetries $\delta^{\sf P1}$ observed in the detector {\sf P1}, just upstream of magnet {\sf TP1}.
A separate study showed that if the normalization to detector {\sf P1} were not done as in Eq.~(\ref{eq:subtraction}), then the asymmetries $\delta$ would have been roughly the sum of the asymmetries $\delta$ and $\delta^{\sf P1}$ given in Table\:\ref{tab:Asymmetries}. 
\begin{table*}
  \caption{Best-fit asymmetries $\delta$ of Eq.~(\ref{eq:asym1}) (with respect to the polarity of magnet {\sf TP1}) measured in the central CsI crystal for all settings of spectrometer {\sf D2}, together with the $\chi^2$ per degree of freedom of the fits.
Also listed are the asymmetries $\delta^{\sf P1}$ observed in detector
{\sf P1}.}%
  \label{tab:Asymmetries}
  \centering
 \vspace{1mm}
\footnotesize
  \begin{tabular*}{11cm}{l@{\extracolsep\fill}ccccc}
    \hline
    \hline
    I$_{S}$ &  E$_{tot}$(e$^{\pm}$) & $\delta \pm \Delta\delta$ & $\chi^{2} / n_{\rm df}$ & $\delta^{\sf P1} \pm \Delta\delta^{\sf P1}$  \\
(A) & (MeV) & (\%) & & (\%) \\
    \hline
    100              & 4.59 & 0.689  $\pm$    0.165    &    166.4   /       203 & $-$0.115  $\pm$   0.017  \\
    120              & 5.36 & 0.961  $\pm$    0.083    &    167.9   /       184 & $-$0.066  $\pm$   0.020  \\
    140(1)           & 6.07 & 1.197  $\pm$    0.145    &    404.1   /       276 & $-$0.107  $\pm$   0.110  \\
    140(2)           & 6.07 & 1.079  $\pm$    0.063    &    274.9   /       217 & $-$0.055  $\pm$   0.019  \\
    140(1+2)         & 6.07 & 1.132  $\pm$    0.061    &    628.2   /       490 & $-$0.095  $\pm$   0.034  \\
    160(e$^{+}$)      & 6.72 & 0.923  $\pm$    0.080    &    257.9   /       166 & $-$0.035  $\pm$   0.023  \\
    160(e$^{-}$)      & 6.72 & 0.938  $\pm$    0.051    &    167.2   /       144 & $+$0.382   $\pm$   0.022  \\
    180              & 7.35 & 0.886  $\pm$    0.196    &    59.8    /       63  & $-$0.108  $\pm$   0.063  \\
    180(ff)          & 7.35 & 0.995  $\pm$    0.140    &    174.2   /       101 & $-$0.081  $\pm$   0.055  \\
    \hline
    \hline
  \end{tabular*}
\end{table*}

\subsection{Systematic Uncertainties}\label{sec:syserr}

\begin{table}
    \centering
    \caption{Estimates of maximum systematic uncertainties of the asymmetries $\delta$ (Table\:\ref{tab:Asymmetries}) from various sources.}
    \label{tab:Es_Syst_Asym}
 \vspace{1mm}
\footnotesize
    \begin{tabular*}{8cm}{l@{\extracolsep\fill}lc}
      \hline
      \hline
      & Source &  Relative \\
       &        & uncertainty \\
      \hline
      a)  &      Non-stat. fluctuations per cycle pair & 7\,\% \\
      b)  &      Outlier rejection ($\pm 3\sigma$) & 8\,\% \\
      c)  &      Beam-current cut & 5\,\% \\
      d)  &      Cycle pairing & 9\,\% \\
      e)  &      Background correction & 7\,\% \\
      f)  &      Stray-field-induced asymmetry in {\sf P1} & 5\,\,\% $(e^+)$ \\
          &                                          & 12\,\% $(e^-)$ \\
      \hline
      \hline
    \end{tabular*}
  \end{table}

The calculation of the asymmetries $\delta$ assumed that the measurements of energies $E^+$ and $E^-$ in the central CsI crystal for polarities + and $-$ of magnet {\sf TP1} were performed under identical conditions. 
Differences in those conditions could lead to systematic uncertainties in the asymmetries, whether or not corrections were made for the differences.
The various sources of systematic uncertainty on the measurements of the electron/positron asymmetries $\delta$ that were considered are listed in Table\:\ref{tab:Es_Syst_Asym}.   
Estimates of the systematic uncertainties were made separately for each data set reported in Table\:\ref{tab:Asymmetries}, and the maximum uncertainty for any data set is listed in Table\:\ref{tab:Es_Syst_Asym}.
These uncertainties were later combined with those on the analyzing power $A_{e^\pm}$ and on the polarization $P^{\rm Fe}_{e^-}$ of the iron in magnet {\sf TP1} to evaluate the systematic uncertainties in the polarization $P_{e^\pm}$ (Sec.~\ref{sec:posi_pol}). 
\begin{list}{}{\leftmargin 0.05\columnwidth \labelwidth 0.95\columnwidth}
\item[a)]
A measure of non-Gaussian fluctuations in the data is that the $\chi^{2}$ per degree of freedom for the fits to the
asymmetries $\delta$ was larger than 1 in some cases.
Since for each energy point the individual asymmetries $\delta_i$ were statistically independent, a $\chi^{2}$ test can be performed on the statistical nature of the fluctuations. 
To obtain $\chi^{2}/{\rm ndf} \leq 1$ in all cases the uncertainties on the $\delta_i$ must be scaled up by as
much as 7\%, which was taken as the maximum systematic uncertainty due to non-Gaussian statistics.

\item[b)]
The best-fit asymmetries $\delta$ were also calculated without making the cut on ``outliers'', as described in Sec.~\ref{sec:asym_det}.
The resulting changes in the $\delta$s were up to 8\%, which was taken as the systematic uncertainties associated with
the ``outlier'' cut.

\item[c)]
  The influence of the quality cuts on the beam current ($\pm 3\sigma$, Sec.~\ref{sec:ana_data_sel}) and on the background-subtracted energy distributions ($\pm 2\sigma$, Sec.\:\ref{sec:background}) was investigated. 
Variation of these cuts between zero and their nominal values led to changes in the determined asymmetry of up to 5\,\%,
which was taken as the systematic uncertainties associated with these cuts.

\item[d)]
The combination of cycles into pairs is arbitrary, in principle, although our practice of combining adjacent cycles with ``+'' and ``$-$'' magnet polarity may have reduced effects of long-term changes. 
Studies in which cycles were paired at random showed relative changes in the asymmetries of up to 9\,\%, which was taken as the systematic uncertainties associated with cycle pairing.

\item[e)]
To study whether the backgrounds in signal (undulator-on) and background (undulator-off) events were the same,  
target-out data (Sec.\:\ref{sec_und_back}) were used to compare the undulator-on and -off energies in the central CsI crystal.  
The contribution to the systematic uncertainty in the asymmetry from this was thereby estimated to be at most 7\%.

\item[f)]
The relative systematic uncertainty due to the renormalization
(\ref{eq:subtraction}) of the CsI-crystal energies to the  
signal in detector {\sf P1} was taken to be 0.3 of the ratio of the
{\sf P1} asymmetry to the asymmetry $\delta$. 
\end{list}

Two other systematic effects entered in the determination of the final positron polarization: the uncertainty of the electron polarization in the core of the analyzer magnet of about 3\,\% (Sec.~\ref{sec_analyzer_magnet}), and the systematic uncertainty of 7\,\% in the simulation of the analyzing power (Sec.~\ref{sec:pos_power}).

\subsection{Asymmetry Results}\label{sec:asym_result}

\begin{figure}[htbp]
  \begin{center}
    \includegraphics[width=8cm]{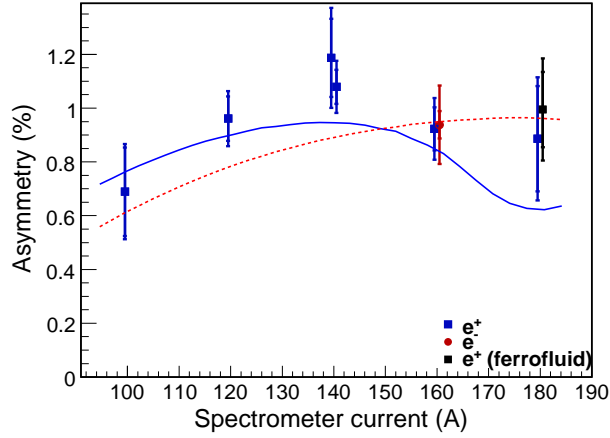}  
    \caption[The asymmetries $\delta$ and their total uncertainties measured with the central crystal of the CsI calorimeter.]{The asymmetries $\delta$ and their total uncertainties (statistical and systematic combined in quadrature) measured at different spectrometer settings with the central crystal of the CsI calorimeter. 
The solid curve (positrons) and dashed curve (electrons)  are from simulations described in Sec.~\ref{sec:delta_sim}.
}
    \label{fig:sum_asy} 
  \end{center}
\end{figure}

The  results of the asymmetry analysis are summarized in Table\:\ref{tab:sum_asy}. 
Only the asymmetries for the central CsI crystal (Fig.\:\ref{fig:sum_asy}) are reported here, as these are the most significant with respect to the combination of statistical and systematic errors.
However, within these uncertainties the measured asymmetries are consistent with expectations from simulation
for all crystals (Sec.~\ref{sec:delta_sim}).   

\subsection{Positron Polarization}\label{sec:posi_pol} 

The longitudinal polarization of the positrons (electrons) was determined from the relation
\begin{linenomath}
  \begin{equation}
    P_{e^\pm} =\frac{\delta}{A_{e^\pm}P_{e^-}^{\rm Fe}}\, ,
    \label{eq_pol}
  \end{equation}
\end{linenomath}
where the analyzing power $A_{e^\pm}(E)$ was determined for different positron energies $E$ by simulation (Sec.\:\ref{sec_simulation}) and the effective polarization of the electrons in the iron of magnet {\sf TP1} was $P_{e^-}^{\rm Fe} = 0.0695\,\pm\,0.0021$ (Table~\ref{tab_analmag_two} and Sec.~\ref{sec:syserr}).
The results are listed in Table\:\ref{tab:sum_asy}.
\begin{table*}
  \caption[Asymmetry ($\delta$), analyzing power ($A$), and polarization ($P$) determined from the central CsI crystal
at different spectrometer settings.]{Asymmetry ($\delta$), analyzing power ($A$), and polarization ($P$) determined from the central CsI crystal
at different spectrometer settings. 
    The systematic uncertainties are the combinations of all systematic effects described in Sec.\:\ref{sec:syserr}.}
  \label{tab:sum_asy}
  \centering
 \vspace{1mm}
\footnotesize
  \begin{tabular*}{13cm}{c@{\extracolsep\fill}cr@{\extracolsep{2pt}}c@{\extracolsep{2pt}}c@{\extracolsep{2pt}}c@{\extracolsep{2pt}}l@{\extracolsep\fill}cr@{\extracolsep{2pt}}c@{\extracolsep{2pt}}c@{\extracolsep{2pt}}c@{\extracolsep{2pt}}l}
    \hline
    \hline
    $I_{\rm S}$ & $E_{\rm tot}$ & \multicolumn{5}{c}{$\delta \pm \Delta \delta_{\rm stat} \pm \Delta \delta_{\rm syst}$ } & $A \pm \Delta A_{\rm stat}$  &  \multicolumn{5}{c}{$P \pm \Delta P_{\rm stat} \pm \Delta P_{\rm syst}$ } \\
(A) & (MeV) &&& (\%) &&& (\%) &&& (\%) \\
    \hline
    100        &   4.59 &      0.69    &       $\pm$   &       0.17    &       $\pm$   &       0.06    &       0.1498  $\pm$   0.0016  &       66.22   &       $\pm$   &       15.90   &       $\pm$   &       7.87    \\
    120        &   5.36 &       0.96    &       $\pm$   &       0.08    &       $\pm$   &       0.06    &       0.1563  $\pm$   0.0015  &       88.50   &       $\pm$   &       7.70    &       $\pm$   &       8.62    \\
    140(1)        &  6.07 &     1.20    &       $\pm$   &       0.15    &       $\pm$   &       0.12    &       0.1616  $\pm$   0.0014  &       106.56  &       $\pm$   &       12.98   &       $\pm$   &       13.14   \\
    140(2)        &  6.07 &     1.08    &       $\pm$   &       0.06    &       $\pm$   &       0.07    &       0.1616  $\pm$   0.0014  &       96.10   &       $\pm$   &       5.69    &       $\pm$   &       9.80    \\
    140(1+2)       &  6.07 &     1.13    &       $\pm$   &       0.06    &       $\pm$   &       0.08    &       0.1616  $\pm$   0.0014  &       100.75  &       $\pm$   &       5.54    &       $\pm$   &       10.23    \\
    160(e$^{+}$)        &  6.72 &     0.92    &       $\pm$   &       0.08    &       $\pm$   &       0.08    &       0.1651  $\pm$   0.0013  &       80.47   &       $\pm$   &       7.02    &       $\pm$   &       9.40    \\
    160(e$^{-}$)        &  6.72 &     0.94    &       $\pm$   &       0.05    &       $\pm$   &       0.14    &       0.1528  $\pm$   0.0014  &       88.29   &       $\pm$   &       4.86    &       $\pm$   &       14.55    \\
    180        &  7.35 &     0.89    &       $\pm$   &       0.20    &       $\pm$   &       0.12    &       0.1686  $\pm$   0.0013  &       75.58   &       $\pm$   &       16.72   &       $\pm$   &       11.60   \\
    180(ff)        &   7.35 &    1.00    &       $\pm$   &       0.14    &       $\pm$   &       0.13    &       0.1686  $\pm$   0.0013  &       84.91   &       $\pm$   &       11.99   &       $\pm$   &       12.69   \\
    \hline
    \hline
  \end{tabular*}
\end{table*}
Both the statistical and systematical uncertainties in the measured polarizations were determined by the usual error propagation from the uncertainties of the quantities entering\:\eqref{eq_pol}, $\delta$, $A_{e^+}$, and $P_{e^-}^{\rm Fe}$.

\begin{figure}
  \centering
  \includegraphics[width=8cm]{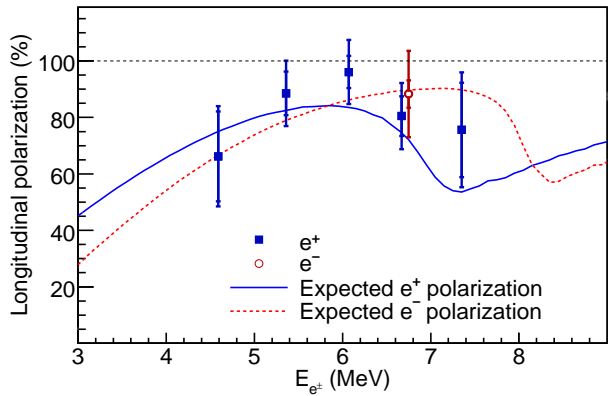} 
  \caption[Longitudinal polarization for positrons and electrons as
  deduced from the asymmetries in the central CsI crystal.]{Longitudinal polarization for positrons and electrons as deduced from the asymmetries (\ref{eq:asym1})
in the central CsI crystal only. 
The measurements are compared to the predictions from simulations.}
  \label{fig:DegPolarization_central}
\end{figure}

Comparison in Fig.\:\ref{fig:DegPolarization_central} of the measured polarization as a function of energy with simulation (Sec.~\ref{sec:delta_sim}) demonstrates good agreement with expectations.
Figure\:\ref{fig:DegPolarization_central} does not include the result of the first run at $I_{\rm S} = 140$\, A during which the beam
conditions were somewhat unstable and the resulting uncertainty large, nor the run with ferrofluid in
the undulator, which altered the distributions of photon and positron energy and polarization somewhat.
%
The one measurement of electron polarization also agrees with expectations and thus confirms that systematic effects, which could be different for electrons and positrons, were under control in the analysis.

\section{Summary}

Experiment E166 has successfully demonstrated a method for producing polarized positrons suitable for the next generation of linear colliders. 
The method uses a high-energy electron beam in conjunction with a short-period, helical undulator to produce circularly polarized photons with energies of several MeV. 
The photons are converted in a thin target to generate longitudinally polarized positrons. 
The tools, techniques, and methodologies developed for the experiment are directly applicable to the design of polarized positron sources for the ILC and for CLIC.

The experiment was carried out with a one-meter-long, 400-period, pulsed helical undulator installed in the Final Focus Test Beam (FFTB) at SLAC. 
A low-emittance, 10\,Hz, 46.6\,GeV electron beam passing through this undulator generated circularly polarized photons in the MeV range, with a cutoff energy of the first harmonic at about 7.9\,MeV.

These polarized photons were then converted to polarized positrons (and electrons) via pair production in a thin tungsten target. 
The experiment measured the flux and polarization of the undulator photons, and the spectrum and polarization of the positrons generated in the production target. 
The photon  polarization was determined by measuring the transmission asymmetry in polarized iron. 
To determine the positron polarization the same method was applied to photons generated by the positrons in a ``reconversion target''. 

After installation and commissioning of the experiment in the FFTB beam line in 2004, data were collected in two run periods of about 4 weeks each in June and September 2005.

For a detailed comparison of the results with predictions, the \geant\ simulation package has been extended to include all relevant polarization effects of electrons, positrons, and photons in matter.  
Thus, experiment E166 directly tested the validity of the software used to simulate the physics of polarized pair production in matter and the subsequent polarization-dependent electromagnetic interaction processes in finite thicknesses of matter. 
This resulting improvement in the simulations leads to greater confidence in the proposed designs of polarized positron sources for the next generation of linear colliders.

Positron polarization was measured at five energy settings of the analyzing spectrometer. 
In addition, an electron polarization measurement was done at a single energy setting by reversing the polarity of the spectrometer. 
Over the measured energy range of 4--8\,MeV, the positron (and electron) polarization was about 80\,\% with a relative measurement error of about 10\,\% to 15\,\%, dominated by the systematic uncertainties. 
The measured polarization values agree well with expectations from detailed
simulations.

The E166 experiment sucessfully demonstrated the technique of
undulator-based production of polarized positrons. The observed
performance of undulator, collection, and analyzing systems as well as
the photon and positron polarization measurements serve to
validate the underlying design methodologies.  
The modeling tools developed for E166 are now being used 
to design polarized positron sources for the next generation of
linear colliders.  

\section*{Acknowledgments}

The authors gratefully acknowledge the general support of the DESY research
division and the particular contributions of Y.~Holler and
A.~Petrov at DESY/Hamburg and M.~Jablonski from Humboldt University;
and the support and efforts of the entire SLAC staff, and in
particular the assistance of S.~Anderson, A.~Baker, L.~Bentson,
B.~Brugnoletti, F.~Gaudreault, H.~Imfeld, J.~Minister, M.~Racine,
R.~Rogers, N.~Spencer, K.~Traeger, and H.~Vincke. 



\end{document}